\documentclass[mnsc,nonblindrev]{informs3noheader}

\OneAndAHalfSpacedXI 

\newcommand{\bE}{\mathbb{E}}

\newcommand{\bI}{\mathbbm{1}}

\newcommand{\ALG}{\mathsf{ALG}}
\newcommand{\OPT}{\mathsf{OPT}}

\newcommand{\EXP}{{\mathbb{E}}}

\newcommand{\cS}{\mathcal{S}}
\newcommand{\cA}{\mathcal{A}}
\newcommand{\Val}{\mathsf{Val}}
\newcommand{\cO}{\mathcal{S}^{\mathsf{Alg}}}
\newcommand{\oS}{\overline{S}}

\newcommand{\ssx}{x^*}

\newcommand{\tU}{\tilde{U}}

\newcommand{\type}{\mathsf{Type}}
\newcommand{\avail}{\mathsf{Avail}}
\newcommand{\offered}{\mathsf{Offer}}
\newcommand{\Toffered}{\mathsf{True\hbox{-}offer}}
\newcommand{\accept}{\mathsf{Accept}}
\newcommand{\imatch}{\mathsf{IMatch}}
\newcommand{\cmatch}{\mathsf{CMatch}}
\newcommand{\seen}{\mathsf{Seen}}
\newcommand{\timeout}{\mathsf{Timeout}}

\newcommand{\edge}{\mathrm{edge}}
\newcommand{\vertex}{\mathrm{vertex}}

\graphicspath{{Figures/}}

\usepackage{natbib,bbm,multirow,multicol}
 \bibpunct[, ]{(}{)}{,}{a}{}{,}%
 \def\bibfont{\small}%

\usepackage{algorithm}
\usepackage[noend]{algpseudocode}
\usepackage{subfig}
\usepackage{multirow}
\usepackage{array}

\usepackage[dvipsnames]{xcolor}

\TheoremsNumberedThrough     
\ECRepeatTheorems

\EquationsNumberedThrough    

\begin{document}


\RUNAUTHOR{Fata, Ma, and Simchi-Levi}

\RUNTITLE{Multi-stage and Multi-customer Assortment Optimization with Inventory Constraints}

\TITLE{Multi-stage and Multi-customer Assortment Optimization with Inventory Constraints}

\ARTICLEAUTHORS{
\AUTHOR{Elaheh Fata}
\AFF{Department of Aeronautics and Astronautics, Massachusetts Institute of Technology, \texttt{efata@mit.edu}} 
\AUTHOR{Will Ma}
\AFF{Decision, Risk, and Operations Division, Graduate School of Business, Columbia University, \texttt{wm2428@gsb.columbia.edu}}
\AUTHOR{David Simchi-Levi}
\AFF{Institute for Data, Systems, and Society, Department of Civil and Environmental Engineering, and Operations Research Center, Massachusetts Institute of Technology, \texttt{dslevi@mit.edu}}
} 

\ABSTRACT{
We consider an assortment optimization problem where a customer chooses a single item from a sequence of sets shown to her, while
limited inventories constrain the items offered to customers over time.
In the special case where all of the assortments have size one, our problem captures the online stochastic matching with timeouts problem.
For this problem, we derive a polynomial-time approximation algorithm which earns at least $1-\ln(2-1/e)$, or 0.51, of the optimum.
This improves upon the previous-best approximation ratio of 0.46, and furthermore, we show that it is tight.
For the general assortment problem, we establish the first constant-factor approximation ratio of 0.09 for the case that different types of customers value items differently, and an approximation ratio of 0.15 for the case that different customers value each item the same.
Our algorithms are based on rounding an LP relaxation for multi-stage assortment optimization, and improve upon previous randomized rounding schemes to derive the tight ratio of $1-\ln(2-1/e)$.
}



\maketitle

%


\section{Introduction}\label{sec:intro}
In the assortment optimization problem, a firm decides on a set of items to offer to a customer for sale.
One of the key considerations is cannibalization, where the firm might not offer an item that serves as a substitute for a different, more profitable item.
In the traditional assortment optimization problem, the customer is presented with a single assortment, and chooses an item to purchase (or no item at all) based on all her options. With the advent of e-commerce, more sophisticated interactions with the customer that occur over multiple \textit{stages} have become possible.
This has led to a booming literature for modeling these interactions, which we review in Section~\ref{sec::relatedWork}.
In this work, we closely follow the model from \citet{liu2019managing,liu2019assortment}, which we now describe.
An arriving customer is interested in purchasing a specific category of items, e.g.\ a shirt.
The e-tailer shows her an assortment of shirts on the first web/mobile page.
If the customer sees a shirt she is satisfied with, then she purchases it and leaves the system.
Otherwise, she proceeds to the next page, where the e-tailer shows her a new assortment of shirts. The different assortments shown to a customer can be overlapping or non-overlapping, and both are actively used in different online advertising campaigns. Some advertisers, such as video games developers, might prefer to offer an item at most once to a customer as they believe if the customer is not interested in that video game then she would not purchase it regardless of the other video games that accompany it in that stage of advertising. On the other hand, as shown by~\citet{chernev'06} and \citet{chernev'12}, for some category of products, such as jewelry items, advertisers might prefer to show an item multiple times to a customer via different assortments. This is because these advertisers believe that the assortment through which an item is presented can affect the possibility that a customer purchases that item. The customer then leaves the system once she either purchases an item, or runs out of patience to view more pages.

While the aforementioned literature has led to insightful models backed by empirical evidence,
it focuses on the optimization problem for a single customer, and ignores the original revenue management problem of how to control a firm's limited item inventories over time (see \citet{talluri2006theory}).
In this work, we study a multi-stage, multi-customer assortment optimization problem, where limited inventories are offered to multiple customers over a finite time horizon, and furthermore can be offered to those customers in multiple stages.
This captures both demand-facing and supply-facing considerations for an e-tailer, where the assortments are optimized to reduce both the cannibalization within a single customer's demand, and also the ``cannibalization'' of inventory supplies that are better suited for another customer.

We also study the special case where all of the assortments have size 1.
In this case, the firm shows the items to the customer one-at-a-time,
which gives the firm increased control and is also natural in some applications such as online dating.
This special case corresponds to the \textit{online stochastic matching with timeouts} problem, which has been studied in a stream of work \citep{bansal2012lp,adamczyk2015improved,brubach2017attenuate} originating from \citet{chen2009approximating}.

In our model, we assume that the arrival process is \textit{known and stationary}, although we discuss where this assumption can be relaxed.
That is, there is a finite universe of customer types, and the probability distribution for the type of an arriving customer is known and unchanging over time.
This is the simplest arrival process to consider, which is also assumed throughout the work in online stochastic matching.
In our case, a customer type can be interpreted as a customer segment (e.g.\ females from LA aged 25--35),
and is assumed to have a known choice model (estimated from historical data) as well as a patience level indicating the number of stages over which assortments can be offered to that customer type.
Different customer types are also allowed to have different willingness-to-pay for the same item.
The firm's objective is to maximize the total revenue it expects to earn over the time horizon, through a policy which dynamically computes assortments to offer while respecting inventory constraints. 

This problem can be solved using dynamic programming, but that is impractical due to the curse of dimensionality; in fact, \citet{liu2019assortment} show that even the single-customer problem is NP-hard.
Therefore, we seek polynomial-time approximation algorithms which have performance guarantees relative to the optimal dynamic program.
In fact, our guarantees are relative to a stronger Multi-stage Choice-based Deterministic Linear Program (MCDLP) benchmark which we introduce.
The MCDLP provides an upper-bound on the revenue of any feasible policy, and generalizes the existing CDLP by allowing for assortments to be offered in multiple stages with the constraint that these assortments must respect patience levels.

\subsection{Contributions}
\label{subsec:contribution}

We derive two main sets of algorithms both based on rounding an optimal MCDLP solution.
First, in the special case of online stochastic matching with timeouts, our algorithm imitates a scaled copy of the LP solution using a randomized rounding scheme.
We improve previous schemes by increasing the likelihood that items with lower sales probabilities are offered first, and then use the framework of \citet{brubach2017attenuate} to bound the probability that each item has positive remaining inventory.
Under some additional technical assumptions (see Section~\ref{sec::single-item}),
this leads to an approximation ratio of $1-\ln(2-1/e)\simeq0.51$ relative to the MCDLP, improving the best-known guarantee from 0.46 (due to \citet{brubach2017attenuate}) to 0.51.
We also improve the best-known impossibility result from $1-1/e$ to $1-\ln(2-1/e)$ through the analysis of a differential equation, showing that our approximation ratio of $1-\ln(2-1/e)$ is \textit{tight}, and closes the gap for online stochastic matching with timeouts.

We then make appropriate modifications to this algorithm so that it can offer assortments of items in each stage of the problem as opposed to a single item. However, in the general assortment problem, rounding the MCDLP solution is more challenging, because the LP can suggest fractionally-intersecting assortments for a single customer, effectively having ``multiple chances'' to sell her the same item over multiple stages. As discussed earlier, the possibility of showing an item multiple times to a customer is desirable in some settings while not favorable in some other settings. We show that in the settings for which it is allowed to offer an item repeatedly to a customer, the algorithm proposed to address the online stochastic matching with timeouts problem can be modified to offer assortments of items to customers. We prove that this algorithm has an approximation guarantee of 0.51, under some technical assumptions. 

To address the settings in which it is not desirable to show an item multiple times to a customer, our previous random ordering is not possible to maintain, we instead consider the assortments in a uniformly random order, like in \citep{bansal2012lp}.
We discard from each assortment the items which have already been offered, and show that this leads to a constant-factor approximation ratio of 0.09. Moreover, for the case that items are valued homogeneously across customers we improve this guarantee by making a slight modification to this algorithm. Table~\ref{table:results} summarizes the results of these algorithms.


\begingroup
\renewcommand{\arraystretch}{1.6}
\begin{table}[]
	\caption{Summary of the main approximation algorithms in the work. The second column demonstrates the approximation guarantees provided in this work and whether they are tight. The last column specifies the arrival model of customers. The results in the first three rows hold under some technical assumptions that are further explained in Sections~\ref{sec::single-item} and~\ref{sec::generalAsst}. \label{table:results}}
	\begin{tabular}{|l|c|c|}
		\hline
		\multicolumn{1}{|c|}{\textbf{\begin{tabular}[c]{@{}c@{}} Problem \\\end{tabular}}}                                       & \textbf{Approx. guarantee}                                                          & \textbf{Arrival model}                                                \\ \hline
		Online stochastic matching with timeouts                                                                                                                                 & \begin{tabular}[c]{@{}c@{}}$ 1-\ln(2-\frac{1}{e})\simeq 0.51$ \\ (optimal)\end{tabular} & Known and stationary                                                  \\ \hline
		\begin{tabular}[c]{@{}l@{}}Multi-stage and multi-customer assortment \\ optimization with repeated offerings\end{tabular}                                              & \begin{tabular}[c]{@{}c@{}}$1-\ln(2-\frac{1}{e})\simeq 0.51$ \\ (optimal)\end{tabular}    & Known and stationary                                                  \\ \hline
		\begin{tabular}[c]{@{}l@{}}Multi-stage and multi-customer assortment \\ optimization without repeated offerings\end{tabular}                                          & 0.09                                                                                      & \begin{tabular}[c]{@{}c@{}}Known, stationary,\\  and Integral\end{tabular} \\ \hline
		\begin{tabular}[c]{@{}l@{}}Multi-stage and multi-customer assortment \\ optimization without repeated offerings;  \\ items priced homogeneously across customers\end{tabular} & 0.15                                                                                      & Known                                                                 \\ \hline
	\end{tabular}
\end{table}
\endgroup

In the special case where all of the patience levels equal 1, our problems reduce to the classical online stochastic matching/assortment problems where the firm only gets one chance to make an offering to each customer.
Under stationary arrivals, the tight approximation ratio relative to the LP is $1-1/e$ in both of these cases (see \citet{brubach2017attenuate} and \citet{ma2018dynamic}, respectively).
Our work, and specifically our impossibility result, shows that both of these problems are substantially more challenging when there are multiple chances to interact with each customer.
We derive a tight approximation ratio for the special case of online stochastic matching with timeouts, and the first constant-factor approximation ratios for the general assortment problem.

Moreover, this work studies heuristics to solve the MCDLP's when repeated offerings are permitted and not allowed. To do so, we use the column generation framework and we show that for general choice models the column generation subproblem cannot be solved in polynomial-time, unless P=NP. However, for the special case of MNL choice model, when assortments shown to a customer can overlap the column generation subproblem can be solved efficiently, and if they cannot overlap then the column generation subproblem can be closely approximated. To do so, we develop a fully polynomial-time approximation scheme (FPTAS) for the column generation subproblem and show that this method results in an FPTAS for the MCDLP. 

Lastly, in Section~\ref{sec::simul} we run simulations on the publicly accessible hotel data set by~\citet{bodea'09}. To estimate customer choice model, we use the product availability and booking information and we consider a stochastic arrival of customers with identical arrival rates. There are four different rooms that can be offered to customers upon their arrival, those are, King room, Queen room, Suite and Two-double room. Furthermore, we try out different patience levels for the customers to observe their effects on the performance of the algorithms. Therefore, we have a multi-stage and multi-customer assortment problem where each room can be offered at multiple prices (those are, rack rate, discounted rate, etc.). On this data set, we compare the performance of our algorithms for the multi-stage and multi-customer assortment problem to various benchmarks. These simulations show that in different settings of the problem, in the two extremes where the number of rooms is very small compared to the number of arriving customers or when the there is an abundant number of rooms available to be offered, other benchmark heuristics might perform very well. However, in the regime that lies in between these two extremes, our algorithms outperform these benchmarks. This is significant as most practical cases lie in such regimes and our simulations show a strong performance of our algorithms for them.

\subsection{Comparison of Model to Related Literature} \label{sec::relatedWork}
We compare our modeling assumptions to those in the literature, first focusing on the aspect of multi-stage interactions with the same customer.
Our model for sequential assortment offerings is inspired by \citet{flores2019assortment} and \citet{liu2019assortment}, where the customer leaves upon \textit{accepting} an offered item.
This model was originally used by \citet{liu2019managing} for healthcare scheduling systems, where the patient leaves upon accepting an appointment.
By contrast, in other applications where the firm is selling a sequence of ``add-ons'', the customer leaves upon \textit{declining} an offer \citep{chen2016assortment,xu2018assortment}.
We use the former model because it naturally coincides with the online stochastic matching with timeouts model when the assortments are restricted to have size 1.

The works above focus specifically on the Multi-nomial Logit (MNL) choice model, and derive refined results using the structure of MNL.
By contrast, we are the first to attack these multi-stage assortment problems using the generality of the MCDLP, and our results hold for the very large class of \textit{substitutable} choice models.
An assumption made in our work is that the number of stages a customer is willing to interact is known when she arrives; this is also assumed in the works above (and in some cases, assumed to always equal two).
Other models have been proposed when a single customer views a variable number of pages; see \citep{gallego2018approximation}, and also \citep{wang2017impact} who models the ``search cost'' paid by the customer to view more pages.

Regarding the second aspect of our work---inventory-constrained assortment optimization, we provide the first constant-factor approximation guarantee in the multi-stage, multi-customer setting.
\citet{chen2016assortment} have previously established a constant-factor guarantee for a firm selling add-ons, but in their case there are only two stages and an assortment is offered in only one of the stages.
We should mention that \citet{lei2018randomized} have recently studied an inventory-constrained assortment optimization problem where the item display order affects each customer's choice.

Finally, we point out that in our model the next customer does not arrive until the previous customer has finished making her decisions across the stages.
This is a modeling assumption which is also made throughout the previous literature (see \citet{liu2019managing,chen2016assortment}), as well as in online stochastic matching with timeouts.
Nonetheless, performance should only improve if multiple customers are progressing through different stages at the same time, since the online algorithm has gained information about future customer arrivals.
We believe that formally modeling this behavior would be an interesting future work.

\section{Problem Description}\label{sec::problem-def}
An e-commerce platform is selling a set of $n$ different items. Suppose there exists $m$ customer types that can arrive to the platform. Each item $i$ can have an initial inventory level $b_i$ which is a non-negative integer. For the sake of simplicity, we assume that there is a single copy of each item, i.e., $b_i=1$ for all $i\in [n]$\footnote{For a positive integer $a$, the set $\{1,\ldots,a\}$ is denoted by $[a]$.}. This assumption is without loss of generality, as an item with multiple units of inventory can be split into separate items. Furthermore, for each customer type, item $i$ has a selling revenue $r_{ij}\ge 0$\footnote{We assume the more general case of \emph{type-dependent revenues} for items. Of course our methods and formulations can be used for the cases that the selling revenue of each item is the same for all customer types.}. We consider a finite selling horizon, consisting of $T$ discrete time-steps, over which there is no inventory replenishment.

During each time-step $t=1,\ldots,T$, a customer of one of the $m$ types could arrive, we refer to whom as customer $t$. The probability that customer $t$ has type $j$ is given by $q_{tj}$, where $\sum_jq_{tj}\le1$;
the inequality can be strict to model a time-step $t$ which could have no customer arrival.
Each type $j$ is endowed with its own choice model, given by a \textit{choice function} $p_j(\cdot,\cdot)$ such that for every \textit{assortment} of items $S$ and every $i\in S$,
\begin{align} \label{eqn::choiceProbs}
p_j(i,S)=\Pr[\text{a type-$j$ customer purchases item $i$ when offered subset $S$}].
\end{align}
Note that $\sum_{i\in S}p_j(i,S)\le1$ for all types $j$ and assortments $S$;
the inequality can be strict to model the fact that the customer could purchase nothing. We denote the probability of making no purchase by a customer of type $j$ when assortment $S$ is offered to her by $p_j(0,S)$\footnote{Let index 0 refer to the ``no purchase" option.}.
We assume that assortment $S$ must lie in some downward-closed family $\cS$, over which the probabilities (\ref{eqn::choiceProbs}) can be input in a computationally-efficient manner and satisfy a \textit{substitutability} condition. All of these are standard assumptions in the literature on inventory-constrained assortment optimization. Moreover, we assume that all random realizations (customer arrivals and purchase choices) are independent from each other.
This includes the multiple purchase choices made by a customer, which is why we impose that the same customer cannot see the same item through different assortments.

The novel aspect of this work is that we allow for multiple assortments to be shown to the same customer.
Specifically, each type $j$ has a \textit{patience level} $\ell_j$, a positive integer denoting the number of assortments that the customer is willing to view before ``timing out'' and leaving.
When a type $j$ customer arrives, the platform shows her a sequence of up to $\ell_j$ \textit{non-intersecting} assortments, which she views in order, stopping once she either makes a purchase or reaches the end of the sequence. We refer to each time an assortment is offered to a customer by a \emph{stage}. If $\ell_j=1$ for all $j$, then this is the classical setting where the platform has only one chance to sell to each customer.

\subsection{Online Stochastic Matching with Timeouts}
The online stochastic matching with timeouts problem considers a special case of the previous model.  It makes the following simplifying assumptions:
\begin{enumerate}
	\item $q_{tj}=q_j$ for all $t$ and $j$, i.e.\ the customer arrival probabilities are stationary over time;
	\item $\cS=\{S\subseteq[n]:|S|\le1\}$, i.e.\ a customer can be shown only one item at a time.
\end{enumerate}
Since $|S|\le1$, we can further simplify notation by letting $p_{ij}$ denote $p_j(i,\{i\})$, the probability that a type $j$ customer purchases item $i$ when shown.
The non-intersection property reduces to the constraint that a particular customer cannot be shown an item she previously rejected.

\section{Tight Result for Online Stochastic Matching with Timeouts}
\label{sec::single-item} 
In each time-step, a customer of one of the types $1,\ldots, m$ arrives. To a customer of type $j$ we can offer at most $\ell_j$ items. If the customer does not purchase any of these items, she loses interest and leaves the system. The goal is to offer sequences of items to arriving customers that maximize the total expected revenue. To measure the performance of an online algorithm for an instance of a problem, it is conventional to compare the online algorithm's collected revenue with that of the offline optimal solution for that instance. Lemma~\ref{lem::lp-ub} shows that the following linear program provides an upper-bound on the revenue of the offline optimal solution for any instance of the problem. Therefore, instead of comparing the total expected revenue of an online algorithm with that of the offline optimal solution, we make the comparison with the objective value of LP\ref{lp:LP}. We refer to LP\ref{lp:LP} simply as the LP for the rest of this work.
\begin{subequations}
	\label{lp:LP}
	\begin{align}
	\max\sum_{j=1}^mTq_j\sum_{i=1}^nr_{ij}p_{ij}x_{ij} \nonumber \\
	\sum_{j=1}^mTq_jp_{ij}x_{ij} &\le1 &\forall i=1,\ldots,n \label{constr::inv} \\
	\sum_{i=1}^np_{ij}x_{ij} &\le1 &\forall j=1,\ldots,m \label{constr::sellOne} \\
	\sum_{i=1}^nx_{ij} &\le\ell_j &\forall j=1,\ldots,m \label{constr::timeout} \\
	0\le x_{ij} &\le1 &\forall i=1,\ldots,n;\ \forall j=1,\ldots,m \label{constr::01}
	\end{align}
\end{subequations}
where decision variable $x_{ij}$ denotes the probability that item $i$ is offered to a customer of type $j$ and $p_{ij}x_{ij}$ is the probability that item $i$ is matched to a customer of type $j$. Constraint~\eqref{constr::inv} ensures that no item is purchased more than once, its inventory. Constraint~\eqref{constr::sellOne} states that at most one item is sold to a visiting customer, and~\eqref{constr::timeout} ensures that the number of items offered to a customer is no more than her patience limit. Let $\OPT$ be the optimal objective function of the LP and for an online algorithm let $\EXP[\ALG]$ denote the total expected revenue of that algorithm on the same instance of the problem. Our goal is to devise an online algorithm that achieves a constant approximation guarantee, formally defined below.

\begin{definition}[$\alpha$-approximation Algorithm]
\label{def:approx} For a maximization problem we say that an algorithm has an approximation guarantee of $\alpha$ if for any instance of the problem  it provides a solution in polynomial-time whose total expected revenue, $\EXP[\ALG]$, is within an $\alpha$ factor of the optimum, $\OPT$. In other words, 
\begin{align*}
\alpha\OPT\le \EXP[\ALG]\le \OPT.
\end{align*}
\end{definition}
Theorem~\ref{thm::0.51} summarizes our main result for the online stochastic matching with timeouts problem.

\begin{theorem}\label{thm::0.51}
	Suppose that $\sum_{i=1}^n p_{ij}\le 1$ or $\ell_j\ge n$ for all customer types $j$.
	Then there is a polynomial-time algorithm whose expected revenue is at least $(1-\ln(2-1/e))\cdot\OPT$, which implies an approximation ratio of $1-\ln(2-1/e)\simeq0.51$.
\end{theorem}

In the rest of this section we discuss the algorithm and prove that it achieve this guarantee. The online stochastic matching with timeouts problem can be divided into two distinct subproblems: the \emph{offline subproblem} that handles which items should be offered to a visiting customer, and the \emph{online subproblem} that manages how series of customer arrivals should be handled in order to have a desired approximation ratio. To address these two subproblems, our algorithm consists of two steps: an \emph{offline black-box} to address the offline subproblem and an \emph{attenuation framework} to take care of the online subproblem. In a nutshell, the offline black-box decides which items to be offered to an arriving customer and the attenuation framework modifies this decision to provide the desired bounds on the performance of the overall algorithm. We will discuss these two steps in more details later in this section. With this intuition in mind we introduce the following definitions.

\begin{definition}\label{def:single-item-event}
	For each $t\in[T]$, item $i\in[n]$, and type $j\in[m]$ let us define the following events:
	\begin{itemize}
		\item $\type_t(j)$: the type of customer $t$ realizes to $j$;
		\item $\avail_t(i)$: item $i$ is still available at the \textit{start} of time-step $t$;
		\item $\offered_t(i,j)$: the algorithm (pre-attenuation) intends to offer $i$ to customer $t$, who has type $j$;
		\item $\Toffered_t(i,j)$: the algorithm (post-attenuation)  offers item $i$ to customer $t$, whose type is $j$;
		\item $\accept_t(i,j)$: customer $t$, with type $j$, would have purchased item $i$ if truly offered (i.e., offered post-attenuation).
	\end{itemize}
\end{definition}
With these definitions, in the following lemma we establishes an upper-bound on the expected revenue of any algorithm for the problem, with the proof provided in Appendix~\ref{apx-sec::single-item}.
\begin{lemma}
	\label{lem::lp-ub}
	For any instance of the online stochastic matching with timeouts problem, the total expected revenue of any algorithm is upper-bounded by the optimal value of the LP. The expectation is taken with respect to the purchase choice of customer types for each item as well as the random selections of the algorithm (in case it is a randomized algorithm).
\end{lemma}

As discussed earlier, we propose an offline black-box to address the offline problem, which will be later used in the algorithm for the overall problem. Suppose customer $t$ is of type $j$. We denote the set of items that are still available when this customer arrives and for which $x_{ij}>0$ by $U_j^t$. Our offline black-box offers a subset of at most $\ell_j$ items $i\in U_j^t$ to customer $t$ such that each such item $i$ is offered with at least a certain probability. Lemma~\ref{lem::focrsGeneral} discusses the black-box in details.

\begin{lemma}[Black-box Randomized Procedure] \label{lem::focrsGeneral}
	Let $A$ be a set of coins.  Each coin $i\in A$ can be flipped at most once and lands on ``heads'' independently with probability $p_i$.
	We can flip the coins in any (possibly randomized) order, and must stop once we get a ``heads'', or have flipped $\ell$ coins, where $\ell$ is a positive integer.
	
	Let $(x_i)_{i\in A}$ be any vector of weights in $[0,1]^{|A|}$ satisfying $\sum_{i\in A}p_ix_i\le1$ and $\sum_{i\in A}x_i\le\ell$.
	Then there exists a randomized procedure for flipping the coins such that the probability of any coin $i$ being flipped, before the process is stopped, is at least
	\begin{align} \label{eqn::boundInLemma}
	\frac{1-e^{-w_i}}{w_i}\cdot x_i,
	\end{align}
	where $w_i=\frac{1}{1-p_i}\sum_{i'\neq i}p_{i'}x_{i'}$ (or $w_i$ is understood to be 1 if $p_i=1$) if $\sum_{i\in A}p_i\le 1$ and $w_i=\frac{1}{1-p_ix_i}\sum_{i'\neq i}p_{i'}x_{i'}$ (or $w_i$ is understood to be 1 if $p_ix_i=1$) if $\ell\ge |A|$.
\end{lemma}

\proof{Proof.}The elements of $(x_i)_{i\in A}$ are fractional, hence they do not clearly determine whether a coin $i$ should be flipped. We use a rounding procedure introduced by~\citet{GKPS} on $x_{i}$'s to make them integral, i.e., 0 or 1. If $x_{i}$ is rounded up to 1 then coin $i$ would be among the coins that can be flipped and otherwise it would not be. We refer to this rounding procedure by GKPS in the rest of this work. The following theorem states the main properties of the GKPS rounding and we refer the interested readers to~\citep{GKPS} for a thorough discussion on it.
	
	\begin{theorem}[\citet{GKPS}]
		\label{thm:GKPS}
		Let $(z_i)_{i\in A}$ be any vector of weights in $[0,1]^{|A|}$. The GKPS algorithm is a randomized algorithm that in polynomial-time creates an integral solution $(Z_{i})_{i\in A}\in\{0,1\}^{|A|}$ that guarantees the following three properties:
		\begin{enumerate}
			\item Marginal distribution: For each coin $i$, $\Pr[Z_{i}=1]=z_{i}$.
			\item Degree preservation: $\sum_{i\in A} Z_{i}\le \lceil{\sum_{i\in A} z_{i}}\rceil\le \ell$.
			\item Negative correlation: For any subset of coins $S$ and any $b\in \{0,1\}$, $\Pr[\bigwedge_{i\in S}(Z_{i}=b)]\le \prod_{i\in S}\Pr[Z_{i}=b]$.
		\end{enumerate}
	\end{theorem}
	
	Using the GKPS rounding, we can round vector $(x_i)_{i\in A}$ to get $(X_i)_{i\in A}$. For each coin $i$, $X_{i}\in\{0,1\}$, satisfying the marginal distribution, degree preservation and negative correlation properties mentioned in Theorem~\ref{thm:GKPS}. With the help of this rounding procedure, we first prove the bound for the case that $\ell\ge |A|$ and then discuss the case $\sum_{i\in A} p_i\le 1$.
	
	\noindent\textbf{The case of }{$\mathbf{\ell\ge |A|}$:} The randomized procedure we run works as the following: Firstly, the GKPS rounding procedure is used on $(x_i)_{i\in A}$, let $\tilde{U}$ denote the set of coins rounded up by GKPS. Secondly, for each rounded coin $i$, i.e., $i\in \tilde{U}$, a random variable $Y_i$ is picked from the interval $[0,1]$, IID and uniformly at random. The coins in $\tilde{U}$ are then ordered independently according to the weights $p_ix_i$ and random variables $Y_i$, that is, in an increasing order of $Y_i/(1-p_ix_i)$, and flipped accordingly until the process stops by seeing a heads or running out of the $\ell$ flip chances, see Algorithm~\ref{alg:BB}.
	
	\begin{algorithm}
		\caption{Black-box}\label{alg:BB}
		\textbf{INPUT:} $\ell$, $A$, $x_i$ and $p_i$ for all $i\in A$
		\begin{algorithmic}[1]
			\State Apply the GKPS rounding to $(x_i)_{i\in A}$. Let $\tilde{U}$ be the set of coins that are rounded up by the GKPS process.
			\State For each coin $i\in \tilde{U}$ pick a number $Y_i$ uniformly at random and IID from $[0,1]$.
			\State Flip coins $i\in \tilde{U}$ in an increasing order of $\frac{Y_i}{1-p_{i}}$ (if $\sum_{i\in A}p_i\le 1$) and $\frac{Y_i}{1-p_{i}x_{i}}$ (if $\ell\ge |A|$) until a ``heads" comes or $\ell$ coins are flipped.
		\end{algorithmic}
	\end{algorithm}	
	Consider any coin $i\in A$ and suppose it is rounded up by the GKPS process. Assuming $Y_i=y$, a coin $i'$ that has also passed through the GKPS process is flipped before $i$ if $\frac{Y_{i'}}{1 - p_{i'}x_{i'}}\le \frac{y}{1 - p_{i}x_{i}}$. 
	We first provide an upper-bound on the probability that coin $i'$ is flipped before coin $i$. To do this, we divide the problem into two cases: $y\ge 2(1- p_{i}x_{i})$, and otherwise, $y< 2(1 -p_{i}x_{i})$. We begin with the former case. If $y\ge 2(1-p_{i}x_{i})$, then 
	\begin{align}
	\frac{1-\exp(-\frac{yp_{i'}x_{i'}}{1- p_{i}x_{i}})}{p_{i'}x_{i'}} &\ge \frac{1-\exp(-2p_{i'}x_{i'})}{p_{i'}x_{i'}}\ge 2(1-p_{i'}x_{i'})\label{eq:RHS},
	\end{align}
	where the last inequality uses the fact that $1-2p_{i'}x_{i'}+2p_{i'}^2{x_{i'}}^2\ge \exp(-2p_{i'}x_{i'})$, 
	derived using the Taylor expansion of $\exp(-2p_{i'}x_{i'})$. Moreover, since $\sum_{i\in A}p_{i}x_{i}\le 1$, we have that $(1-p_{i}x_{i})\ge p_{i'}x_{i'}$, which in addition to $y\ge 2(1-p_ix_i)$ and $y\le 1$ concludes that 
	$p_{i'}x_{i'}\le \frac{1}{2}$. This, in combination with~\eqref{eq:RHS} implies that $\frac{1-\exp(-\frac{yp_{i'}x_{i'}}{1- p_{i}x_{i}})}{p_{i'}x_{i'}} \ge 1,$ making $\frac{1-\exp(-\frac{yp_{i'}x_{i'}}{1- p_{i}x_{i}})}{p_{i'}x_{i'}}$ a potential upper-bound on the probability of coin $i'$ getting flipped before coin $i$. Now, consider the case that $y< 2(1 -p_{i}x_{i})$. We have that 
	\begin{align}
	\frac{1-\exp(-\frac{yp_{i'}x_{i'}}{1- p_{i}x_{i}})}{p_{i'}x_{i'}} = \frac{1-\exp(-\frac{yp_{i'}x_{i'}}{1- p_{i}x_{i}})}{\frac{yp_{i'}x_{i'}}{1-p_{i}x_{i}}}\cdot\frac{y}{1-p_{i}x_{i}}&\ge (1-\frac{yp_{i'}x_{i'}}{2(1-p_{i}x_{i})})\cdot \frac{y}{1-p_{i}x_{i}}\label{eq:taylor}\allowdisplaybreaks\\
	&\ge (1-p_{i'}x_{i'})\cdot \frac{y}{1-p_{i}x_{i}}\label{eq:case-assm},\allowdisplaybreaks
	\end{align}
	where for~\eqref{eq:taylor} we used a similar Taylor expansion to the one discussed earlier and assumption $y< 2(1 -p_{i}x_{i})$ was used in~\eqref{eq:case-assm}. Finally, recall that coin $i'$ is flipped before $i$ if $Y_{i'}\le ({1 - p_{i'}x_{i'}})\cdot \frac{y}{1 - p_{i}x_{i}}$, where $Y_{i'}$ is picked uniformly at random in $[0,1]$. Therefore, the probability of flipping $i'$ before $i$ is at most $({1 - p_{i'}x_{i'}})\cdot \frac{y}{1 - p_{i}x_{i}}$, which in combination with~\eqref{eq:case-assm} makes $\frac{1-\exp(-\frac{yp_{i'}x_{i'}}{1- p_{i}x_{i}})}{p_{i'}x_{i'}}$ a valid upper-bound for the probability of coin $i'$ getting flipped before $i$.
	
	We can now bound the probability that a coin $i$ is flipped before the process is stopped for the case that $\ell\ge |A|$. Note that a coin $i$ is flipped only if it passes through the GKPS rounding, which happens with probability $x_{i}$ by the marginal distribution property of GKPS. After the GKPS process, we flip each coin $i\in\tilde{U}$, sequentially according to its $Y_i/(1-p_ix_i)$ value, as described in Algorithm~\ref{alg:BB}, until a coin comes heads, happening with probability $p_{i}$ for each $i\in\tilde{U}$. Observe that by the degree preservation property of the GKPS rounding, there are at most $\ell$ coins in $\tilde{U}$. Thus, even if all the coins in $\tilde{U}$ are flipped we have no more than $\ell$ total flips. The key property of Algorithm~\ref{alg:BB} is to flip each coin in  $\tilde{U}$ in a suitable random order so that $i$ is flipped with a probability at least $x_i\cdot{(1-e^{-w_i})}/{w_i}$. Ranking coins based on their $Y_i/(1-p_ix_i)$ provides us with this property. We use $I^{i',i}$ to denote the indicator variable that coin $i'$ was flipped before coin $i$ and we so far know that $\Pr[I^{i',i}|i,i'\in \tilde{U},Y_i=y]\le \frac{1-\exp(-\frac{yp_{i'}x_{i'}}{1-p_{i}x_{i}})}{p_{i'}x_{i'}}$. Below we provide a lower-bound on the probability that coin $i$ is flipped conditioned on $i$ being rounded up by the GKPS process:
	\begin{align}
	&\Pr[i \mathrm{~flipped}|i\in \tilde{U}] \ge \Pr[\bigcap_{i'\neq i} i' \mathrm{~not~heads~before~}i|i\in \tilde{U}]\nonumber\allowdisplaybreaks\\
	&= \prod_{i'\neq i} \Pr[i'\mathrm{~not~heads~before~}i|i\in \tilde{U}]\label{eq:single-item-main2}\allowdisplaybreaks\\
	&= \int_0^1 \prod_{i'\neq i} \Pr[i'\mathrm{~not~heads~before~}i|i\in \tilde{U}, Y_i=y]dy\nonumber\allowdisplaybreaks\\
	&= \int_0^1 \prod_{i'\neq i} (1-\Pr[i'\mathrm{~heads~before~}i|i\in \tilde{U}, Y_i=y])dy\nonumber\allowdisplaybreaks\\
	&= \int_0^1 \prod_{i'\neq i} (1-\Pr[i'\in \tilde{U}|i\in \tilde{U}, Y_i=y]\Pr[I^{i',i}|i,i'\in \tilde{U}, Y_i=y]\Pr[i'\mathrm{~flips~heads}|i,i'\in \tilde{U}, Y_i=y, I^{i',i}])dy\nonumber\allowdisplaybreaks\\
	&\ge \int_0^1 \prod_{i'\neq i} (1-x_{i'}\frac{1-\exp(-\frac{yp_{i'}x_{i'}}{1-p_{i}x_{i}})}{p_{i'}x_{i'}}p_{i'})dy\label{eq:single-item-main6}\allowdisplaybreaks\\
	&= \int_0^1 \prod_{i'\neq i} (1-(1-\exp(-\frac{yp_{i'}x_{i'}}{1-p_{i}x_{i}})))dy\nonumber\allowdisplaybreaks\\
	&= \int_0^1 \prod_{i'\neq i} \exp(-\frac{yp_{i'}x_{i'}}{1-p_{i}x_{i}})dy\nonumber\allowdisplaybreaks\\
	&= \int_0^1 \exp(-\sum_{i'\neq i}\frac{yp_{i'}x_{i'}}{1-p_{i}x_{i}})dy\nonumber\allowdisplaybreaks\\
	&= \frac{1}{\sum_{i'\neq i}\frac{p_{i'}x_{i'}}{1-p_{i}x_{i}}}(1-\exp(-\sum_{i'\neq i}\frac{p_{i'}x_{i'}}{1-p_{i}x_{i}}))=\frac{1-\exp(-w_i)}{w_i}\nonumber.\allowdisplaybreaks
	\end{align}
	For~\eqref{eq:single-item-main2}, we used the fact that as there is enough patience to flip all coins when $\ell\ge|A|$, the probability that coins $i'\neq i$ are not flipped heads is independent. Inequality~\eqref{eq:single-item-main6} uses $\Pr[I^{i',i}|i,i'\in \tilde{U}, Y_i=y]\le \frac{1-\exp(-\frac{yp_{i'}x_{i'}}{1-p_{i}x_{i}})}{p_{i'}x_{i'}}$ and $\Pr[i'\in \tilde{U}|i\in \tilde{U}, Y_i=y]\le \Pr[i'\in \tilde{U}]=x_{i'}$ which is derived from independence of $Y_i$ from whether $i,i'\in \tilde{U}$ and, more importantly, the fact that $\Pr[i'\in \tilde{U}|i\in \tilde{U}]\le \Pr[i'\in \tilde{U}]$. The latter is derived from the negative correlation property of the GKPS rounding as \begin{align*}
	\Pr[i'\in \tilde{U}|i\in \tilde{U}]=\frac{\Pr[(i'\in \tilde{U}) \cap (i\in \tilde{U})]}{\Pr[i\in \tilde{U}]}\le \frac{\Pr[i'\in \tilde{U}] \Pr[i\in \tilde{U}]}{\Pr[i\in \tilde{U}]}=\Pr[i'\in \tilde{U}].
	\end{align*}
	Therefore, 
	\begin{align*}
	\Pr[i\mathrm{~flipped}]&=\Pr[i\mathrm{~flipped}|i\in \tilde{U}]\Pr[i\in \tilde{U}]= \Pr[i\mathrm{~flipped}|i\in \tilde{U}]\cdot x_{i} \ge \frac{1-\exp(-w_i)}{w_i}\cdot x_{i},
	\end{align*}
	providing the desired bound for the case that $\ell\ge |A|$. 
	
	\noindent\textbf{The case of }{$\mathbf{\sum_{i\in A}p_i\le 1}$:} This case is similar to the previous case with some small yet essential modifications. Similar to the former case, we first run the GKPS process on $(x_i)_{i\in A}$ and choose a random variable $Y_i$, IID and uniformly at random for each coin that is rounded up, i.e., $i\in \tilde{U}$. This time, we rank coins in an increasing order of $Y_i/(1-p_i)$ and flip them in this order until the process stops, see Algorithm~\ref{alg:BB}. Therefore, assuming $Y_i=y$, coin $i'$ is flipped before coin $i$ if $\frac{Y_{i'}}{1 - p_{i'}}\le \frac{y}{1 - p_{i}}$. 
	We consider the following two cases: $y\ge 2(1- p_{i})$, and otherwise, $y< 2(1 -p_{i})$ and we begin with the former one: If $y\ge 2(1-p_{i})$, then 
	\begin{align}
	\frac{1-\exp(-\frac{yp_{i'}}{1- p_{i}})}{p_{i'}} &\ge \frac{1-\exp(-2p_{i'})}{p_{i'}}\ge 2(1-p_{i'})\label{eq:RHS-small}\allowdisplaybreaks.
	\end{align}
	Assumption $\sum_{i\in A}p_{i}\le 1$ implies $p_{i}+p_{i'}\le 1$, which in addition to $y\ge 2(1-p_{i})$ and  $y\le 1$ gives $p_{i'}\le \frac{1}{2}$.
	This, in combination with~\eqref{eq:RHS-small} concludes that $\frac{1-\exp(-\frac{yp_{i'}}{1- p_{i}})}{p_{i'}} \ge 1,$ making $\frac{1-\exp(-\frac{yp_{i'}}{1- p_{i}})}{p_{i'}}$ a possible upper-bound on the probability that coin $i'$ gets flipped before coin~$i$ for this case of the problem. Now let us consider the case $y< 2(1 -p_{i})$. Here,  
	\begin{align}
	\frac{1-\exp(-\frac{yp_{i'}}{1- p_{i}})}{p_{i'}} 
	&\ge (1-\frac{yp_{i'}}{2(1-p_{i})})\cdot \frac{y}{1-p_{i}}\label{eq:taylor-small} \allowdisplaybreaks\\
	&\ge (1-p_{i'})\cdot \frac{y}{1-p_{i}}\label{eq:case-assm-small}, \allowdisplaybreaks
	\end{align}
	where~\eqref{eq:taylor-small} uses a Taylor expansion similar to the one discussed for~\eqref{eq:taylor} and $y< 2(1 -p_{iv})$ is used for~\eqref{eq:case-assm-small}. Since the probability of flipping $i'$ before $i$ is at most $({1 - p_{i'}}).\frac{y}{1 - p_{i}}$, we have $({1-\exp(-\frac{yp_{i'}}{1- p_{i}})})/{p_{i'}}$ is a valid upper-bound for the probability of the event of interest. We are now equipped to prove the lower-bound on the probability that $i$ is flipped before the process stops.
	\begin{align}
	\Pr[i\mathrm{~flipped}| i\in \tilde{U}]&\ge \EXP_{\tilde{U}}[\int_0^1\prod_{i'\in \tilde{U}\backslash\{i\}}(1-\frac{1-\exp(-\frac{yp_{i'}}{1- p_{i}})}{p_{i'}}p_{i'})dy|i\in \tilde{U}]\label{eq:lem-intro} \allowdisplaybreaks\\
	&= \EXP_{\tilde{U}}[\int_0^1\exp(-\sum_{i'\in \tilde{U}\backslash\{i\}}\frac{yp_{i'}}{1- p_{i}})dy|i\in \tilde{U}]\nonumber \allowdisplaybreaks\\
	&= \EXP_{\tilde{U}}[\frac{1}{\sum_{i'\in \tilde{U}\backslash\{i\}}\frac{p_{i'}}{1- p_{i}}}(1-\exp(-\sum_{i'\in \tilde{U}\backslash\{i\}}\frac{p_{i'}}{1- p_{i}}))|i\in \tilde{U}]\nonumber \allowdisplaybreaks\\
	&\ge\frac{1}{ \EXP_{\tilde{U}}[\sum_{i'\in \tilde{U}\backslash\{i\}}\frac{p_{i'}}{1- p_{i}}|i\in \tilde{U}]}(1-\exp(- \EXP_{\tilde{U}}[\sum_{i'\in \tilde{U}\backslash\{i\}}\frac{p_{i'}}{1- p_{i}}|i\in \tilde{U}]))\label{eq:lem-jens} \allowdisplaybreaks\\&\ge\frac{1}{\sum_{i'\neq i}\frac{p_{i'}x_{i'}}{1- p_{i}}}(1-\exp(- \sum_{i'\neq i}\frac{p_{i'}x_{i'}}{1- p_{i}}))\label{eq:lem-f-dec} \allowdisplaybreaks\\&\ge \frac{1-\exp(-w_i)}{w_i}\nonumber. \allowdisplaybreaks
	\end{align}
	Every coin that is flipped before $i$ has to come tails so that $i$ gets a chance to get flipped before the process is stopped, this is captured by~\eqref{eq:lem-intro}. Moreover, the expectation with respect to $\tU$ in~\eqref{eq:lem-intro} is due to the fact that only coins that are rounded up by GKPS, which is a random process, can get flipped. Inequality~\eqref{eq:lem-jens} uses Jensen's inequality and the fact that function $(1-\exp(-z))/z$ is convex, which will be later proved in Lemma~\ref{lem:f(x)}. To show~\eqref{eq:lem-f-dec} holds, we prove and use the following two claims: $\EXP_{\tilde{U}}[\sum_{i'\in \tilde{U}\backslash\{i\}}{p_{i'}}/{(1- p_{i})}|i\in \tilde{U}]\le \sum_{i'\neq i}{p_{i'}x_{i'}}/{(1- p_{i})}$ and, function $(1-\exp(-z))/z$ is decreasing. To prove the former claim, observe that by the negative correlation and marginal distribution of the GKPS process, respectively, we know that $\EXP_{\tilde{U}}[\bI(i'\in \tilde{U}\backslash\{i\})|i\in \tilde{U}]\le \EXP_{\tilde{U}}[\bI(i'\in \tilde{U}\backslash\{i\})]\le x_{i'}$, where $\bI(.)$ is the indicator function. Thus, $\EXP_{\tilde{U}}[\sum_{i'\in \tilde{U}\backslash\{i\}}{p_{i'}}|i\in \tilde{U}]\le \sum_{i'\neq i}{p_{i'}x_{i'}}$, proving the former claim. Since $(1-\exp(-z))/z$ is decreasing (see Lemma~\ref{lem:f(x)}), replacing $\EXP_{\tilde{U}}[\sum_{i'\in \tilde{U}\backslash\{i\}}{p_{i'}}|i\in \tilde{U}]$ with its upper-bound would not increase~\eqref{eq:lem-jens}. 
	Subsequently, it can be concluded that
	\begin{align*}
	\Pr[i\mathrm{~flipped}]\ge \frac{1-\exp(-w_i)}{w_i}\cdot x_{i},
	\end{align*}
	for this case.
	
	Finally, for both the case $\ell\ge |A|$ and ${\sum_{i\in A}p_i\le 1}$, in order for $({1-\exp(-w_i)})/{w_i}\cdot x_{i}$ to be a valid lower-bound on the probability of coin $i$ getting flipped before the process stops it is required that $({1-\exp(-w_i)})/{w_i}\cdot x_{i}\le 1$. As $x_i\le 1$, it is sufficient to have $({1-\exp(-w_i)})/{w_i}\le 1$. Defining $f(w_i)=({1-\exp(-w_i)})/{w_i}$, Lemma~\ref{lem:f(x)} later shows that $f(z)\le 1$ for all $z\in [0,1]$ and addresses this concern.
\Halmos\endproof

\begin{corollary}\label{cor::atmost1}
For both cases in Lemma~\ref{lem::focrsGeneral} we have that $w_i\le 1$.
\end{corollary}
\proof{Proof.}
We begin with the case that $\ell\ge |A|$. By Lemma~\ref{lem::focrsGeneral}, for this case $w_i=\sum_{i'\neq i}\frac{p_{i'}x_{i'}}{1-p_{i}x_{i}}$. Recall that the weight vector $(x_i)_{i\in A}$ satisfies $\sum_{i\in A}p_{i}x_{i}\le 1$. In particular, the weight vector satisfies $p_{i}x_{i}+\sum_{i'\neq i}p_{i'}x_{i'}\le 1$, implying that $w_i=\sum_{i'\neq i}\frac{p_{i'}x_{i'}}{1-p_{i}x_{i}}\le 1$. For the second case, since $\sum_{i\in A} p_i\le 1$, it holds that $\sum_{i'\neq i}\frac{p_{i'}}{1- p_{i}} \le 1$. Moreover, $(x_i)_{i\in A}\in[0,1]^{|A|}$, thus $w_i=\sum_{i'\neq i}\frac{p_{i'}x_{i'}}{1- p_{i}} \le \sum_{i'\neq i}\frac{p_{i'}}{1- p_{i}} \le 1$.\Halmos\endproof

\begin{remark}\label{rmk::w-compare}
In Lemma~\ref{lem::focrsGeneral} we showed that $w_i^{(1)}=\sum_{i'\neq i}p_{i'}x_{i'}/{(1-p_ix_i)}$ and $w_i^{(2)}=\sum_{i'\neq i}p_{i'}x_{i'}/{(1-p_i)}$ for the cases $\ell\ge|A|$ and $\sum_{i\in A} p_i\le 1$, respectively. It can be easily seen that $w_i^{(1)} \le w_i^{(2)}$, which combined with the fact that $f(z)=(1-\exp(-z))/z$ is a decreasing function (see Lemma~\ref{lem:f(x)}) implies that $f(w_i^{(1)})\ge f(w_i^{(1)})$. In other words, Lemma~\ref{lem::focrsGeneral} provides a tighter lower-bound on the probability of any coin $i\in A$ being flipped before the process is stopped for the case $\ell\ge|A|$ compared to the case $\sum_{i\in A} p_i\le 1$.
\end{remark}

\begin{remark}
	When we use Lemma~\ref{lem::focrsGeneral}, $A$ will be a random set, hence $w_i$ will be a random variable.
	The attenuation framework of \citet{brubach2017attenuate} uses a black-box randomized procedure where the guarantee on a coin $i$ being flipped depends on the random variable $\tilde{w}_i=\sum_{i'\neq i:i\in A}p_{i'}x_{i'}$ instead.
	Their guarantee is $(1-\tilde{w}_i/2)\cdot x_i$, which is a weaker guarantee than ours in that $1-\tilde{w}_i/2\le(1-e^{-\tilde{w}_i})/\tilde{w}_i$.
	Therefore, we modify their attenuation framework using our random variable $w_i$, which leads to our tight approximation ratio.
\end{remark}

\begin{remark}\label{rmk::p*x}
Unlike the case of $\sum_{i\in A} p_i\le1$, if $\sum_{i\in A} p_i>1$ then ranking coins based on solely $p_i$'s (e.g., ranking in an increasing order of $Y_i/(1-p_i)$) would  not attain a lower-bound similar to that of Lemma~\ref{lem::focrsGeneral} on the probability of coin $i$ getting flipped before the process gets stopped. We show this through a counter-example. Suppose $\ell=2$, $|A|=3$ and $p_1=1-\epsilon$, $p_2=0$, $p_3=1-\epsilon$ and $x_1=1$, $x_2=1-\epsilon$, $x_3=\epsilon$ for some small $0.5>\epsilon>0$. Observe that vector $(x_i)_{i\in A}$ satisfies $\sum_{i\in A} p_ix_i\le 1$ and $\sum_{i\in A} x_i\le \ell$, as stated in Lemma~\ref{lem::focrsGeneral}. By the marginal distribution property of the GKPS rounding algorithm, coin 1 is in $\tilde{U}$ with probability $x_1=1$ (i.e., coin 1 is always in $\tilde{U}$). Moreover, by the degree preservation property of GKPS, at most $\ell$ coins can be in $\tilde{U}$ (2 coins in this example). We are interested in finding the probability that coin 3 gets flipped before the stop of the process. Conditioned on coin $3\in \tilde{U}$, we have that $\tilde{U}=\{1,3\}$; therefore, $\sum_{i\in \tilde{U}} p_i=p_1+p_3=2-2\epsilon>1$. If we order the coins in $\tilde{U}$ solely based on $p_i$'s then, since $p_1=p_3$, coins 1 and 3 have the same priority in expectation, resulting in $\Pr[3\mathrm{~flipped}|3\in \tilde{U}]=0.5<1-1/e\le (1-\exp(w_3))/w_3$ for $w_3\in[0,1]$, as our offline black-box would guarantee by Lemma~\ref{lem::focrsGeneral}.
\end{remark}

The offline subproblem was addressed by the offline black-box algorithm, discussed in Lemma~\ref{lem::focrsGeneral}. In the rest of this section we study the online subproblem. As discussed earlier in this section, attenuation framework is used to address the online subproblem. \citet{brubach2017attenuate} provide three different attenuation frameworks: the \emph{edge-attenuation}, the \emph{vertex-attenuation} and the \emph{combined edge and vertex-attenuation}. The last one is the most powerful one, giving~\citet{brubach2017attenuate} the approximation ratio of 0.46 for the online stochastic matching with timeouts problem. Thus, we only study the performance of our online algorithm when the combined edge and vertex-attenuation framework (called the attenuation framework henceforth) is used alongside the offline black-box. Before discussing this framework, in the following lemma let us review a few properties of function $f(z)=\frac{1}{z}(1-\exp(-z))$ introduced in the proof of Lemma~\ref{lem::focrsGeneral}. Lemma~\ref{lem:f(x)} is proven in Appendix~\ref{apx-sec::single-item}.

\begin{lemma}
	\label{lem:f(x)}
	$f(z)=\frac{1}{z}(1-\exp(-z))$ is a decreasing and convex function with finitely bounded first derivatives in $[0,1]$. Moreover, for any $z\in [0,1]$, we have $f(z)\le \lim_{z\rightarrow 0}f(z)\le 1$.
\end{lemma}

The attenuation framework is the combination of the edge-attenuation and the vertex-attenuation frameworks, hence its role is twofold. Lemma~\ref{lem::focrsGeneral} established lower-bounds on the probability of each item getting offered to the customers. In reality, some items might be offered to some customers with probabilities higher than these lower-bounds. These over-performing items cause other items to have poor performances, lowering the approximation ratio of the algorithm. Edge-attenuation solves this issue by weakening the performance of the over-performing items through forcing the probability that each item gets offered to be equal to the bounds established in Lemma~\ref{lem::focrsGeneral}. This allows the lowest performing items, which determine the approximation ratio of the algorithm, to perform better and in turn improving the approximation ratio.

Furthermore, as more customers visit the platform, more items get purchased, thus, less items are available. This makes $w_i$'s in Lemma~\ref{lem::focrsGeneral} smaller as time goes on. Since $f(w_i)$ is a decreasing function, the probability of offering an available item $i$ increases with time. Thus, the available items are offered with higher probabilities as time goes on, resulting in smaller availability probabilities of items with time. Vertex-attenuation allows us to take advantage of the decrease of the availability probability of items and ensures that, at the beginning of each time-step, the probability of items being available  is uniform and decreasing over time. This leads to a higher probability of offering the available items, improving the performance of the low-performing items and ameliorating the algorithm's approximation ratio.  

We remark that for the attenuation framework, we assume that it is possible to obtain an accurate estimate of probabilities of interest via Monte-Carlo simulations. This relies on the results by~\citet{adamczyk2015improved} and~\citet{Ma'14} that demonstrate that boundedness of the first derivative of purchase probabilities  results in the boundedness of the accumulated error of the simulations. For an item $i$ and a customer of type $j$, the purchase probability is $f(w_i)\cdot \ssx_{ij}p_{ij}$ and Lemma~\ref{lem:f(x)} shows that the derivative of $f(w_i)$ is finitely bounded. Therefore, the simulation errors can be manipulated such that the ratio of the total expected revenue of our algorithm to the optimum of the LP only loses an additive factor of $\epsilon=o(1)$\footnote{$o(1)$ is the set of all functions that are asymptotically smaller than constants; thus,  $\lim_{n\rightarrow\infty} o(1)=0$.}. 

Algorithm~\ref{alg:full} gathers the offline black-box (discussed in Lemma~\ref{lem::focrsGeneral}) and the attenuation framework, and Theorem~\ref{thm::attenFramework} discusses the attenuation framework and its properties.  Before that, we introduce the following notation that comes handy in the algorithm.

\begin{definition}\label{def:gamma}
For $t\in[T]$, let $\gamma_1=1$ and $\gamma_t$ be recursively computable by $\gamma_t=\gamma_{t-1}-\frac{1-e^{-\gamma_{t-1}}}{T}$.
\end{definition}

\begin{algorithm}
	\caption{Online Algorithm}\label{alg:full}
	\begin{algorithmic}[1]
		\State Before any customers arrive, solve the LP to get $x^*$.
		\For{time-steps $t=1,\ldots, T$}
		\State Let customer $t$ be of type $j$, $U^t_j$ be the set of available items such that for each $i\in U^t_j$ we have $\ssx_{ij}>0$, and $x^*(U_j^t)$ be the elements of $x^*$ that correspond to type $j$ and items in $U_j^t$.
		\State Run the offline black-box (Algorithm~\ref{alg:BB}) on $x^*(U_j^t)$ as a subroutine and apply edge-attenuation  to ensure each item $i\in U^t_j$ is offered to customer $t$ with probability $\ssx_{ij}\cdot(1-e^{-\gamma_t})/\gamma_t$.		
		\State  Apply vertex-attenuation to each item so that they are available with probability equal to $\gamma_{t+1}=\gamma_{t}-({1-\exp(-\gamma_{t})})/{T}$ at time-step $t+1$.
		\EndFor
	\end{algorithmic}
\end{algorithm}

\begin{theorem}[Modified Attenuation Framework] \label{thm::attenFramework}
	Consider any time-step $t\in[T]$.
	Suppose there exist attenuation factors $a^{\vertex}_{t'}(i),a^{\edge}_{t'}(i,j)\in[0,1]$ for all items $i\in[n]$, customer types $j\in[m]$, and time-steps $t'<t$ on which Algorithm~\ref{alg:full} can be run until the start of time-step $t$, at which point for all items $i$,
	\begin{align*}
	\Pr[\avail_t(i)]=\gamma_t.
	\end{align*}
	Then there exist attenuation factors $a^{\vertex}_t(i),a^{\edge}_t(i,j)\in[0,1]$ for all $i\in[n],j\in[m]$ on which Algorithm~\ref{alg:full} can be run during time-step $t$, so that for all items $i$ and types $j$,
	\begin{align*}
	\Pr[\accept_t(i,j)] &=(1-e^{-\gamma_t})q_jp_{ij}\ssx_{ij}; \allowdisplaybreaks\\
	\Pr[\avail_{t+1}(i)] &=\gamma_{t+1}. \allowdisplaybreaks
	\end{align*}
\end{theorem}

The proof of Theorem~\ref{thm::attenFramework} is provided in Appendix~\ref{apx-sec::single-item}.


\proof{Proof of Theorem~\ref{thm::0.51}} In the proof of Theorem~\ref{thm::attenFramework} we showed that at time-step $t$, $\Pr[\accept_t(i,j)]=(1-e^{-\gamma_t})q_jp_{ij}\ssx_{ij}$. Thus, for the total expected revenue of Algorithm~\ref{alg:full}, denoted by $\EXP[\ALG]$, we have
\begin{align}
\EXP[\ALG]=\sum_{t=1}^T\sum_{j=1}^m\sum_{i=1}^n  r_{ij}\Pr[\accept_t(i,j)]&= \sum_{t=1}^T (1-e^{-\gamma_t}) \sum_{j=1}^m q_j \sum_{i=1}^n  \ssx_{ij}r_{ij}p_{ij}.\nonumber
\end{align}
On the other hand, \begin{align*}
\OPT= \sum_{j=1}^m Tq_j \sum_{i=1}^n  \ssx_{ij}r_{ij}p_{ij}.
\end{align*}
Hence, to bound the approximation ratio of Algorithm~\ref{alg:full} it suffices to bound the ratio between $\EXP[\ALG]$ and $\OPT$, i.e., finding a lower-bound for \begin{align*}
\frac{\sum_{t=1}^T (1-e^{-\gamma_t})}{T}.
\end{align*}
We take care of this in the following:
\begin{align}
\frac{\sum_{t=1}^T  (1-e^{-\gamma_t})}{T}
&= \frac{1}{T}\sum_{t=1}^T T(- \gamma_{t+1}+\gamma_t) \label{eq:{O}_v^i2} \allowdisplaybreaks\\ 
&= \sum_{t=1}^T (-\gamma_{t+1}+\gamma_t) \nonumber \allowdisplaybreaks\\
&= \gamma_1-\gamma_{T+1}\label{eq:{O}_v^i5}\\
&\ge 1-\ln(2-\frac{1}{e})\label{eq:{O}_v^i6}.
\end{align}
We use the recursive formula of $\gamma_t$ from~\eqref{eq:gamma-recurs} in~\eqref{eq:{O}_v^i2}, and in~\eqref{eq:{O}_v^i5} the telescopic property of $\sum_{t=1}^T (-\gamma_{t+1}+\gamma_t)$ was of use. Lastly, for~\eqref{eq:{O}_v^i6} we use Lemma~\ref{lem:g+h}. Therefore, Algorithm~\ref{alg:full} has an approximation ratio of at least $(1-\ln(2-\frac{1}{e}))$.
\Halmos\endproof

In Lemma~\ref{lem:g+h}, we provide an upper-bound on $\gamma_{T+1}$, as required in the proof of Theorem~\ref{thm::0.51}. The proof of this lemma is available in Appendix~\ref{apx-sec::single-item}. 

\begin{lemma}\label{lem:g+h}
	For all $T\in \mathbb{N}$ \footnote{$\mathbb{N}$ denotes the set of positive integers.} 
	we have $\gamma_{T+1}\le h(1) = \ln(2-\frac{1}{e})$, where $h(z) = \ln((e-1)\exp(-z)+1)$.
\end{lemma}

\subsection{Tightness of Approximation Ratio}
\label{subsec::ub-proof}
\citet{brubach2017attenuate} showed that there are instances of the problem for which no online algorithm can provide an approximation ratio better than $1-\frac{1}{e}\simeq 0.63$ for the online stochastic matching with timeouts problem. We improve this bound and show that in fact there exist instances of this problem for which no online algorithm can obtain an approximation ratio larger than $1-\ln(2-1/e)$. This in addition to Theorem~\ref{thm::0.51} shows that Algorithm~\ref{alg:full} has a tight approximation ratio. The proof of the following lemma is provided in Appendix~\ref{apx-sec::single-item}.

\begin{lemma}
	\label{lem:ub-proof}
	There exist instances of the online stochastic matching with timeouts problem for which no algorithm can have an approximation ratio larger than $1-\ln(2-1/e)$.
\end{lemma}

\begin{remark}\label{rmk:tightness}
The example discussed in Lemma~\ref{lem:ub-proof} satisfies both assumptions in Lemma~\ref{lem::focrsGeneral}, those are, $\ell_j\ge n$ or $\sum_{i=1}^n p_{ij}\le 1$ for all types $j$. Therefore, even in these two restricted cases of the problem no online algorithm can have an approximation ratio better than the one provided in Theorem~\ref{thm::0.51} in the worst case. Furthermore, one might think that if some items have the same expected revenues then it would be possible to have a better approximation ratio for the problem. However, in the example provided in Lemma~\ref{lem:ub-proof} the expected revenues of all items are the same, yet the best approximation ratio for any algorithm on that instance is no better than what was established by Theorem~\ref{thm::0.51}.
\end{remark}

\section{Result for General Assortments} 
\label{sec::generalAsst}
In this section, we extend the online stochastic matching with timeouts problem to the multi-stage and multi-customer assortment offering setting. In particular, once a customer arrives, based on her type and the remaining inventory of items, the platform can decide to show her an assortment of items $S\in \cS$ as apposed ot a single item. Here, $\cS$ denotes the set of all feasible assortments, set by the platform, that can be customized according to the constraints specific to that platform, such as, assortment size. We assume that $\{0\}\in \cS$, meaning that there exists the option of not offering any assortment to a customer at some stage. As discussed in Section~\ref{sec::problem-def}, customer types are assumed to be associated with general choice models that determine the purchase probability of each item in an offered assortment. Moreover, we assume the substitutability assumption on choice models, which is defined formally below (see~\citet{golrezaei'14} for further discussions). Our goal is to design an online algorithm that maximizes the total revenue over the time horizon, while respecting the inventory constraints.

\begin{definition}[Substitutability]
\label{def:substitutability}
The substitutability condition is a mild condition on choice models and for all customer types $j$ and assortments $S$ and items $i\neq i'$ it implies that $p_j(i,S)\ge p_j(i,S\cup\{i'\})$.
\end{definition}

Before discussing the results for the multi-stage and multi-customer assortment optimization problem we remark that other than the substitutability assumption we impose no constraints on choice models and assortments that can be offered. In some application of this problem, only assortments of a limited size are permitted to be offered to customers. That is, there exists a constant $k$ such that assortments with size larger than $k$ are not allowed to be offered. Let $\cS^k$ denote the set of all assortments with size less than or equal to $k$. Solving the assortment optimization problem in this case can be done by using the set of assortments $\cS^k$ instead of $\cS$. In fact it can be easily seen that doing so all algorithms discussed in this section provide the same approximation guarantees they had on $\cS$.

In the rest of this section, we study the multi-stage and multi-customer assortment optimization problem in two cases. Section~\ref{subsec:assort-repeated} studies the case that an item can be offered multiple times to a customer through different assortment; whereas in Section~\ref{subsec:no-repeat-hetro} an approximation algorithm for the case that items can be offered at most once to each customer is provided. This algorithm is then improved in Section~\ref{subsec:assort-sameprice} for the special case that items are valued homogeneously across customer types. Finally, Section~\ref{subsec:MCDLP_gap} discusses the integrality gap of the MCDLP associated to the case that only a single-customer visits the platform and an item cannot be offered to her more than once.

\subsection{General Assortments with Repeated Items Offerings}
\label{subsec:assort-repeated}
In this section we allow for the same item to be offered multiple times to the same customer. The customer makes choices with the same probabilities as if she had not seen the item before.  This can be justified by \cite{chernev'06} and \citet{chernev'12} that showed that the relative attractiveness of the items available in an assortment can affect the items' purchase provability. More specifically, even though a customer is not inclined to purchase an item, e.g., a pair of headphones, in one stage of offerings, it is possible that she is willing to purchase it in a later stage if it is accompanied with trusted brands, e.g., specific cellphone brand. 

Below we demonstrate the generalization of the LP to the case that a general assortment of items can be offered to a customer at each stage. Observe that this formulation does not avoid multiple offerings of an item to a customer through different assortments. That is, if an assortment $S_1\ni i$ is first offered to a customer of type $j$ and no item of it, including item $i$, is purchased by that customer then $i$ can be offered to the customer through another assortment $S_2\ni i$. 
\begin{subequations}
	\label{lp:assort-repeated}
	\begin{align}
	\max\sum_{j=1}^m Tq_j \sum_{S\in\cS}x_j(S)\sum_{i\in S}r_{ij}p_j(i,S) \nonumber\allowdisplaybreaks\\
	\sum_{j=1}^m Tq_j \sum_{S\in\cS:S\ni i}x_j(S)p_j(i,S) &\le 1 &\forall i=1,\ldots,n \label{constr::inv-assort-repeated} \allowdisplaybreaks\\
	\sum_{S\in\cS}x_j(S)\sum_{i\in S}p_j(i,S) &\le1 &\forall j=1,\ldots,m \label{constr::sellOne-assort-repeated} \allowdisplaybreaks\\
	\sum_{S\in\cS}x_j(S) &\le\ell_j &\forall j=1,\ldots,m \label{constr::timeout-assort-repeated} \allowdisplaybreaks\\
	0\le x_j(S)&\le 1 &\forall j=1,\ldots,m;\ \forall S\in\cS \label{constr::01-assort-repeated} \allowdisplaybreaks
	\end{align}
\end{subequations}

We refer to the above LP as the MCDLP-R, as it is a multi-stage choice-based deterministic LP that allows repeated offerings of items to customers. In this section, let $\OPT$ denote the optimal value of the MCDLP-R. With a proof similar to that of Lemma~\ref{lem::lp-ub} the following lemma can be shown about $\OPT$.

\begin{lemma}
	\label{lem::lp-ub-assort-repeated}
	The optimal value of the MCDLP-R provides an upper-bound on the total expected revenue of any algorithm for any instance of the multi-stage and multi-customer assortment optimization problem when an item can be presented multiple times to a customer.
\end{lemma}

We show that a modification of Algorithms~\ref{alg:BB} and~\ref{alg:full} results in a 0.51 approximation guarantee when items can be offered more than once to a customer. 

\begin{theorem}\label{thm:assort-0.51}
	Suppose that $\sum_{S\in\cS}\sum_{i\in S} p_j(i,S)\le 1$ or $\ell_j\ge |\cS|$ for all customer types $j$.
	Then there is a polynomial-time algorithm whose expected revenue is at least $(1-\ln(2-1/e))\cdot\OPT$, which implies an approximation ratio of $1-\ln(2-1/e)\simeq0.51$ for the multi-stage and multi-customer assortment optimization with the platform is allowed to show repeated items to customers.
\end{theorem}

The proof of Theorem~\ref{thm:assort-0.51} as well as the supplementary lemmas are provided in Appendix~\ref{apx-subsec:assort-repeated}.

\subsection{General Assortments with No Repeated Items Offerings}
\label{subsec:no-repeat-hetro}
In this section, we study the multi-stage and multi-customer optimization problem when items cannot be offered multiple times to customers through different assortments. This is a reasonable assumption in the settings where if a customer is not interested in purchasing an item, then changing the other items accompanying it would not change the decision of the customer. 

Similar to the assortment optimization case studied in Section~\ref{subsec:assort-repeated}, we use a linear program to provide an upper-bound on the total expected revenue of the offline optimal solution when items cannot be offered repeatedly. The linear program below is the extension of the MCDLP-R, where Compared to the MCDLP-R here we have a new constraint,~\eqref{constr::asst-norepeat-overlap}, to ensure that the assortments shown to each customer are non-overlapping. This is to capture the assumption that a customer's disinterest in an item would not get changed by the combination of the items that accompany it. We refer to LP\ref{lp:assort-norepeat} as the MCDLP-NR as it does not repeatedly offer items to customers.   
\begin{subequations}
	\label{lp:assort-norepeat}
	\begin{align}
	\max\sum_{t=1}^T\sum_{j=1}^m q_{tj} \sum_{S\in\cS}x_j(S)\sum_{i\in S}r_{ij}p_j(i,S) \nonumber\allowdisplaybreaks\\
	\sum_{t=1}^T\sum_{j=1}^m q_{tj}\sum_{S\in\cS:S\ni i}x_j(S)p_j(i,S) &\le 1 &\forall i=1,\ldots,n \label{constr::asst-norepeat-inv} \allowdisplaybreaks\\
	\sum_{S\in\cS}x_j(S)\sum_{i\in S}p_j(i,S) &\le1 &\forall j=1,\ldots,m \label{constr::asst-norepeat-sellOne} \allowdisplaybreaks\\
	\sum_{S\in\cS}x_j(S) &\le\ell_j &\forall j=1,\ldots,m \label{constr::asst-norepeat-timeout} \allowdisplaybreaks\\
	\sum_{S\ni i} x_j(S)&\le 1  &\forall i=1,\ldots,n;\ \forall j=1,\ldots,m \label{constr::asst-norepeat-overlap} \allowdisplaybreaks\\
	x_j(S)&\ge 0 &\forall j=1,\ldots,m;\ \forall S\in\cS \label{constr::asst-norepeat-01} \allowdisplaybreaks
	\end{align}
\end{subequations}

This section studies the case that revenue of an item can be different for different customer types and Section~\ref{subsec:assort-sameprice} discusses the problem when items have homogeneous revenues across different customer types. For the sake of this section, we impose an additional technical assumption on the customers arrival rates, that is, arrival rates are integral, meaning the expected number of arrivals of each type is integral. This is not a strict assumptions and has been used by~\citet{bansal2012lp} and~\citet{adamczyk2015improved} previously. Due to the integrality of arrival rates of all types, without loss of generality, one can always split a customer type in such a way that arrival rates of each of these new types is unit, i.e., $Tq_j=1$ for all customer types $j$. Thus, we assume $T=m$, $q_j=1/m$ for all $j$, where $m$ is the number of types that might arrive. This assumption allows to cancel $T$ and $q_j$ out and omit them from the objective function and constraint~\eqref{constr::asst-norepeat-inv} of the MCDLP-NR.

Let the optimal objective function of the MCDLP-NR be denoted by $\OPT$. In the following, we provide an algorithm for the multi-stage and multi-customer assortment optimization with no repeated item offerings problem and show that it achieves a constant approximation ratio. Algorithm~\ref{alg:asst1} is a modification of the algorithm designed by~\citet{bansal2012lp} which has a constant approximation ratio for the case of single item offerings, as opposed to assortments. We modify the algorithm to offer assortments and ensure that a customer would not see an item multiple times through different assortments. Theorem~\ref{thm:main--assort-norepeat} discusses the approximation ratio of Algorithm~\ref{alg:asst1}, the proof of which is later provided in this section.

\begin{algorithm}
	\caption{Approximation Algorithm for General Assortments with No Repeated Offerings}\label{alg:asst1}
	\begin{algorithmic}[1]
		\State Before any customer arrives, solve the MCDLP to get values $x^*$.
		\For{time-steps $t=1,\ldots, T$}
		\State Let the arriving customer be of type $j$.  
		\If{customer $t$ is first customer of type $j$ arriving} 
		\State Let $\pi$ be a uniformly at random order of assortments $S\in\cS$.
		\While{less than $\ell_j$ offers are
			made and customer $t$ has not purchased any item}
		\State One by one, offer each
		assortment $S$ along $\pi$ (after removing its sold-out items and items previously seen by the customer) independently with probability $x^*_j(S)/\alpha$.
		\EndWhile
		\Else
		\State Do not offer any assortments to the customer.
		\EndIf
		\EndFor
	\end{algorithmic}
\end{algorithm}
\begin{theorem}\label{thm:main--assort-norepeat}
	There is a polynomial-time algorithm whose expected revenue is at least $0.093\cdot\OPT$, implying an approximation ratio of $0.093$ for the multi-stage and multi-customer assortment optimization problem.
\end{theorem}

Lemma~\ref{lem::lp-ub} can be extended to establish an upper-bound on the total expected revenue of any algorithm for the multi-stage and multi-customer assortment optimization problem in Lemma~\ref{lem::assort-lp-ub}. The proof of this Lemma is available in Appendix~\ref{apx-sec::generalAsst-proofs}.

\begin{lemma}\label{lem::assort-lp-ub}
	For any instance of the multi-stage and multi-customer assortment optimization problem, the expected revenue of any algorithm is upper-bounded by $\OPT$. 
\end{lemma}

Before analyzing the performance of Algorithm~\ref{alg:asst1}, let us define a few notations. Let $t_j$ denote the time-step in which the first customer of type $j$ arrives, and if no such customer arrives let $t_j=\infty$. Thus, if a customer of type $j$ arrives, then $t_j$ has a value between 1 and $T$; otherwise, it is $\infty$. 
We also use $t_j$ to refer to the first customer of type $j$ arriving to the platform, if any. Furthermore, we use $t_j^S$ to denote the stage at which assortment $S$ is offered to customer $t_j$, where if $S$ is offered to customer $t_j$ then $t_j^S$ has a value between 1 and $\ell_j$; otherwise, it is $\infty$. Below we introduce a few events, some are special or generalized cases of the events introduced in Definition~\ref{def:single-item-event}.
\begin{definition}\label{def:assort-event}
	For each $i\in[n]$, $j\in[m]$, and $S\in\cS$ let us define the following events:
	\begin{itemize}
		\item $\type(j)$: a customer of type $j$ arrives, i.e., $t_j<\infty$;
		\item $\offered_{t}(S,j)$: the algorithm offers assortment $S$ to customer $t$, who has type $j$;
		\item $\offered_{t}((i,S),j)$: the algorithm offers assortment $S$ to customer $t$ that has type $j$, such that $i\in S$ and $i$ is not removed from $S$ by the algorithm;
		\item $\accept_{t}(S,j)$: customer $t$, with type $j$, purchases an item from assortment $S$;
		\item $\imatch_{j}(i)$: item $i$ is already matched before the first arrival of type $j$, i.e., before $t_j$;
		\item $\seen_{S}(i,j)$: item $i$ is already shown to customer $t_j$ before $S$ is offered to her, i.e., before $t_{j}^S$;
		\item $\timeout_{S}(j)$: customer $t_j$ is already timed out when assortment $S$ is offered to her;
		\item $\cmatch_{S}(j)$: customer $t_j$ is already matched when assortment $S$ is offered to her.
	\end{itemize}
\end{definition}

Let $\EXP[\ALG]$ denote total expected revenue of Algorithm~\ref{alg:asst1}. Lemma~\ref{lem:main-with-alpha} establishes a lower-bound on $\EXP[\ALG]$ that will be later used to prove Theorem~\ref{thm:main--assort-norepeat}.

\begin{lemma}\label{lem:main-with-alpha}
	The total expected revenue of Algorithm~\ref{alg:asst1} is at least $(1-\frac{1}{e})\frac{1}{\alpha}(1-\frac{3}{2\alpha}-\frac{2}{3\alpha^2})\cdot \OPT$, that is $\EXP[\ALG]\ge (1-\frac{1}{e})\frac{1}{\alpha}(1-\frac{3}{2\alpha}-\frac{2}{3\alpha^2})\OPT$.
\end{lemma}
\proof{Proof.} We can break the total expected revenue collected by Algorithm~\ref{alg:asst1} into the sum of the revenues the algorithm collects from each customer $t_j$. In other words, we have $\EXP[\ALG]=\EXP[\sum_{j=1}^m R_j]$, where $R_j$ denotes the revenue collected from $t_j$. This is due to the fact that Algorithm~\ref{alg:asst1} offers assortments and so collects revenue only from the first customer of each type arriving. 
Similarly, let $R_j^i$ be the part of $R_j$ that comes from purchasing item $i$ by customer $t_j$, if any. In the following we focus on bounding $\EXP[R_j^i]$.
	
The customer arrival rates are uniformly $1/m$ and the number of time-steps is $T=m$; therefore, $\Pr[\type(j)]=1-(1-\frac{1}{m})^m\ge 1-\frac{1}{e}$. Thus, $\EXP[R_j]\ge \EXP[R_j|\type(j)]\cdot(1-\frac{1}{e})$. In the following we bound $\EXP[R_j|\type(j)]$, hence it is safe to assume that a customer of type $j$ arrives. Recall that Algorithm~\ref{alg:asst1} removes the already seen and already matched items from assortments before offering them. Therefore, even if an item $i$ is in an assortment $S$, it is not necessarily shown to a customer when $S$ is offered to her. Removing the already seen and matched items without negatively affecting the purchase probabilities of other items in that assortment is made possible by the substitutability assumption. This assumption states that removing an item from an assortment does not decrease the purchase probabilities of the other items in that assortment, i.e., for each customer type $j$, assortment $S$ and item $i\neq i'$ we have $p_j(i,S)\le p_j(i,S\backslash\{i'\})$. 
We establish a lower-bound on the probability of $\offered_{t_j}((i,S),j)$ that will be later useful to lower-bound $\EXP[R_j|\type(j)]$. 
\begin{align*}
&\Pr[\offered_{t_j}((i,S),j)|\type(j)]=1-\Pr[\imatch_{j}(i)\cup \seen_{S}(i,j)\cup \timeout_{S}(j) \cup \cmatch_{S}(j) |\type(j)]\frac{\ssx_j(S)}{\alpha} \nonumber \allowdisplaybreaks\\&\ge 1-(\Pr[\imatch_{j}(i)|\type(j)]-\Pr[\seen_{S}(i,j)|\type(j)]-\Pr[\timeout_{S}(j)\cup \cmatch_{S}(j)|\type(j)])\frac{\ssx_j(S)}{\alpha}. \allowdisplaybreaks
\end{align*}
By Lemma~\ref{lem:M_i} we have $\Pr[\imatch_{j}(i)|\type(j)]\le \frac{1}{2\alpha}$.  Lemma~\ref{lem:D_i,S} shows that $\Pr[\seen_{S}(i,j)|\type(j)]\le \frac{1}{2\alpha}$. Finally, Lemma~\ref{lem:O_S-M'_S} proves that $\Pr[\timeout_{S}(j)\cup \cmatch_{S}(j)|\type(j)]\le \frac{1}{2\alpha}+\frac{2}{3\alpha^2}$. Therefore,
\begin{align*}
\Pr[\offered_{t_j}((i,S),j)|\type(j)]\ge \frac{1}{\alpha}(1-\frac{3}{2\alpha}-\frac{2}{3\alpha^2})\ssx_j(S),
\end{align*} 
resulting in
\begin{align*}
\EXP[R_j^i|\type(j)]&=\sum_{S\ni i} r_{ij}p_j(i,S) \Pr[\offered_{t_j}((i,S),j)|\type(j)] \\&\ge \frac{1}{\alpha}(1-\frac{3}{2\alpha}-\frac{2}{3\alpha^2})\sum_{S\ni i} \ssx_j(S) r_{ij}p_j(i,S).
\end{align*}
Since $\EXP[R_j|\type(j)]=\sum_{i=1}^n \EXP[R_j^i|\type(j)]$,
we have 
\begin{align*}
\EXP[R_j|\type(j)]\ge \frac{1}{\alpha}(1-\frac{3}{2\alpha}-\frac{2}{3\alpha^2})\sum_{i=1}^n \sum_{S\ni i} \ssx_j(S) r_{ij}p_j(i,S).
\end{align*}
Moreover, we showed that $\Pr[\type(j)] \ge 1-\frac{1}{e}$, which gives us $\EXP[R_j]\ge (1-\frac{1}{e})\frac{1}{\alpha}(1-\frac{3}{2\alpha}-\frac{2}{3\alpha^2}) \sum_{i=1}^n \sum_{S\ni i} \ssx_j(S) r_{ij}p_j(i,S)= (1-\frac{1}{e})\frac{1}{\alpha}(1-\frac{3}{2\alpha}-\frac{2}{3\alpha^2}) \sum_{S\in\cS} \ssx_j(S) \sum_{i \in S} r_{ij} p_j(i,S)$. Therefore,
\begin{align*}
\EXP[\ALG]=\sum_{j=1}^m\EXP[R_j]\ge  (1-\frac{1}{e})\frac{1}{\alpha}(1-\frac{3}{2\alpha}-\frac{2}{3\alpha^2})\cdot \OPT.
\end{align*}
\Halmos\endproof

In the following we discuss Lemmas~\ref{lem:M_i},~\ref{lem:D_i,S} and~\ref{lem:O_S-M'_S} that were used in the proof of Lemma~\ref{lem:main-with-alpha}.

\begin{lemma}
	\label{lem:M_i} 
	For any item $i$ and customer type $j$,  $\Pr[\imatch_{j}(i)|\type(j)]\le \frac{1}{2\alpha}$.
\end{lemma}
\proof{Proof.} For any customer type $j'\neq j$, let $I^{j',j}$ denote the indicator variable that a customer of type $j'$ appeared before the first arrival of a customer of type $j$, that is $t_{j'}<t_j$. Thus,
\begin{align}
\Pr[\imatch_{j}(i)|\type(j)] &= \sum_{j'\neq j}\Pr[I^{j',j}\cap\accept_{t_{j'}}(i,j')|\type(j)]\nonumber \allowdisplaybreaks\\
&= \sum_{j'\neq j}\Pr[I^{j',j}|\type(j)]\Pr[\accept_{t_{j'}}(i,j')|I^{j',j},\type(j)]\nonumber \allowdisplaybreaks\\
&\le \sum_{j'\neq j}\Pr[I^{j',j}|\type(j)]\frac{\sum_{S\ni i} \ssx_{j'}(S)p_{j'}(i,S)}{\alpha}\label{eq:lem:M_i-3} \allowdisplaybreaks\\
&\le \frac{1}{2} \sum_{j'\neq j}\frac{\sum_{S\ni i} \ssx_{j'}(S)p_{j'}(i,S)}{\alpha}\label{eq:lem:M_i-4} \allowdisplaybreaks\\
&\le \frac{1}{2\alpha}\label{eq:lem:M_i-5}. \allowdisplaybreaks
\end{align}
An assortment $S$ is offered to a customer of type $j'$ with probability at most $\frac{\ssx_{j'}(S)}{\alpha}$; hence, an item $i\in S$ is purchased by a type $j'$ customer with probability at most $\frac{\ssx_{j'}(S)p_{j'}(i,S)}{\alpha}$. Union bound on all the assortments that contain $i$ results in~(\ref{eq:lem:M_i-3}). Moreover,~(\ref{eq:lem:M_i-4}) and~(\ref{eq:lem:M_i-5}) are due to the symmetry in customer arrivals (as $q_j$ is the same for all types $j$) and constraint~(\ref{constr::asst-norepeat-inv}) in the MCDLP, respectively.
\Halmos\endproof

\begin{lemma}
	\label{lem:D_i,S} 
	For any item $i$ and assortment $S$ that contains $i$ and for any customer type $j$, we have $\Pr[\seen_{S}(i,j)|\type(j)]\le \frac{1}{2\alpha}$. 
\end{lemma}
\proof{Proof.} Item $i\in S$ is already seen by a customer if an assortment $S'\ni i$ was offered to that customer before $S$. More formally, for any assortment $S'\neq S$, let $I^{S',S}$ denote the indicator variable that $S'$ is ordered before $S$ on $\pi$. Given that a customer $t_j$ arrives, the probability that $t_j$ has already seen item $i$ when $S\ni i$ is offered to her is the sum of the probabilities that another assortment $S'\ni i$ was offered to $t_j$ before $S$. By conditioning on $S'$ appearing before $S$ on $\pi$ we have
\begin{align}
\Pr[\seen_{S}(i,j)|\type(j)]
&= \sum_{S'\ni i, S'\neq S}\Pr[I^{S',S}|\type(j)]\Pr[\offered_{t_j}(S',j) |I^{S',S},\type(j)]\nonumber \allowdisplaybreaks\\
&\le \sum_{S'\ni i, S'\neq S}\Pr[I^{S',S}|\type(j)]\frac{ \ssx_j(S')}{\alpha}\label{eq:lem:D_i,S-3} \allowdisplaybreaks\\
&\le \frac{1}{2} \sum_{S'\ni i, S'\neq S}\frac{ \ssx_j(S')}{\alpha}\label{eq:lem:D_i,S-4} \allowdisplaybreaks\\
&\le \frac{1}{2\alpha}\label{eq:lem:D_i,S-5}, \allowdisplaybreaks
\end{align}
where~\eqref{eq:lem:D_i,S-3} follows as Algorithm~\ref{alg:asst1} offers assortments to a customer independently from those that were previously offered to her,~\eqref{eq:lem:D_i,S-4} is due to the symmetry in the assortment ordering ($\pi$ is a uniformly at random ordering of assortments) and~\eqref{eq:lem:D_i,S-5} holds by constraint~(\ref{constr::asst-norepeat-overlap}) in the MCDLP.
\Halmos\endproof

\begin{lemma}
	\label{lem:O_S-M'_S} 
	For any assortment $S$ and customer type $j$, $\Pr[\timeout_{S}(j)\cup \cmatch_{S}(j)|\type(j)]\le \frac{1}{2\alpha}+\frac{2}{3\alpha^2}$.
\end{lemma}
\proof{Proof.} We divide the problem into two cases: (i) $\ell_j=1$, and (ii) $\ell_j\ge 2$. Let us focus on case (i) first. If $\ell_j=1$, then $\cmatch_{S}(j)\subseteq \timeout_{S}(j)$ as there is only one chance of offering assortments to the arrived customer; thus, $\Pr[\timeout_{S}(j)\cup \cmatch_{S}(j)|\type(j)]=\Pr[\timeout_{S}(j)|\type(j)]$. Let $U$ be the random variable indicating the number of assortments offered to customer $t_j$ right before $S$ is about to be offered to her.
\begin{align}
\EXP[U] &= \sum_{S'\neq S} \Pr[I^{S',S}\cap\offered_{t_j}(S',j)]
\\&\le \sum_{S'\neq S} \Pr[I^{S',S}]\frac{\ssx_j(S')}{\alpha}\label{eq:U1-1} \allowdisplaybreaks\\
&=\sum_{S'\neq S}\frac{\ssx_j(S')}{2\alpha}\nonumber \allowdisplaybreaks\\
&\le \frac{\ell_j}{2\alpha}\label{eq:U1-3},
\end{align}
where~\eqref{eq:U1-1} comes from the fact that Algorithm~\ref{alg:asst1} offers an assortment $S'$ with probability at most $\frac{\ssx_j(S')}{\alpha}$. Inequality~\eqref{eq:U1-3} is from constraint~(\ref{constr::asst-norepeat-timeout}) in the MCDLP. Using Markov's inequality we have
\begin{align*}
\Pr[\timeout_{S}(j)|\type(j)]=\Pr[U\ge \ell_j]\le \frac{\EXP[U]}{\ell_j}\le \frac{\frac{\ell_j}{2\alpha}}{\ell_j}=\frac{1}{2\alpha}. \allowdisplaybreaks
\end{align*}

Now, consider the case that $\ell_j\ge 2$. By union bound $\Pr[\timeout_{S}(j)\cup \cmatch_{S}(j)|\type(j)]\le \Pr[\cmatch_{S}(j)|\type(j)] + \Pr[\timeout_{S}(j)|\type(j)]$. We bound the two terms $\Pr[\cmatch_{S}(j)|\type(j)]$ and $\Pr[\timeout_{S}(j)|\type(j)]$ separately, starting with the former. Bounding $\Pr[\cmatch_{S}(j)|\type(j)]$ is similar to the above discussion with some modifications. Let $U$ be the random variable indicating the number of assortments of which $t_j$ purchases any items right before $S$ is about to be offered to her (assuming $t_j$ can purchase items for any number of stages).
\begin{align}
\EXP[U] = \sum_{S'\neq S} \Pr[I^{S',S}\cap\accept_{t_j}(S',j)]
&\le \sum_{S'\neq S} \Pr[I^{S',S}]\frac{\ssx_j(S')}{\alpha}\sum_{i\in S'}p_j(i,S')\label{eq:U1} \allowdisplaybreaks\\
&=\sum_{S'\neq S}\frac{\ssx_j(S')}{2\alpha}\sum_{i\in S}p_j(i,S')\nonumber \allowdisplaybreaks\\
&\le \frac{1}{2\alpha}\label{eq:U3},
\end{align}
where~\eqref{eq:U1} is by union bound because if an assortment $S'$ is offered to a customer of type $j$, then the customer purchases an item of the assortment with probability at most $\sum_{i\in S'}p_j(i,S')$. The last inequality is from constraint~(\ref{constr::asst-norepeat-sellOne}) in the MCDLP. Using Markov's inequality we have
\begin{align*}
\Pr[\cmatch_{S}(j)|\type(j)]=\Pr[U\ge 1]\le \frac{\EXP[U]}{1}\le \frac{1}{2\alpha}. \allowdisplaybreaks
\end{align*}
We finish the proof with upper-bounding $\Pr[\timeout_{S}(j)|\type(j)]$ when $\ell_j\ge 2$. Suppose $\ell_j=a$, 
\begin{align}
\Pr[\timeout_{S}(j)|\type&(j)] \le \sum_{\substack{\{S_1,\ldots,S_a\}, \nonumber\\S_1,\ldots,S_a\neq S}} \Pr[(I^{S_1,S}\cap\ldots\cap I^{S_a,S})\cap(\offered_{t_j}(S_1,j)\cap\ldots\cap\offered_{t_j}(S_a,j))|\type(j)] \allowdisplaybreaks \nonumber\\
&\le \frac{1}{a!}\sum_{\substack{S_1,\ldots,S_a, \\S_1,\ldots,S_a\neq S}} \Pr[(I^{S_1,S}\cap\ldots\cap I^{S_a,S})\cap\left(\offered_{t_j}(S_1,j)\cap\ldots\cap\offered_{t_j}(S_a,j)\right)|\type(j)] \allowdisplaybreaks \nonumber\\
&\le \frac{1}{a!}\sum_{\substack{S_1,\ldots,S_a, \\S_1,\ldots,S_a\neq S}} \Pr[I^{S_1,S}\cap\ldots\cap I^{S_a,S}|\type(j)]\prod_{k=1}^a\frac{\ssx_j(S_k)}{\alpha} \allowdisplaybreaks \nonumber\\
&= \frac{1}{(a+1)!}\sum_{\substack{S_1,\ldots,S_a, \\S_1,\ldots,S_a\neq S}} \prod_{k=1}^a\frac{\ssx_j(S_k)}{\alpha} \allowdisplaybreaks \nonumber\\
&\le \frac{1}{(a+1)!}(\sum_{S'\in\cS}\frac{\ssx_j(S')}{\alpha})^a \le \frac{1}{(a+1)!}(\frac{a}{\alpha})^a. \allowdisplaybreaks \label{eq:timeout-bound}
\end{align}
Note that the first summation is over unordered assortments, while it changes to ordered assortments in the next step, hence we have a factor of $\frac{1}{a!}$. In the third inequality we used the way Algorithm~\ref{alg:asst1} works that a considered assortment $S_k$ is offered to $t_j$ with probability $\frac{\ssx_j(S_k)}{\alpha}$. The next summation comes from the fact that assortments $(S_1,\ldots,S_a)$ and $S$ appear uniformly at random on $\pi$, and the last inequality is from constraint~(\ref{constr::asst-norepeat-timeout}) in the MCDLP. Finally,~\citet{bansal2012lp} showed that $\frac{1}{(a+1)!}(\frac{a}{\alpha})^a\le \frac{2}{3\alpha^2}$, proving that $\Pr[\timeout_{S}(j)|\type(j)]\le \frac{2}{3\alpha^2}$ when $\ell_j\ge 2$.
\Halmos\endproof

\proof{Proof of Theorem~\ref{thm:main--assort-norepeat}.} Lemmas~\ref{lem::assort-lp-ub} and \ref{lem:main-with-alpha}, and setting $\alpha=\frac{3+\sqrt{17}}{2}$ proves the approximation ratio of 0.093 for Algorithm~\ref{alg:asst1}. 
\Halmos\endproof

In the following theorem, we discuss a variant of the multi-stage multi-customer assortment optimization problem where patience levels are \emph{non-deterministic}. As opposed to constant patience levels that the patience level of each customer type is known, here patience level of customers is only known in expectation. In other words after seeing each assortment, a type $j$ customer decides to leave the platform with probability $p^{\text{out}}_j$ or choose to see the next assortment with probability $1-p^{\text{out}}_j$. Theorem~\ref{thm:undeterm-patient} discusses the approximation ratio of Algorithm~\ref{alg:asst1} for this case of the problem, with the proof provided in Appendix~\ref{apx-sec::generalAsst-proofs}.

\begin{theorem}
\label{thm:undeterm-patient}
	Algorithm~\ref{alg:asst1} has an approximation ratio of $0.093$ for the multi-stage and multi-customer assortment optimization problem with non-deterministic patience levels.
\end{theorem}

\begin{remark}
GKPS-based algorithms can have arbitrarily bad approximation ratios for the multi-stage and multi-customer assortment optimization problem when items cannot be offered repeatedly to customers. To see it consider the instance of the problem discussed in Section~\ref{subsec:MCDLP_gap}. In this instance of the problem there exists one customer type, $M$ assortments and $M(M-1)/2$ items such that each item appears in exactly two assortments. Section~\ref{subsec:MCDLP_gap} shows that the optimum is 1. On the other hand, since any pair of assortments has a shared item, any GKPS-based algorithm can contain at most one assortment to respect LP constraint \eqref{constr::asst-norepeat-overlap}. Therefore, the total expected revenue of any such algorithm is at most $\sum_{i\in S} p(i,S)=2/(M(M-1))\cdot (M-1)=2/M$ for the rounded up assortment $S$. So, the approximation ratio of the algorithm is $2/M$, which goes to zero as $M$ goes to infinity.		
\end{remark}

\subsection{General Assortments with Homogeneous Item Revenues Across Customers and No Repeated Offerings}
\label{subsec:assort-sameprice}
In the previous section, we discussed a 0.09-approximation algorithm for the problem of offering assortments to customers, while refraining from showing an item multiple times to a single customer. For this algorithm to work, there is no restrictions on the revenue acquired from selling an item to different types of customers. In this section, we study a scenario that is more restricted yet practical in many settings, that is, the case when the revenue of an item $i$ does not depend on the type of the customer it is presented to, that is, $r_{ij}=r_i$ for all customer types $j$. We show that a slight modification of Algorithm~\ref{alg:asst1} has an improved performance for this case of the problem. 

Note that here, we don not require the assumption that all customer types have the same stationary arrival rate of $q_j=1/m$, where $T=m$. In fact we show that even for the non-stationary and heterogeneous customer arrivals, the modified Algorithm~\ref{alg:asst1} provides a 0.15 approximation guarantee for the problem. Thus, the linear program that provides an upper-bound on any algorithm for this case of the problem is similar to the MCDLP-NR with the slight change of $r_{ij}=r_i$ for all types $j$. This modified linear program is referred to as the MCDLP-NRS to highlight that no repeated items can be shown to a customer and items are valued the same across customer types. In this section $\OPT$ is used to denote the optimal objective value of the MCDLP-NRS. We use a slight modification of  Algorithm~\ref{alg:asst1}, where assortments are offered to a customer $t$, whether she is the first arrival of her type or not. We then use the MCDLP-NRS to prove the approximation guarantee of this algorithm in the following theorem, with the proof available in Appendix~\ref{apx-sec::generalAsst-proofs}.

\begin{theorem}
\label{thm:assort-sameprice}
When item revenues are homogeneous across customer types, there is a polynomial-time algorithm whose expected revenue is at least $0.15\cdot\OPT$, implying an approximation ratio of 0.15 for this case of the multi-stage and multi-customer assortment optimization problem. 
\end{theorem}

\subsection{Integrality Gap for Single-Customer MCDLP-NRS}
\label{subsec:MCDLP_gap}
In this section we focus on the \emph{single-customer} multi-stage assortment optimization problem, as studied in~\citet{liu2019assortment}. In other words, here only a single customer visits the platform and the problem is what assortments of items and in what order should be offered to her. Consider the LP relaxation for this problem:
\begin{subequations}
\begin{align}
\max\sum_{S\in\cS}x(S)\sum_{i\in S}r_ip(i,S)\nonumber \\
\sum_{S\in\cS}x(S)\sum_{i\in S}p(i,S) &\le1 \label{constr::singleCust::asst-sellOne} \allowdisplaybreaks\\
\sum_{S\in\cS}x(S) &\le\ell \label{constr::singleCust::asst-timeout} \allowdisplaybreaks\\
\sum_{S\ni i} x(S)&\le 1  &i=1,\ldots,n \label{constr::singleCust::ass-overlap} \allowdisplaybreaks\\
x(S)&\ge 0 &S\in\cS \label{constr::singleCust::ass-01} \allowdisplaybreaks
\end{align}
\end{subequations}
\citet{segev'19} provided a PTAS for the single-customer multi-stage assortment problem via dynamic programming, where they compared their algorithm's total expected revenue with that of the optimal solution. In contrast, in this work we use the optimum of different MCDLP's as upper-bounds on the total expected revenue of the optimal solutions of different versions of the problem. The following theorem shows that, when compared to the MCDLP-NRS, the multi-stage assortment optimization problem has an integrality gap of 0.527, even when a single customer visits the platform. The proof of this theorem can be found in Appendix~\ref{apx-sec::generalAsst-proofs}. Note that our proof of Theorem~\ref{thm:assort-sameprice} shows that the integrality gap for a single customer is at worst 1/6.

\begin{theorem} \label{thm::mcdlpIntegGap}
	There exists an instance of the multi-stage assortment problem where the fraction of the optimal MCDLP-NRS value obtained by any multi-stage assortment algorithm is at most $1-e^{-3/4}\approx0.527$.
\end{theorem}

\section{Solving the MCDLP}
\label{sec:mcdlp}
In Section~\ref{sec::generalAsst}, we studied different versions of the assortment problem and for each of them we showed that there exists an MCDLP that provides an upper-bound on the total expected revenue of any algorithm. We later used the solution to those MCDLP's to design approximation algorithms for the different versions of the multi-stage and multi-customer assortment optimization problem. 

The MCDLP's has exponential number of decision variables, as there can be $2^n$ different assortments in $\cS$; thus, solving the MCDLP's can take a very long time.~\citet{gallego'04} suggested the use of column generation technique to overcome this issue. The main idea of the column generation technique is to solve a reduced LP with only a limited number of assortments $\cS'$ and check the corresponding dual LP to check whether the exists any assortment in $\cS$ for which the resulted dual variables constitute a positive reduced cost. In that case, the assortment with positive reduced cost is then added to $\cS'$ and the process is iterated. Otherwise, the current solution is optimal and no further assortments are needed to be added to $\cS'$. In other words, decision variables are added to the MCDLP as needed, as opposed to having a decision variable for each assortment in $\cS$, which might run into long time-runs. In Appendix~\ref{apx-sec:mcdlp} we discuss the column generation algorithm and how its approximation ratio translates into that of the MCDLP-NR. Moreover, we provide an FPTAS for the column generation subproblem when the choice model is MNL.

\section{Simulations on Hotel Data Set of~\citet{bodea'09}}\label{sec::simul}
In this section, we test our algorithms designed to address the assortment planning problem on the publicly accessible hotel data gathered by \citet{bodea'09} and compare their performances with those of several benchmark algorithms. Based on the data, here we consider a more general setting of the assortment planning problem, that is, a multi-price multi-stage multi-customer assortment optimization problem. In this problem, an item $i$ can be offered to a customer $t$ of type $j$ at different prices, $r^k_{ij}$ and $r^{k'}_{ij}$, which are called price $k$ and $k'$, respectively. The probability that a customer of type $j$ purchases an item $i$ offered as a part of an assortment $S$ with price $k$ can be different if this item was offered as a part of the same assortment with a different price $k'$. Due to this additional dimension in the problem, the choice function is modified as
\begin{align} \label{eqn::choiceProbs-pricing}
p_j(i,k,S)=\Pr[\text{a type-$j$ customer purchases item $i$ when offered subset $S$ with price $k$}].
\end{align}

For clarity, we use the term \emph{product} to refer to an (item, price) combination, e.g., product $(i,k)$. Therefore, upon the arrival of a customer $t$ with type $j$, for any assortment $S$ and product $(i,k)\in S$, we are given the probability that the customer purchases product $(i,k)$, that is, $p_j(i,k,S)$. Having these probabilities, the goal is to determine which assortment $S$ should be offered to customer $t$ at each stage, that is, which products and at what prices among their permitted prices should be shown to the customer. We note that the assortments offered to a customer can be limited so that they lie in an arbitrary downward-closed family, for instance, one can constrain the assortments offered to a customer to not offer an item at different prices simultaneously. The following LP captures the problem when items can be offered at multiple prices, which we refer to as MMCDLP-NR to signify items can be offered at multiple prices and it is not allowed to offer repeated products to the same customer:
\begin{subequations}
	\label{lp:assort-norepeat-pricing}
	\begin{align}
	\max\sum_{t=1}^T\sum_{j=1}^m q_{tj} \sum_{S\in\cS}x_j(S)\sum_{(i,k)\in S}r^k_{ij}p_j(i,k,S) \nonumber\allowdisplaybreaks\\
	\sum_{t=1}^T\sum_{j=1}^m q_{tj}\sum_{S\in\cS:S\ni (i,k)}x_j(S)p_j(i,k,S) &\le 1 &\forall i=1,\ldots,n \label{constr::asst-norepeat-inv-pricing} \allowdisplaybreaks\\
	\sum_{S\in\cS}x_j(S)\sum_{(i,k)\in S}p_j(i,k,S) &\le1 &\forall j=1,\ldots,m \label{constr::asst-norepeat-sellOne-pricing} \allowdisplaybreaks\\
	\sum_{S\in\cS}x_j(S) &\le\ell_j &\forall j=1,\ldots,m \label{constr::asst-norepeat-timeout-pricing} \allowdisplaybreaks\\
	\sum_{S\ni (i,k)} x_j(S)&\le 1  &\forall i=1,\ldots,n;\ \forall j=1,\ldots,m \label{constr::asst-norepeat-overlap-pricing} \allowdisplaybreaks\\
	x_j(S)&\ge 0 &\forall j=1,\ldots,m;\ \forall S\in\cS \label{constr::asst-norepeat-01-pricing} \allowdisplaybreaks
	\end{align}
\end{subequations}
Note that constraint~\eqref{constr::asst-norepeat-overlap-pricing} is written so that it allows the same item to be offered to a customer at different prices in different offering stages. For the case of hotel room offering, this assumption is sensible. However, it is possible to rewrite this constraint so that once an item is offered to a customer, it is not allowed to be offered to the same customer again at a different price. In this section we use the hotel data set to test the algorithms developed for the multi-price multi-stage multi-customer assortment problem in the setting where items are not allowed to be shown repeatedly to a customer through different offering stages, as studied in Section~\ref{sec::generalAsst}.

\subsection{Experimental Setup}
\label{subsec:exp-setup}
From the hotel data set collected by \citet{bodea'09} we consider the occupancies taking place in the 5-week period from March 11th 2007 to April 15th 2007. The hotel rooms available to be offered to customers are merged into the 4 following categories: King rooms, Queen rooms, Suites and Two-double rooms. Different categories of rooms have their own distinct inventory of rooms. Moreover, there are two fare classes for each room, those are, regular rates and discounted rates. Hence there are 8 different combinations constituted by these 4 rooms categories and 2 fare classes. We refer to each of these 8 combinations as a \emph{product}. Note that a customer might choose a room offered at its regular rate to its discounted rate as oftentimes higher fares are packaged with extra offers, such as free breakfast, higher speed internet, etc.

Furthermore, 1315 different customer types appear in the data set, which are set based on the 8 different features available in the data: (i) whether the booking was done on the hotel website, (ii) whether the booking was done via a travel agency, (iii) party size, (iv) membership level, (v) VIP level, (vi) number of days booking was done in advance, (vii) whether booking was done over the weekend, (viii) whether the customer checked-in over the weekend\footnote{Among these, features (i), (ii), (vii) and (viii) are binary features, feature (iii) is at least 1, feature (iv) is a number from 0 to 3, feature (v) is a number from 0 to 2, and feature (vi) is a non-negative number.}. Thus, these customer types capture the heterogeneity in customers' preferences. For each customer type, the data set provides the assortment of products offered to the customer as well as the product picked by her. Note that the hotel data set was collected using a platform that offers assortments in a single stage. To make the data set compatible to the multi-stage setting, which is the focus of this paper, we synthetically attach patience levels to customers, as further explained in Section~\ref{subsec:exp-results}. Moreover, each occupancy date for a customer is considered as a separate instance of the problem. This simplifying assumption allows for the multi-price multi-stage multi-customer assortment optimization problem to be studied for each day separately without complications coming from new rooms becoming available when customers depart. 

On the available 8 products and for each of these 1315 customer types we estimate a Multinomial Logit (MNL) choice model. The MNL choice model is useful for two reasons. Firstly, it has been assessed to perform reasonably well on this data set \citep{vanRyzin'15}, and secondly, in Section~\ref{sec:mcdlp} we showed that there is an FPTAS to solve MCDLP under MNL. 

Next, we discuss the limitations for our analysis on the hotel data set. First, the data set does not provide the number of available rooms in each room category. In the experiments we test a wide range of initial room availabilities to observe the effect of the starting room capacities on the performance of the algorithms. Moreover, The data set does not provide any information on the customers that do not make a purchase after an assortment of hotel rooms are offered to them. While this is a standard challenge in choice modeling, we address it by trying different assumptions on the weight of the no-purchase option in MNL model for each customer type. In general, setting larger weights for the no-purchase option causes the myopically revenue-maximizing assortments to offer rooms at their lower prices. This creates a nontrivial tradeoff between offering lower prices which maximizes immediate revenue, v.s. offering higher prices which conserve inventory. Therefore, in many of our experiments we set the no-purchase option to have a large weight.

Having the data set and the choice model, we are now ready to discuss a test instance. As mentioned earlier, an instance is defined for a single occupancy date to ensure the inventory of room does not replenish. Once a customer arrives, her features and patience level are revealed; therefore, her type and purchase probabilities are given. The goal is to show personalized assortments of products, which are (room,fare) combinations, in different stages to the customer until she purchases a product or runs out of patience. Below we summarize the test instances:
\begin{table}[]
	\caption{Room categories' fares and inventory percentages}
	\label{table:price-inv}
	\begin{center}
	\begin{tabular}{|c|c|c|c|}
		\hline
		\multicolumn{1}{|l|}{Room Category} & \multicolumn{1}{l|}{Low Fare} & \multicolumn{1}{l|}{High Fare} & \multicolumn{1}{l|}{Percentage of Rooms} \\ \hline
		King                                & \$307                         & \$361                          & 52\%                                     \\
		Queen                               & \$304                         & \$361                          & 15\%                                     \\
		Suite                               & \$384                         & \$496                          & 13\%                                     \\
		Two-Double                          & \$306                         & \$342                          & 20\%                                     \\ \hline
	\end{tabular}
	\end{center}
\end{table}
\begin{itemize}
	\item The total number of customer types is $m=1315$. To comply with the assumptions made in Section~\ref{subsec:no-repeat-hetro}, we assume that the rate of customer arrivals for all customer types is the same and it is equal to $1/m$, i.e, the total number of customers that might arrive on a day is 1315. 
	\item The total number of products is the same for all instances and is equal to 8.
	\item As shown in Table~\ref{table:price-inv}, rooms can be offered at two prices and they are identical for all instances. In the experiments, we manipulate the room prices and increase the high fares and add heterogeneity to room prices for different customer types to demonstrate the strengths and weaknesses of algorithms more clearly. These manipulations are explained in Section~\ref{subsec:exp-results}.  
	\item The hotel data set only provides the relative inventories of the different categories of rooms; see Table~\ref{table:price-inv}. To address this, we consider different starting inventory of rooms. Similar to~\citep{golrezaei'14,Ma-SimchiLevi'17}, this is done through setting different \emph{loading factors}, which is defined as the ratio of total arriving customers to the total initial inventory of rooms. Section~\ref{subsec:exp-results} further discusses different loading factors used in the tests.
	\item The hotel data is collected by offering assortments to customers in a single stage. That is, all customers have a patience level of 1. To be able to account for different patience levels, we synthetically assign patience levels to different customer types.
\end{itemize}

Knowing the experiment setup, in the next section we discuss the different algorithms used on the data set.

\subsection{Algorithms Compared}
\label{subsec:exp-algs}

As mentioned earlier, we constrained the algorithms for the experiments to not offer repeated products to a customer in different stages. On different instances of the problem we compare 4 algorithms, listed below:
\begin{enumerate}
	\item \textbf{Greedy:} At each stage, the algorithm myopically offers the assortment of available products which maximizes the immediate expected revenue, irrespective of inventory.
	\item \textbf{Conservative:} At each stage, the algorithm offers rooms only at their high prices. This algorithm chooses the assortment of rooms from the rooms that are not sold out using a greedy algorithm that maximizes the immediate expected revenue.
	\item \textbf{Algorithm~\ref{alg:asst1}:} This is the 9\% approximation algorithm from Section~\ref{subsec:no-repeat-hetro}, with the difference that here assortments are defined over products and not items (i.e., rooms in the experiments). Moreover, we tested different values for $\alpha$ and it was evident that for these experiments $\alpha=1$ provides the best performance. In general, $\alpha=1$ performs more greedily than $\alpha=\frac{3+\sqrt{17}}{2}$, which is the optimal value of $\alpha$ from our approximation algorithm. The reason that $\alpha=1$ works better in the experiments is the limited number of products that can be offered to customers. 
	\item \textbf{Modified Algorithm~\ref{alg:asst1}:} In Section~\ref{subsec:assort-sameprice}, we discussed a modification of Algorithm~\ref{alg:asst1} for cases that all customers have the same rewards for each item and showed that the approximation guarantee of this algorithm was 15\%. As opposed to Algorithm~\ref{alg:asst1}, this modified algorithm offers assortments of items to customers even if it is not the first time a customer of that type has visited the platform. We test this algorithm on the hotel data set even when the room fares are synthetically manipulated so that rooms are offered at different prices to different customers. Similarly, $\alpha$ is set to 1 here, as opposed to $\alpha=3$ from our approximation algorithm.
\end{enumerate} 
There are two extremes to the problem. One extreme is when there exists an infinite initial room inventories and a limited number of customers arriving on the platform, i.e., a loading factor of 0. The other extreme is where the loading factor approaches infinity and there exists too many customers relative to rooms. It can be easily seen that when the loading factor is 0 the greedy algorithm is the optimal approach as it extracts the most expected revenue for each arriving customer. On the other hand, when the loading factor is $\infty$ the conservative algorithm has the best performance as it collects the maximum expected revenue from each unit of room inventory. The problems becomes challenging when the loading factor lies in-between these two extremes, where most practical assortment planning problems also lie. In this region, our algorithms select the assortments to be offered to customers by balancing the revenue collected per unit of inventory and the revenue collected per customer.

\subsection{Results}
\label{subsec:exp-results}
On every instance, we run the 4 algorithms discussed in Section~\ref{subsec:exp-algs} and consider the average revenues over the 35 runs. The performance of these algorithms is measured as a percentage of MMCDLP-NR upper-bound introduced in Section~\ref{sec::simul}. 


In these experiments, we synthetically introduce heterogeneity in room fares for different customer types. To do so, for each customer type we pick a random Gaussian number for each room fare, while ensuring the low fares are smaller than the high fares. More specifically, for each room category and fare, we draw a random number from a Gaussian distribution whose mean is the original room fare (see Table~\ref{table:price-inv}) and whose standard deviation is the square root of the original room fare. While different methods and parameters can be used to add this heterogeneity, we chose these parameters to be close to real world scenarios, that is, room fares are different for different customer types but not considerably different.

As discussed earlier, the number of customer types and the number customer arrivals are equal. This means that each customer type arrives once on the platform \emph{in expectation}, as we assumed that all customer types have the same stationary rate of arrival of $1/m$. Naturally, it does not mean that each customer type arrives exactly once on the platform. Furthermore, we assign different patience levels to customers to see how the four algorithms behave when they are allowed to offer assortments to customers in different number of stages. Moreover, in different settings, a platform might be interested in offering assortments of at most a certain size to avoid overwhelming customers or giving them too many choices. To observe how this affects the performance of algorithms, we also try different upper-bounds on the size of assortments that can be offered. Furthermore, we try out a range of loading factors from 1 to 7 to see its impact on different algorithms' performances. 

Lastly, we are also interested in seeing how the gap between the low and high fares of rooms might affect algorithms' performances. To do this, we first scale the original high fares of rooms by a \emph{scale factor} and then use that in the Gaussian distribution discussed earlier to introduce heterogeneity in high fares for rooms for different customer types. Moreover, as discussed earlier, the weight of no-purchase option in MNL model cannot be estimated from the data. Accordingly, we shift the mean utilities in the MNL model so that for each customer type the weights of the no-purchase option and the most preferable purchase option is equal. In the settings that the scale factor is greater than 1 (i.e., where greater fare differentiation is induced in the model), the mean utility of the no-purchase option is also multiplied by the scale factor for each customer type. This is to ensure that the revenue-maximizing assortments still contain both high and low fares, hence, keeping the problem instances nontrivial.

Figure~\ref{fig:hetero-with-pricing-sc=2} depicts the performance of the 4 algorithms on the hotel data set for uniform patience levels of 1 to 4 and two different bounds on the size of allowed assortments, which are 1 and 4. Moreover, the scale factor is 2 here. Appendix~\ref{sec:apx-sec::simul} demonstrates the desirable performance of our algorithms on some additional problem settings on the hotel data set.

\begin{figure}[H]
	\centering
	\begin{tabular}{c c}
		\hbox
		{\subfloat[Max. assortment size=1, patience=1]{\includegraphics[width=0.4\textwidth]{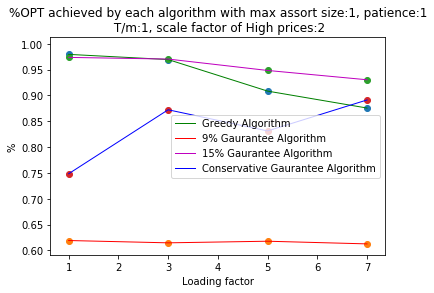}}} & 
		\subfloat[Max. assortment size=4, patience=1]{\includegraphics[width=0.4\textwidth]{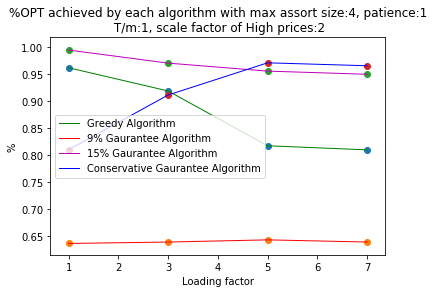}}\\
		{\subfloat[Max. assortment size=1, patience=2]{\includegraphics[width=0.4\textwidth]{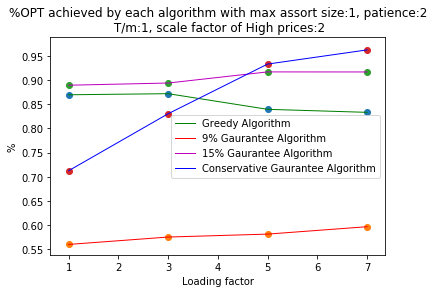}}} & 
		\subfloat[Max. assortment size=4, patience=2]{\includegraphics[width=0.4\textwidth]{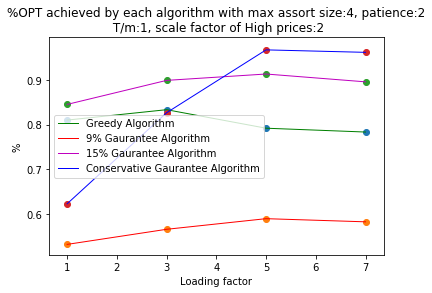}}		
	\end{tabular}
	\caption{Performance of Greedy, Conservative, Algorithm~\ref{alg:asst1} (i.e., the 9\% algorithm) and Modified Algorithm~\ref{alg:asst1} (i.e., the 15\% algorithm) on hotel data set with heterogeneous room fares for different patience levels, sizes of permissible assortments for scale factor=2 over different loading factors.}
\end{figure}

\begin{figure}[H]
	\centering
	\ContinuedFloat 
	\begin{tabular}{c c}
		\hbox
		{\subfloat[Max. assortment size=1, patience=3]{\includegraphics[width=0.4\textwidth]{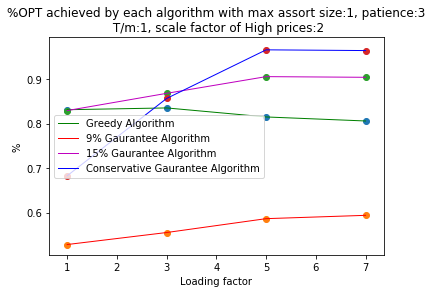}}} & 
		\subfloat[Max. assortment size=4, patience=3]{\includegraphics[width=0.4\textwidth]{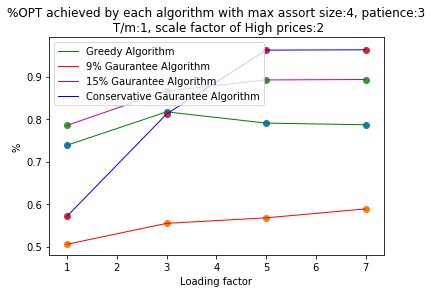}}\\
		{\subfloat[Max. assortment size=1, patience=4]{\includegraphics[width=0.4\textwidth]{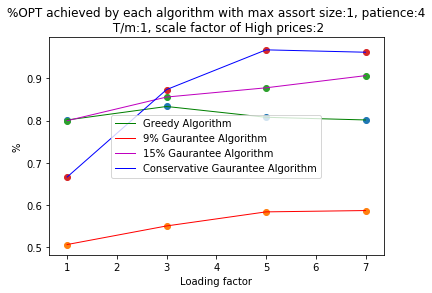}}} & 
		\subfloat[Max. assortment size=4, patience=4]{\includegraphics[width=0.4\textwidth]{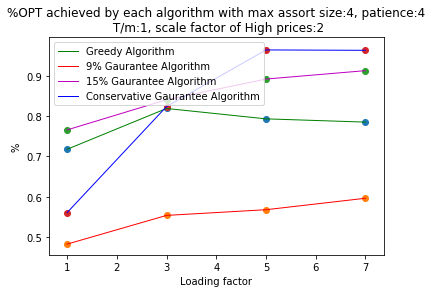}}
	\end{tabular}
	\caption{Continued: Performance of Greedy, Conservative, Algorithm~\ref{alg:asst1} (i.e., the 9\% algorithm) and Modified Algorithm~\ref{alg:asst1} (i.e., the 15\% algorithm) on hotel data set with heterogeneous room fares for different patience levels, sizes of permissible assortments for scale factor=2 over different loading factors.}
	\label{fig:hetero-with-pricing-sc=2}
\end{figure}
\vspace{-0.5cm}
Figure~\ref{fig:hetero-with-pricing-sc=2} shows that the performance of Algorithm~\ref{alg:asst1} (i.e., the 9\% algorithm) is not as good as other algorithms tested on the data. This was expected as Algorithm~\ref{alg:asst1}  offers assortments only to the first customer arrival of each type; whereas, this is not the case for the other three  algorithms. Moreover, it can be seen that in some cases, for low loading factors the Greedy algorithm outperform the other algorithms; on the other extreme, for large loading factors, the Conservative algorithm outperform all the other methods. However, it is important to observe that in cases where the loading factor is in-between these two extremes the Modified Algorithm~\ref{alg:asst1} (i.e., the 15\% algorithm) has the best performance of all. This is significant as a low loading factor means that there is not many customer arrivals to the hotel compared to its number of rooms on any day. In such scenario, the best strategy would be to follow the Greedy algorithm for offering assortments of rooms to customers. On the other hand, a high  loading factor means that the hotel gets full everyday, in which case it would be best to always offer rooms at their high prices as the Conservative algorithm does. Most hotels usually lie between these two extremes, that is, depending on occasions they are sometimes full and other times empty. In such cases, these experiments show that the Modified Algorithm~\ref{alg:asst1} (i.e., the 15\% algorithm) has the best performance.

\section{Conclusions}
\label{sec:conclusion}
In this paper, we studied the problem of offering assortments of items with limited inventories at different stages to customers of different types that have limited patience levels. Customers appear on the platform in a stochastic online fashion, that is, even though there is no information on the type of the future customer arrivals, their arrival rate is known. We studied the problem when only a single item (i.e., assortments of size 1) and general assortments can be offered to customers. We tested our algorithms on real world data and showed their significant improvements compared to the benchmark methods.


\bibliographystyle{informs2014} 
\bibliography{bibliography} 


\clearpage

\begin{APPENDICES}

\section{Deferred Proofs from Section~\ref{sec::single-item}} \label{apx-sec::single-item}
\proof{Proof of Lemma~\ref{lem::lp-ub}.}
To prove this, we need to show that the item offering strategy of any online algorithm satisfies the constraints of the LP on any instance of the problem. If so, the offering strategy is a feasible solution to the LP; therefore, the total revenue of the algorithm is bounded by the optimal value of the LP. 

Let $\bar{x}_{ij}$ be the probability that algorithm $A$ offers item $i$ to a customer of type $j$, that is, $\bar{x}_{ij}=\Pr[\offered_t(i,j)]$; thus, $0\le \bar{x}_{ij}\le 1$ satisfying constraint~\eqref{constr::01}. Clearly, no algorithm can offer more than $\ell_j$ items to a customer of type $j$. Therefore, for each type $j$ and time-step $t$
\begin{align*}
\sum_{i=1}^n \offered_t(i,j)\le \ell_j.
\end{align*}
Taking the expectation on algorithm's offering strategy we have that any valid strategy satisfies constraints~\eqref{constr::timeout} of the LP, that is $\sum_{i=1}^n \bar{x}_{ij}\le \ell_j$ for all customer types $j$. A customer purchases either 0 or 1 of the items offered to her, thus for any type $j$ and time-step $t$,
\begin{align*}
\sum_{i=1}^n \accept_t(i,j) \le 1.
\end{align*}
By taking expectation of above and conditioning on whether each item $i$ was offered, we have
\begin{align*}
\sum_{i=1}^n \Pr[\accept_t(i,j)|\offered_t(i,j)]\Pr[\offered_t(i,j)] = \sum_{i=1}^n p_{i,j}\bar{x}_{ij}\le 1.
\end{align*}
Therefore, any valid item offering strategy satisfies constraint~\eqref{constr::sellOne}. Similarly, since we are assuming there exists only a single copy of each item, each one of the items is matched to at most one customer. In other words, an item $i$ is matched at most at one of the time-steps $1,\ldots,T$ and the type of the customer arriving at that time-step is one among $1,\ldots,j$. Therefore, for each item $i$, 
\begin{align*}
\sum_{t=1}^T\sum_{j=1}^m \accept_t(i,j)\cdot\type_t(j)\le 1,
\end{align*}
and taking the expectation of it and using the tower property of conditional expectation we have
\begin{align*}
\sum_{t=1}^T\sum_{j=1}^m \Pr[\accept_t(i,j)|\offered_t(i,j)\cap\type_t(j)]\Pr[\offered_t(i,j)|\type_t(j)]\Pr[\type_t(j)] = \sum_{j=1}^m Tp_{ij}\bar{x}_{ij}q_j\le 1.
\end{align*}
Thus, any valid offering strategy satisfies constraint~\eqref{constr::inv} as well. This proves that $\bar{x}_{ij}$ is a feasible solution of the LP, and it follows that the revenue of each online algorithm is upper-bounded by the optimal value of the LP.
\Halmos\endproof

\proof{Proof of Lemma~\ref{lem:f(x)}.} We first find  $\lim_{z\rightarrow 0}f(z)$. Observe that both $z$ and $1-\exp(-z)$ are 0 at $z=0$. Therefore, we use L'H\^opital rule to determine this, which gives $\lim_{z\rightarrow 0}f(z)=\lim_{z\rightarrow 0}\frac{\exp(-z)}{1}=1$, as claimed. Moreover, $f'(z)=\frac{-1+z\exp(-z)+\exp(-z)}{z^2}$ and $f''(z)=\frac{2-2\exp(-z)-2z\exp(-z)-z^2\exp(-z)}{z^3}$. It can easily be checked that $f'(z)<0$ for all $z\in [0,1]$ as the largest value of the numerator of $f'(z)$ takes place when $z=1$ and it is still negative. Therefore, $f(z)$ is a decreasing function on the interval of interest, which in combination with $\lim_{z\rightarrow 0}f(z)=1$ concludes that $f(z)\le 1$ for $z\in[0,1]$. To see whether $f'(z)$ is bounded for $z\in [0,1]$, we first calculate $f'(1)$ and  $\lim_{z\rightarrow 0}f'(z)$ which are $-1+2\exp(-1)$ and $-\frac{1}{2}$, respectively, where the latter was calculated by applying  L'H\^opital rule two times. Since $f'(z)$ is continuous and it is bounded at the boundaries, it is also bounded in the $[0,1]$ interval. The last thing to show is the convexity of $f(z)$ in the given interval. Observe that on the boundaries of $[0,1]$, $\lim_{z\rightarrow 0}f''(z)=\frac{1}{3}>0$, again using L'H\^opital rule, and $f''(1)=2-\frac{5}{e}>0$. Hence, if the numerator of $f''(z)$  never equals 0, we have that $f''(z)\ge 0$ in the given interval, this $f(z)$ being convex. $f''(z)=0$ only if $2-2\exp(-z)-2z\exp(-z)-z^2\exp(-z)=0$. We show this does not take place by the way of contradiction. To do so, we rewrite this as $\exp(z)=1+z+\frac{z^2}{2}$. Note that by the Taylor expansion, $\exp(z)>1+z+\frac{z^2}{2}$, implying that we the numerator of $f''(x)$ and in turn $f''(z)$ can not be 0. Thus, $f''(z)>0$ for $z\in[0,1]$; in other words, for this interval $f(z)$ is convex.\Halmos\endproof

\proof{Proof of Theorem~\ref{thm::attenFramework}.}
Let $\ssx$ be the optimal solution of the LP. Consider an arbitrary item $i$ and type $j$.
Events $\avail_t(i)$ and $\type_t(j)$ are prerequisites for event $\offered_t(i,j)$ to occur, which in turn is a prerequisite for events $\Toffered_t(i,j)$ and $\accept_t(i,j)$ to take place.
Therefore, we can decompose $\Pr[\accept_t(i,j)]$ as
\begin{align} \label{eqn::decomp}
\Pr[\accept_t(i,j)|\Toffered_t(i,j)]&\cdot\Pr[\Toffered_t(i,j)|\offered_t(i,j)]\nonumber\times \allowdisplaybreaks\\&\Pr[\offered_t(i,j)|\avail_t(i)\cap\type_t(j)]\cdot\Pr[\avail_t(i)\cap\type_t(j)]. \allowdisplaybreaks
\end{align}
$\Pr[\accept_t(i,j)|\Toffered_t(i,j)]$ is simply equal to $p_{ij}$, because conditional on customer $t$ having type $j$ and being offered item $i$, her purchase choice is independent from all other events. Similarly, $\Pr[\avail_t(i)\cap\type_t(j)]=\gamma_tq_j$. 

To analyze $\Pr[\offered_t(i,j)|\avail_t(i)\cap\type_t(j)]$, note that the algorithm will offer items during time-step $t$ using the randomized procedure from Lemma~\ref{lem::focrsGeneral}, with $\ell=\ell_j$,  $A=U^t_j=\{i':\avail_t(i')\}$, and $p_{i'}=p_{i'j},x_{i'}=\ssx_{i'j}$ for all $i'\in A$.
By LP constraints (\ref{constr::sellOne})--(\ref{constr::01}), we know that $(x_{i'})_{i'\in A}$ satisfies the required  conditions in Lemma~\ref{lem::focrsGeneral}.
Therefore,  by Lemma~\ref{lem::focrsGeneral},
conditional on $i\in A$, the probability of $i$ being offered to customer $t$ who is of type $j$ is at least $\ssx_{ij}\cdot(1-\exp({-W_t(i,j)})/W_t(i,j)$. If $\ell_j\ge n$ for all types $j$, then $W_t(i,j)$ denotes the random variable
\begin{align}\label{eq:case1}
W_t(i,j)=\frac{1}{1-p_{ij}\ssx_{ij}}\sum_{i'\neq i}p_{i'j}\ssx_{i'j}\bI(\avail_t(i')),\allowdisplaybreaks
\end{align}
where $W_t(i,j)$ is understood to be 1 if $p_{ij}\ssx_{ij}=1$. Moreover, if $\sum_{i'\in [n]} p_{i'j}\le 1$ for all types $j$, then 
\begin{align}\label{eq:case2}
W_t(i,j) =\frac{1}{1-p_{ij}}\sum_{i'\neq i}p_{i'j}\ssx_{i'j}\bI(\avail_t(i')),
\end{align}
where $W_t(i,j)$ is understood to be 1 if $p_{ij}=1$. Going back to finding the probability of an item $i$ being offered, using the tower property of conditional expectation, we get that
\begin{align}
\Pr[\offered_t(i,j)|\avail_t(i)\cap\type_t(j)] &=\bE_{W_t(i,j)|\avail_t(i)\cap\type_t(j)}\big[\Pr[\offered_t(i,j)|\avail_t(i)\cap\type_t(j),W_t(i,j)]\big] \nonumber \allowdisplaybreaks\\
&\ge\bE_{W_t(i,j)|\avail_t(i)\cap\type_t(j)}\left[\ssx_{ij}\cdot\frac{1-\exp(-W_t(i,j))}{W_t(i,j)}\right] \nonumber \allowdisplaybreaks\\&=\ssx_{ij}\cdot \bE_{W_t(i,j)|\avail_t(i)\cap\type_t(j)}\left[f(W_t(i,j))\right]. \label{eqn::towerExp} \allowdisplaybreaks
\end{align}
By Lemma~\ref{lem:f(x)},
$f(z)=(1-e^{-z})/z$ is convex and decreasing, thus, applying Jensen's inequality
\begin{align}
\bE_{W_t(i,j)|\avail_t(i)\cap\type_t(j)}\left[f(W_t(i,j))\right]\ge f(\bE[W_t(i,j)|\avail_t(i)\cap\type_t(j)]). \label{eq:jensen}
\end{align}

We aim to bound the right-hand side of~\eqref{eq:jensen} below. Since conditioning on $\type_t(j)$ has no effect on $\avail_t(i')$ for any $i'\neq i$, by taking conditional expectation of~\eqref{eq:case1} we have
\begin{align*}
\bE[W_t(i,j)|\avail_t(i)\cap\type_t(j)] &=\frac{1}{1-p_{ij}\ssx_{ij}}\sum_{i'\neq i}p_{i'j}\ssx_{i'j}\bE[\bI(\avail_t(i'))|\avail_t(i)].
\end{align*}
At this point, we invoke \citep[Lemma~3.1]{brubach2017attenuate} from their original attenuation framework, which implies that at time-step $t$, item $i$ being available and item $i'$ being available are negatively correlated.
Therefore, we can subsequently derive
\begin{align}
\bE[W_t(i,j)|\avail_t(i)\cap\type_t(j)] &\le\frac{1}{1-p_{ij}\ssx_{ij}}\sum_{i'\neq i}p_{i'j}\ssx_{i'j}\Pr[\avail_t(i')] \nonumber \\ &=\frac{1}{1-p_{ij}\ssx_{ij}}\sum_{i'\neq i}p_{i'j}\ssx_{i'j}\cdot\gamma_t \le(1)\cdot\gamma_t, \label{eqn::expBound}
\end{align}
with the final inequality coming from Corollary~\ref{cor::atmost1}. Likewise, for the case that $\sum_{i'\in [n]}p_{i'j}\le 1$: 
\begin{align}
\bE[W_t(i,j)|\avail_t(i)\cap\type_t(j)]& \le \frac{1}{1-p_{ij}}\sum_{i'\neq i}p_{i'j}\ssx_{i'j}\Pr[\avail_t(i')] \nonumber\allowdisplaybreaks\\
&=\frac{1}{1-p_{ij}}\sum_{i'\neq i}p_{i'j}\ssx_{i'j}\cdot\gamma_t \le(1)\cdot\gamma_t \label{eqn::expBound2}, \allowdisplaybreaks
\end{align}
where again the last inequality comes from Corollary~\ref{cor::atmost1}. 

Combining~\eqref{eqn::towerExp} and~\eqref{eq:jensen} with~(\ref{eqn::expBound}) and (\ref{eqn::expBound2}), we get that for both cases of $\ell_j\ge n$ and $\sum_{i'\in[n]}p_{i'j}\le 1$ it holds that $\Pr[\offered_t(i,j)|\avail_t(i)\cap\type_t(j)]\ge\ssx_{ij}\cdot(1-e^{-\gamma_t})/\gamma_t$; thus,
\begin{align*}
\Pr[\offered_t(i,j)|\avail_t(i)\cap\type_t(j)]\cdot\Pr[\avail_t(i)\cap\type_t(j)] \ge (\ssx_{ij}\cdot\frac{1-e^{-\gamma_t}}{\gamma_t})\gamma_tq_j=(1-e^{-\gamma_t})q_j\ssx_{ij}.
\end{align*}
Clearly there exists an attenuation factor $a^{\edge}_t(i,j)\in[0,1]$ such that $a^{\edge}_t(i,j)\cdot\Pr[\offered_t(i,j)|\avail_t(i)\cap\type_t(j)]\cdot\Pr[\avail_t(i)\cap\type_t(j)]=(1-e^{-\gamma_t})q_j\ssx_{ij}$, where by the nature of the attenuation framework, $a^{\edge}_t(i,j)=\Pr[\Toffered_t(i,j)|\offered_t(i,j)]$. Putting everything in (\ref{eqn::decomp}) together, we have
\begin{align*}
\Pr[\accept_t(i,j)]=(1-e^{-\gamma_t})q_jp_{ij}\ssx_{ij}.
\end{align*}
Defining such an attenuation factor for all $i\in[n]$ and $j\in[m]$, we can deduce that for any item $i$,
\begin{align*}
\Pr[\avail_{t+1}(i)] &=\Pr[\avail_t(i)]-\sum_{j=1}^m\Pr[\accept_t(i,j)] =\gamma_t-(1-e^{-\gamma_t})\sum_{j=1}^mq_jp_{ij}\ssx_{ij} \ge\gamma_t-\frac{1-e^{-\gamma_t}}{T}, \allowdisplaybreaks
\end{align*}
with the inequality coming from LP constraint (\ref{constr::inv}).  Therefore, there exists an attenuation factor $a^{\vertex}_t(i)\in[0,1]$ such that
\begin{align}\label{eq:gamma-recurs}
\Pr[\avail_{t+1}(i)]=\gamma_t-\frac{1-e^{-\gamma_t}}{T}=\gamma_{t+1}, \allowdisplaybreaks
\end{align}
completing the proof and inductively establishing our modified attenuation framework.
\Halmos\endproof

\proof{Proof of Lemma~\ref{lem:g+h}.} First, we remark that for each $t\in [T]$, $\gamma_t$ is not only a function of $t$, but also a function of $T$, which can be clearly seen from~\eqref{eq:gamma-recurs}. To capture these two dependences, for a fixed $T$, we define function $g^T:[0,1]\rightarrow [0,1]$ such that for all $t\in [T]$, $g^T((t-1)/T)=\gamma_t$ and for all $t_1<t<t_1+1$, where $t_1\in[T]$, $g^T((t-1)/T)$ is the linear combination of $g^T((t_1-1)/T)$ and $g^T(t_1/T)$. Defining  $z=(t-1)/T$ and for a fixed $z$, $\{g^T(z)\}_{T\in \mathbb{N}}$ is a sequence of function and we want to study the behavior $g^T(z)$ as $T$ goes ot infinity. Using this notation,~\eqref{eq:gamma-recurs} can be written as:
\begin{align}
\frac{g^T(z+\frac{1}{T})-g^T(z)}{\frac{1}{T}}= -1+\exp(-g^T(z)).\label{eq:g-def}
\end{align}
By taking the limit of the above equation as $T\rightarrow\infty$ we define a new function $h(z)$, i.e., $h(z)=\lim_{T\rightarrow\infty}g^T(z)$. Recall that $\gamma_t$ is the probability that each item is available at time-step $t$ if there are a total of $T$ customer arrivals. Therefore $0\le g^T(z)=\gamma_t\le 1$. Since for such $g^T(z)$'s $(1-\exp(-g^T(z)))$ is positive, recursion~\eqref{eq:g-def} indicates that as $z$ increases $g^T(z)$ decreases as $z$ grows. In addition to this, we here want to prove that as $T$ increases $g^T(z)$ does not decrease. This is something stronger than what the lemma states and from this it can be concluded that for all $t$ and $T$ we have 
\begin{align}
g^T(z)\le \lim_{T'\rightarrow\infty} g^{T'}(z)=h(z),\label{eq:limit-gamma}
\end{align}
which then concludes that $\gamma_{T+1}=g^T(1)\le h(1)$. To show~\eqref{eq:limit-gamma} holds we use induction on $T$, with the inductive hypothesis that if $g^T(z)$ does not decrease with $T$  then $g^T(z+1/T)$ does not decrease $T$ either. The base of this induction is that $g^T(0)$ (i.e., when $t=1$) does not decrease with $T$. This is clearly true as $\gamma_1=g^T(0)=1$ and constant. Now assume that for some $0\le z\le 1$, we have that $g^T(z)$ is not decreasing with $T$. We can formulate $g^T(z+1/T)$ from~\eqref{eq:g-def}, which gives
\begin{align*}
g^T(z+\frac{1}{T})= g^T(z)-\frac{1}{T}(1-\exp(-g^T(z))).
\end{align*}
To show $g^T(z+\frac{1}{T})$ does not decrease with time, we differentiate $g^T(z+\frac{1}{T})$ with respect to $T$:
\begin{align}
\frac{dg^T(z+\frac{1}{T})}{dT}&=\frac{dg^T(z)}{dT}-\frac{\exp(-g^T(z))T\frac{dg^T(z)}{dT}-(1-\exp(-g^T(z)))}{T^2}\nonumber\\&=\frac{\frac{dg^T(z)}{dT}T(T-\exp(-g^T(z)))}{T^2}+\frac{(1-\exp(-g^T(z)))}{T^2}.\label{eq:g-deriv}
\end{align}
Since $g^T(z)\ge 0$, the second term in~\eqref{eq:g-deriv} is clearly positive. For the first term of~\eqref{eq:g-deriv}, by the same reason $(T-\exp(-g^T(z)))\ge 0$; moreover, recall that by the inductive assumption, ${dg^T(z)}/{dT}\ge 0$. Therefore, we have that ${dg^T(z+\frac{1}{T})}/{dT}\ge 0$ and our induction proof concludes. This implies that~\eqref{eq:limit-gamma} holds and consequently, $\gamma_{T+1}=g^T(1)\le h(1)$. 

It remains to find the value of $h(1)$. Recall that $h(z)$ was defined $h(z)=\lim_{T\rightarrow\infty}g^{T}(z)$. Taking the limit of~\eqref{eq:g-def} as $T$ goes to infinity results in $\frac{dh}{dz}$ on the right-hand side. Therefore, in the limit, function $h(z)$ needs to satisfy the following differential equation 
\begin{align}
\frac{dh(z)}{dz}= -1+\exp(-h(z))\label{eq:ode},
\end{align}
with the boundary condition the same as that of $g^T(0)$, that is, $h(0)=1$. We solve this differential equation for $h(z)$. By multiplying both sides by $\exp(h(z))$ and $dz$ we have
\begin{align*}
\exp(h(z))dh(z) = -(\exp(h(z))-1)dz,
\end{align*}
thus 
\begin{align*}
\frac{\exp(h(z))}{\exp(h(z))-1}dh(z) = -dz.
\end{align*}
The above equation is equivalent to
\begin{align*}
d\ln(\exp(h(z))-1) = -dz.
\end{align*}
Integrating the above in combination to the boundary condition $h(0)=1$ concludes that 
\begin{align}
h(z) = \ln((e-1)\exp(-z)+1)\label{eq:h}.
\end{align}
Plugging $z=1$ in~\eqref{eq:h} gives $h(1)=\ln(2-1/e)$, finalizing our claim of $\gamma_{T+1}\le h(1)=\ln(2-\frac{1}{e})$.
\Halmos\endproof

\proof{Proof of Lemma~\ref{lem:ub-proof}.} We prove the lemma by coming up with an instance of the problem for which we can show it is not possible to have an algorithm with an approximation ratio larger than $1-\ln(2-1/e)$. Consider an instance of the problem where there exists $T=n$ time-steps in which customers of $n$ different types may arrive, where $\ell_j=n$ and $q_j=1/n$ for all customer types $j=1,\ldots, n$. Moreover, there are $n$ items and for all pairs of $i$ and $j$, $p_{ij}=\frac{1}{n}$ and $r_{ij}=1$. As $n$ goes to infinity, we show that no online algorithm can have an approximation ratio larger than $1-\ln(2-1/e)$. 

Since the patience level is equal to the total number of items, a visiting customer would not go out of patience even if all the items are shown to her. Therefore, customers' patience is not a concern in this case of the problem. To study the best possible approximation ratio we first need to determine the value of $\OPT$. It can be easily checked that $x_{ij}=1$ for all $i=1,\ldots,n$ and $j=1,\ldots,n$ is a feasible solution to the LP, for which the objective value of the LP is $n$. Therefore, $\OPT\ge n$. Note that it can be easily checked that $\OPT=n$; however, having $\OPT\ge n$ is sufficient for our purpose.

For each time-step $t=1,\ldots,n$, let $N_t$ denote the total number of available items at the beginning of time-step $t$; therefore, we have $N_1=n$. Since $N_{n+1}$ denotes the number of remaining items once the process is over, for any algorithm its total revenue is $\ALG=n-N_{n+1}$. The approximation ratio of an algorithm is the ratio of its total expected revenue to $\OPT$. Denoting the approximation ratio with $\alpha$ we have 
\begin{align}
\alpha=\frac{\EXP[\ALG]}{\OPT}\le \frac{\EXP[n-N_{n+1}]}{n}=1-\EXP[\frac{N_{n+1}}{n}].\label{eq:hardness-apx-ratio}
\end{align}
In what follows we show that in the limit that $n$ goes to infinity ${\EXP[N_{n+1}]}/{n}\ge  \ln(2-1/e)$, and using\eqref{eq:hardness-apx-ratio} conclude that $\alpha\le 1-\ln(2-1/e)$. To do this, first note that since at each time-step at most one item can be purchased by a customer, $N_t-N_{t+1}$ is either 0 or 1. Therefore, $\EXP[N_t-N_{t+1}|N_t]$ is equal to the probability that a purchase is made during time-step $t$. As there are $N_t$ items at the beginning of this time-step and all of them can get the opportunity to be offered to the visiting customer and their purchase probability is $1/n$ we have
\begin{align*}
\EXP[N_t-N_{t+1}|N_t]=1-(1-\frac{1}{n})^{N_t}.
\end{align*}
Taking another expectation of the above equation we have
\begin{align*}
\EXP[N_t-N_{t+1}]=\EXP[1-(1-\frac{1}{n})^{N_t}].
\end{align*}
With simple calculus and taking twice differentiation it can be seen that $1-(1-\frac{1}{n})^{N_t}$ is a concave function. Therefore, using Jensen's inequality and negating the above equation we have 
\begin{align}
\EXP[N_{t+1}-N_{t}]\ge  -1+(1-\frac{1}{n})^{\EXP[N_t]}=-1+\exp(-\EXP[\frac{N_t}{n}])-o(1).\label{eq:EXP-eqn}
\end{align}
Defining function $g^n:[0,1]\rightarrow [0,1]$ such that for all $t\in [n]$, $g^n((t-1)/n)=\EXP[N_t/n]$ and setting $z=(t-1)/n$, inequality~\eqref{eq:EXP-eqn} becomes
\begin{align}
\frac{g^n(z+\frac{1}{n})+g^n(z)}{\frac{1}{n}}\ge  -1+\exp(-g^n(z))-o(1).\label{eq:EXP-eqn2}
\end{align}
Taking the limit of above as $n$ goes to infinity and introducing $b(z)=\lim_{n\rightarrow\infty}g^n(z)$ we have 
\begin{align}
\frac{db(z)}{dz}\ge  -1+\exp(-b(z)),\label{eq:ode-hardness}
\end{align}
with the boundary condition $b(0)=\EXP[N_1/n]=1$. Observe that the way $b(z)$ is defined we have $b(1)=\lim_{n\rightarrow\infty} \EXP[N_{n+1}/n]$. We now show $b(1)\ge \ln(2-1/e)$. Differential inequality~\eqref{eq:ode-hardness} is similar to differential equation~\eqref{eq:ode} (that is, ${dh(z)}/{dz}=  -1+\exp(-h(z))$ with boundary condition $h(0)=1$) discussed in the proof of Lemma~\ref{lem:g+h}, with the difference of having an equation there and an inequality here. By~\citep[Lemma 3.4]{khalil'02}, for all $0<z\le 1$ we have $b(z)\ge h(z)$, in particular, $b(1)\ge h(1)= \ln(2-1/e)$. Hence, taking the limit of~\eqref{eq:hardness-apx-ratio} as $n$ goes to infinity we have
\begin{align*}
\alpha=\frac{\EXP[\ALG]}{\OPT}\le 1-\EXP[\frac{N_{n+1}}{n}]\le 1- \ln(2-1/e)\simeq 0.51.
\end{align*}
Therefore, there exists no algorithm with an approximation ratio larger than $1- \ln(2-1/e)$ for the stochastic matching with timeouts problem.
\Halmos\endproof

\section{Supplements to Section~\ref{sec::generalAsst}}\label{apx-sec::generalAsst}
This section is constituted of two subsections. The first subsection discusses the proof of Theorem~\ref{thm:assort-0.51} as well as the supplementary lemmas for it to prove the 0.51 approximation for the multi-stage and multi-customer assortment optimization when the platform is allowed to show repeated items to customers. In the second subsection, we provide the deferred proofs from Section~\ref{sec::generalAsst}.

\subsection{Supplements to Section~\ref{subsec:assort-repeated}}\label{apx-subsec:assort-repeated}
Our approach here is similar to, in fact a generalization of, the algorithms we had for the online stochastic matching with timeouts problem. That is, the problem is divided into two subproblems: the online subproblem is addressed via an offline black-box and the online problem uses an attenuation framework. Lemma~\ref{lem:bb-assort} discusses the black-box algorithm and Lemma~\ref{lem::attenFramework-assort-repeated} is about the attenuation framework for this assortment optimization problem.

\begin{lemma}[Black-box Randomized Procedure (Assortment Version)] \label{lem:bb-assort}
	Let $\cA$ be a set of assortment of some coins, where these assortments can potentially overlap with one another.  For an assortment $S\in \cA$, when a coin $i\in S$ is flipped it lands on ``heads''  with probability $p(i,S)$ and independent from any other coin in any other assortment. Once an assortment is picked all its coins are flipped at once, which is referred to as flipping the assortment.
	We can flip assortments in any (possibly randomized) order, and must stop once we get a ``heads'', or have flipped $\ell$ assortments, where $\ell$ is a positive integer.
	
	Let $(x(S))_{S\in \cA}$ be any vector of weights in $[0,1]^{|\cA|}$ satisfying $\sum_{S\in \cA}\sum_{i\in S}p(i,S)x(S)\le1$ and $\sum_{S\in \cA}x(S)\le\ell$.
	Then there exists a randomized procedure for flipping the assortments such that the probability of any assortment $S$ being flipped, before the process is stopped, is at least
	\begin{align} \label{eqn:boundInLemma-assort}
	\frac{1-e^{-w(S)}}{w(S)}\cdot x(S),
	\end{align}
	where $w(S)=\frac{1}{1-\sum_{i\in S} p(i,S)}\sum_{S'\neq S}\sum_{i\in S'}p(i,S')x(S')$ (or $w(S)$ is understood to be 1 if $\sum_{i\in S} p(i,S)=1$) if $\sum_{S\in \cA}\sum_{i\in S}p(i,S)\le 1$ and $w(S)=\frac{1}{1-\sum_{i\in S}p(i,S)x(S)}\sum_{S'\neq S}\sum_{i\in S'}p(i,S')x(S')$ (or $w(S)$ is understood to be 1 if $\sum_{i\in S}p(i,S)x(S)=1$) if $\ell\ge |\cA|$.
\end{lemma}

\proof{Proof.} The proof of this lemma is essentially a generalization of that of Lemma~\ref{lem::focrsGeneral} to account for assortments of coins. As the elements of $(x(S))_{S\in \cA}$ are fractional and do not determine whether an assortment $S$ should be flipped, we again use the GKPS procedure to round them to 0 and 1. So, assortment $S$ is among the assortments that can be flipped if and only if $x(S)$ is rounded up to 1 by the GKPS process. By applying the GKPS rounding, vector $(x(S))_{S\in \cA}$ is rounded to $(X(S))_{S\in \cA}$, where each $(X(S))_{S\in \cA}$ satisfies all three properties mentioned in Theorem~\ref{thm:GKPS}. We say an assortments comes heads if when the assortment is flipped at least one of its coins comes heads.

\noindent\textbf{The case of }{$\mathbf{\ell\ge |\cA|}$:} Algorithm~\ref{alg:BB-assort} shows how the randomized algorithm works in this case of the problem.
\begin{algorithm}
	\caption{Black-box (Assortment Version)}\label{alg:BB-assort}
	\textbf{INPUT:} $\ell$, $\cA$, $x(S)$ and $p(i,S)$ for all $S\in \cA$ and $i\in S$
	\begin{algorithmic}[1]
		\State Apply the GKPS rounding to $(x(S))_{S\in \cA}$. Let $\tilde{U}$ be the set of assortments that are rounded up by the GKPS process.
		\State For each assortment $S\in \tilde{U}$ pick a number $Y_S$ uniformly at random and IID from $[0,1]$.
		\State Flip assortments $S\in \tilde{U}$ in an increasing order of $\frac{Y_S}{1-\sum_{i\in S}p(i,S)}$ (if $\sum_{S\in \cA}\sum_{i\in S}p(i,S)\le 1$) and $\frac{Y_S}{1-\sum_{i\in S}p(i,S)x(S)}$ (if $\ell\ge |\cA|$) until a ``heads" comes or $\ell$ assortments are flipped.
	\end{algorithmic}
\end{algorithm}	
Consider any assortment $S\in \cA$ and suppose it is rounded up by the GKPS process. Assuming $Y_S=y$, an assortment $S'$ that has also passed through the GKPS process is flipped before $S$ if $\frac{Y_{S'}}{1 - \sum_{i\in S'}p(i,S')x(S')}\le \frac{y}{1 - \sum_{i\in S}p(i,S)x(S)}$. 
Let us provide an upper-bound on the probability that assortment $S'$ is flipped before $S$. To do this, we divide the problem into two cases: $y\ge 2(1- \sum_{i\in S}p(i,S)x(S))$, and otherwise, $y< 2(1 -\sum_{i\in S}p(i,S)x(S))$. Beginning with the former case we have
\begin{align}
\frac{1-\exp(-\frac{y\sum_{i \in S'}p(i,S')x(S')}{1- \sum_{i\in S}p(i,S)x(S)})}{\sum_{i\in S'}p(i,S')x(S')} &\ge \frac{1-\exp(-2\sum_{i\in S'}p(i,S')x(S'))}{\sum_{i\in S'}p(i,S')x(S')}\nonumber\\&\ge 2(1-\sum_{i\in S'}p(i,S')x(S'))\ge 1\label{eq:RHS-assort},
\end{align}
where~\eqref{eq:RHS-assort} uses the Taylor expansion of $\exp(-2\sum_{i\in S'}p(i,S')x(S'))$ as well as $\sum_{i\in S'} p(i,S')x(S')\le 1/2$, which is implied by $\sum_{S\in\cA}\sum_{i\in S}p(i,S)x(S)\le 1$ in addition to $y\ge 2(1- \sum_{i\in S}p(i,S)x(S))$ and $y\le 1$. This makes $\frac{1-\exp(-\frac{y\sum_{i \in S'}p(i,S')x(S')}{1- \sum_{i\in S}p(i,S)x(S)})}{\sum_{i\in S'}p(i,S')x(S')}$ a potential upper-bound on the probability of assortment $S'$ getting flipped before assortment $S$. Let us now consider the case that $y< 2(1 -\sum_{i\in S}p(i,S)x(S))$. We have that 
\begin{align}
\frac{1-\exp(-\frac{y\sum_{i \in S'}p(i,S')x(S')}{1- \sum_{i\in S}p(i,S)x(S)})}{\sum_{i\in S'}p(i,S')x(S')} &\ge (1-\frac{y\sum_{i\in S'}p(i,S')x(S')}{2({1- \sum_{i\in S}p(i,S)x(S)})})\cdot \frac{y}{1- \sum_{i\in S}p(i,S)x(S)}\label{eq:taylor-assort}\allowdisplaybreaks\\
&\ge (1-\sum_{i\in S'}p(i,S')x(S'))\cdot \frac{y}{1- \sum_{i\in S}p(i,S)x(S)}\label{eq:case2of1-assort},\allowdisplaybreaks
\end{align}
where again~\eqref{eq:taylor-assort} uses a Taylor expansion similar to the one discussed earlier and assumption $y< 2(1 -\sum_{i\in S}p(i,S)x(S))$ was used in~\eqref{eq:case2of1-assort}. Assortment $S'$ is flipped before $S$ if $Y_{S'}\le ({1 - \sum_{i\in S'}p(i,S')x(S')})\cdot\frac{y}{1 - \sum_{i\in S}p(i,S)x(S)}$ and $Y_{S'}$ is picked uniformly at random in $[0,1]$; thus,~\eqref{eq:case2of1-assort} gives the probability of flipping $S'$ before $S$. Therefore, using $I^{S',S}$ to denote the indicator variable that assortment $S'$ was flipped before assortment $S$, we have
\begin{align}
\Pr[I^{S',S}|S,S'\in \tilde{U},Y_S=y]\le \frac{1-\exp(-\frac{y\sum_{i \in S'}p(i,S')x(S')}{1- \sum_{i\in S}p(i,S)x(S)})}{\sum_{i\in S'}p(i,S')x(S')}.\label{eq:assort-case1-bound}
\end{align}

We are now able to bound the probability that an assortment $S$ is flipped before the process is stopped for the case that $\ell\ge |\cA|$. A lower-bound on the probability that $S$ is flipped given it is rounded up by the GKPS process is provided below:
\begin{align}
&\Pr[S \mathrm{~flipped}|S\in \tilde{U}] \ge \Pr[\bigcap_{S'\neq S} S' \mathrm{~not~heads~before~}S|S\in \tilde{U}]\nonumber\allowdisplaybreaks\\
&= \prod_{S'\neq S} \Pr[S'\mathrm{~not~heads~before~}S|S\in \tilde{U}]\label{eq:assort-main2}\allowdisplaybreaks\\
&= \int_0^1 \prod_{S'\neq S} \Pr[S'\mathrm{~not~heads~before~}S|S\in \tilde{U}, Y_S=y]dy\nonumber\allowdisplaybreaks\\
&= \int_0^1 \prod_{S'\neq S} (1-\Pr[S'\mathrm{~heads~before~}S|S\in \tilde{U}, Y_S=y])dy\nonumber\allowdisplaybreaks\\
&= \int_0^1 \prod_{S'\neq S} (1-\Pr[S'\in \tilde{U}|S\in \tilde{U}, Y_S=y]\Pr[I^{S',S}|S,S'\in \tilde{U}, Y_S=y]\Pr[S'\mathrm{~flips~heads}|S,S'\in \tilde{U}, Y_S=y, I^{S',S}])dy\nonumber\allowdisplaybreaks\\
&\ge \int_0^1 \prod_{S'\neq S} (1-x(S')\frac{1-\exp(-\frac{y\sum_{i \in S'}p(i,S')x(S')}{1- \sum_{i\in S}p(i,S)x(S)})}{\sum_{i\in S'}p(i,S')x(S')}\sum_{i\in S'}p(i,S'))dy\label{eq:assort-main6}\allowdisplaybreaks\\
&= \int_0^1 \prod_{S'\neq S} \exp(-\frac{y\sum_{i \in S'}p(i,S')x(S')}{1- \sum_{i\in S}p(i,S)x(S)})dy\nonumber\allowdisplaybreaks\\
&= \int_0^1 \exp(-y\sum_{S'\neq S}\sum_{i
	\in S'}\frac{p(i,S')x(S')}{1-\sum_{i\in S}p(i,S)x(S)})dy\nonumber\allowdisplaybreaks\\
&= \frac{1}{\sum_{S'\neq S}\sum_{i
		\in S'}\frac{p(i,S')x(S')}{1-\sum_{i\in S}p(i,S)x(S)}}(1-\exp(-\sum_{S'\neq S}\sum_{i
	\in S'}\frac{p(i,S')x(S')}{1-\sum_{i\in S}p(i,S)x(S)}))=\frac{1-\exp(-w(S))}{w(S)}\nonumber,\allowdisplaybreaks
\end{align}
where~\eqref{eq:assort-main2} uses the fact that as there is enough patience to flip all assortments when $\ell\ge|\cA|$ as well as the independence among the events of assortments $S'\neq S$ not being flipped heads. Inequality~\eqref{eq:assort-main6} uses~\eqref{eq:assort-case1-bound} and $\Pr[S'\in \tilde{U}|S\in \tilde{U}, Y_S=y]\le \Pr[S'\in \tilde{U}]=x(S')$. Therefore, 
\begin{align*}
\Pr[S\mathrm{~flipped}]&=\Pr[S\mathrm{~flipped}|S\in \tilde{U}]\Pr[S\in \tilde{U}]= \Pr[S\mathrm{~flipped}|S\in \tilde{U}]\cdot x(S) \ge \frac{1-\exp(-w(S))}{w(S)}\cdot x(S).
\end{align*}

\noindent\textbf{The case of }{$\mathbf{\sum_{S\in \cA}\sum_{i\in S}p(i,S)\le 1}$:} Similar to the previous case, first the GKPS process is performed on $(x(S))_{S\in \cA}$, and for an each assortment $S$ that has rounded up by the GKPS process, i.e., $S\in \tilde{U}$, a random variable $Y_S$ is picked IID and uniformly at random in [0,1], see Algorithm~\ref{alg:BB-assort}. For this case, assortments are ranked in an increasing order of $Y_S/(1-\sum_{i\in S}p(i,S))$ and they are flipped in this order until the process stops. So, assuming $Y_S=y$, assortment $S'$ is flipped before $S$ if $\frac{Y_{S'}}{1 - \sum_{i\in S'}p(i,S')}\le \frac{y}{1 - \sum_{i\in S}p(i,S)}$. 
Again, we divide the problem into the following two cases: $y\ge 2(1- \sum_{i\in S}p(i,S))$, and otherwise, $y< 2(1 -\sum_{i\in S}p(i,S))$. Beginning with the former case we have
\begin{align}
\frac{1-\exp(-\frac{y\sum_{i\in S'}p(i,S)}{1- \sum_{i\in S}p(i,S)})}{\sum_{i\in S'}p(i,S')} &\ge \frac{1-\exp(-2\sum_{i\in S'}p(i,S'))}{\sum_{i\in S'}p(i,S')}\ge 2(1-\sum_{i\in S'}p(i,S'))\label{eq:RHS-small-assort}\allowdisplaybreaks.
\end{align}
We have $\sum_{i\in S}p(i,S)+\sum_{i\in S'}p(i,S')\le 1$ as in this case it is assumed $\sum_{S\in \cA}\sum_{i\in S}p(i,S)\le 1$. This in addition to $2(1- \sum_{i\in S}p(i,S))\le y\le 1$ gives $\sum_{i\in S'}p(i,S')\le \frac{1}{2}$.
Therefore, $\frac{1-\exp(-\frac{y\sum_{i\in S'}p(i,S)}{1- \sum_{i\in S}p(i,S)})}{\sum_{i\in S'}p(i,S')}\ge 2(1-\sum_{i\in S'}p(i,S')) \ge 1,$ making $\frac{1-\exp(-\frac{y\sum_{i\in S'}p(i,S)}{1- \sum_{i\in S}p(i,S)})}{\sum_{i\in S'}p(i,S')}$ a possible upper-bound on the probability that assortment $S'$ gets flipped before $S$ for this case of the problem. Next, we consider the case $y< 2(1 -\sum_{i\in S}p(i,S))$, for which we have 
\begin{align}
\frac{1-\exp(-\frac{y\sum_{i\in S'}p(i,S)}{1- \sum_{i\in S}p(i,S)})}{\sum_{i\in S'}p(i,S')}
&\ge \left(1-\frac{y\sum_{i\in S'}p(i,S')}{2(1-\sum_{i\in S}p(i,S))}\right)\cdot \frac{y}{1-\sum_{i\in S}p(i,S)}\label{eq:taylor-small-assort} \allowdisplaybreaks\\
&\ge \left(1-\sum_{i\in S'}p(i,S')\right) \cdot \frac{y}{1-\sum_{i\in S}p(i,S)}\label{eq:case-assm-small-assort}, \allowdisplaybreaks
\end{align}
where~\eqref{eq:taylor-small-assort} and~\eqref{eq:case-assm-small-assort} use a Taylor expansion similar to the one discussed in the proof of Lemma~\ref{lem::focrsGeneral} and $y< 2(1 -\sum_{i\in S}p(i,S))$, respectively. The probability of flipping $S'$ before $S$ is at most $\left(1-\sum_{i\in S'}p(i,S')\right) \cdot {y}/({1-\sum_{i\in S}p(i,S)})$, therefore
\begin{align*}
\Pr[I^{S',S}|S,S'\in \tilde{U},Y_S=y]\le \frac{1-\exp(-\frac{y\sum_{i \in S'}p(i,S')}{1- \sum_{i\in S}p(i,S)})}{\sum_{i\in S'}p(i,S')}.
\end{align*}
With this, we go forward to provide the lower-bound on the probability that assortment $S$ is flipped before the process stops.
\begin{align}
&\Pr[S\mathrm{~flipped}| S\in \tilde{U}]\ge \EXP_{\tilde{U}}[\int_0^1\prod_{i'\in \tilde{U}\backslash\{S\}}(1-\frac{1-\exp(-\frac{y\sum_{i\in S'}p(i,S)}{1- \sum_{i\in S}p(i,S)})}{\sum_{i\in S'}p(i,S')}\sum_{i\in S'}p(i,S'))dy|S\in \tilde{U}]\label{eq:lem-intro-assort} \allowdisplaybreaks\\
&= \EXP_{\tilde{U}}[\int_0^1\exp(-y\sum_{S'\in \tilde{U}\backslash\{S\}}\sum_{i\in S'}\frac{p(i,S')}{1- \sum_{i\in S}p(i,S)})dy|S\in \tilde{U}]\nonumber \allowdisplaybreaks\\
&= \EXP_{\tilde{U}}[\frac{(1-\exp(-\sum_{S'\in \tilde{U}\backslash\{S\}}\sum_{i\in S'}\frac{p(i,S')}{1-\sum_{i\in S} p(i,S)}))}{\sum_{S'\in \tilde{U}\backslash\{S\}}\sum_{i\in S'}\frac{p(i,S')}{1- \sum_{i\in S}p(i,S)}}|S\in \tilde{U}]\nonumber \allowdisplaybreaks\\
&\ge\frac{(1-\exp(- \EXP_{\tilde{U}}[\sum_{S'\in \tilde{U}\backslash\{S\}}\sum_{i\in S'}\frac{p(i,S')}{1- \sum_{i\in S}p(i,S)}|S\in \tilde{U}]))}{ \EXP_{\tilde{U}}[\sum_{S'\in \tilde{U}\backslash\{S\}}\sum_{i\in S'}\frac{p(i,S')}{1- \sum_{i\in S} p(i,S)}|S\in \tilde{U}]}\label{eq:lem-jens-assort} \allowdisplaybreaks\\&\ge\frac{1}{\sum_{S'\neq S}\sum_{i\in S'}\frac{p(i,S')x(S')}{1- \sum_{i\in S}p(i,S)}}(1-\exp(- \sum_{S'\neq S}\sum_{i\in S}\frac{p(i,S')x(S')}{1- \sum_{i\in S} p(i,S)}))\label{eq:lem-f-dec-assort} \allowdisplaybreaks\\&\ge \frac{1-\exp(-w(S))}{w(S)}\nonumber. \allowdisplaybreaks
\end{align}
The reason behind~\eqref{eq:lem-intro-assort} is that, for $S$ to get flipped before the process is stopped, all assortment in $\tU$ that are flipped before it need to come tails. In~\eqref{eq:lem-jens-assort}, Jensen's inequality and convexity of function $(1-\exp(-z))/z$ are used (see Lemma~\ref{lem:f(x)}). The fact that $(1-\exp(-z))/z$ is a decreasing function and $\EXP_{\tilde{U}}[\sum_{S'\in \tilde{U}\backslash\{S\}}\sum_{i\in S'}{p(i,S')}/{(1- \sum_{i\in S}p(i,S))}|S\in \tilde{U}]\le \sum_{s'\neq S}\sum_{i\in S}{p(i,S')x(S')}/{(1- \sum_{i\in S}p(i,S))}$ are used for~\eqref{eq:lem-f-dec-assort}. The former is by Lemma~\ref{lem:f(x)} and a similar proof for the latter is provided in the proof of Lemma~\ref{lem::focrsGeneral}. Therefore, for this case of the problem we have
\begin{align*}
\Pr[S\mathrm{~flipped}]\ge \frac{1-\exp(-w(S))}{w(S)}\cdot x(S).
\end{align*}
\Halmos\endproof

The above lemma addressed the offline subproblem for the case that assortments of items can be offered at each stage of the problem and showing repeated items is allowed. We now shift our focus to the attenuation framework used to address the online subproblem for this problem. Recall that the attenuation framework used in the stochastic matching with timeouts problem consisted of two parts, those are, vertex and edge-attenuation frameworks. In a few words, vertex-attenuation regulated the probability that each item $i$ is available at the beginning of each time-step; whereas, edge-attenuation adjusted the probability each item $i$ is offered to each customer type $j$. For the case that an assortment of items is offered at each stage to a customer, we use a similar attenuation framework to address the online subproblem. The difference is that here the edge-attenuation framework regulates the probability that an an item $i$ along with an assortment $S\ni i$ is offered to a customer type $j$.

Let $U^t$ denote the set of items that are still available  when customer $t$ arrives. Suppose customer $t$ is of type $j$, then let $\cS_j(U^t)$ denote the set of assortments in $\cS$ for which $\ssx_j(S)>0$ with some modifications performed on them: the items that are sold-out are removed from each of these assortments and if an assortment has become empty by doing so, then it is not included in $\cS_j(U^t)$. Moreover, let $\gamma_t$ be as it was defined in Definition~\ref{def:gamma}. Lastly, we use events $\type_t(j)$ and $\avail_t(i)$ as introduced in Definition~\ref{def:single-item-event} and we modify the other events discussed there into $\offered_t(i,S,j)$, $\Toffered_t(i,S,j)$, and $\accept_t(i,S,j)$. These events are natural extensions of similar events in Definition~\ref{def:single-item-event} with the purpose of capturing an item $i$ being offered along an assortment $S$, formally defined below.
\begin{definition}\label{def:assortment-repeat-event}
	For each $t\in[T]$, item $i\in[n]$, assortment $S\in\cS$ and type $j\in[m]$ let us define the following events:
	\begin{itemize}
		\item $\offered_t(i,S,j)$: the algorithm (pre-attenuation) intends to offer item $i$ along assortment $S$ to customer $t$ that is of type $j$;
		\item $\Toffered_t(i,S,j)$: the algorithm (post-attenuation)  offers item $i$ along assortment $S$ to customer $t$, whose type is $j$;
		\item $\accept_t(i,S,j)$: customer $t$, with type $j$, would have purchased item $i$ if truly offered along assortment $S$ (i.e., offered post-attenuation).
	\end{itemize}
\end{definition}
With these in mind let us discuss the overall online algorithm for the multi-stage and multi-customer assortment optimization problem where an item can be offered multiple times to a customer and prove the algorithm's performance.

\begin{algorithm}
	\caption{Online Algorithm for the Assortment Optimization with Repeated Offerings}\label{alg:full-assort-repeated}
	\begin{algorithmic}[1]
		\State Before any customers arrive, solve the MCDLP-R to get $x^*$.
		\For{time-steps $t=1,\ldots, T$}
		\State Suppose the arriving customer be of type $j$. Let $U^t$ be the set of available items when customer $t$ arrives and let the set of assortments $\cS_j(U^t)$ be defined as $\cS_j(U^t)=\{S': S'\subseteq U^t, S'\neq \emptyset,\exists S\in\cS \text{ s.t. } S'\subseteq S \text{ and } \ssx_j(S)>0\}$. Also, let $x^*(\cS_j(U^t))$ be the elements of $x^*$ that correspond to customer type $j$ and assortments in $\cS_j(U^t)$.
		\State Run the offline black-box (Algorithm~\ref{alg:BB-assort}) on $x^*(\cS_j(U^t))$ as a subroutine and apply edge-attenuation to ensure each item $i\in U^t$ along with an assortment $S\ni i$ such that $S\in \cS_j(U^t)$ is offered to customer $t$ with probability $\ssx_j(S)\cdot(1-e^{-\gamma_t})/\gamma_t$.		
		\State  Apply vertex-attenuation to each item so that they are available with probability equal to $\gamma_{t+1}=\gamma_{t}-({1-\exp(-\gamma_{t})})/{T}$ at time-step $t+1$.
		\EndFor
	\end{algorithmic}
\end{algorithm}

\begin{lemma}[Modified Attenuation Framework (Assortment Version)] \label{lem::attenFramework-assort-repeated}
	Consider any time-step $t\in[T]$.
	Suppose there exist attenuation factors $a^{\vertex}_{t'}(i),a^{\edge}_{t'}(i,S,j)\in[0,1]$ for all items $i\in[n]$, customer types $j\in[m]$, time-steps $t'<t$, and assortments $S\ni i$ and $S\in \cS$ on which Algorithm~\ref{alg:full-assort-repeated} can be run until the start of time-step $t$, at which point for all items $i$,
	\begin{align*}
	\Pr[\avail_t(i)]=\gamma_t.
	\end{align*}
	Then there exist attenuation factors $a^{\vertex}_t(i),a^{\edge}_t(i,S,j)\in[0,1]$ for all $i\in[n]$, $j\in[m]$ and $S\ni i$ and $S\in \cS$ on which Algorithm~\ref{alg:full-assort-repeated} can be run during time-step $t$, so that for all items $i$ and types $j$,
	\begin{align*}
	\Pr[\accept_t(i,S,j)] &=(1-e^{-\gamma_t})q_j p(i,S)\ssx_j(S); \allowdisplaybreaks\\
	\Pr[\avail_{t+1}(i)] &=\gamma_{t+1}. \allowdisplaybreaks
	\end{align*}
\end{lemma}

\proof{Proof.} Let $\ssx$ denote the optimal solution of the MCDLP-R. For an item $i$, a customer type $j$, an assortment $S\ni i$, and a time-step $t$, $\Pr[\accept_t(i,S,j)]$ can be decomposed as
\begin{align} \label{eqn::decomp-assort-repeated}
\Pr[\accept_t(i,S,j)&|\Toffered_t(i,S,j)]\cdot\Pr[\Toffered_t(i,S,j)|\offered_t(i,S,j)]\nonumber\times \allowdisplaybreaks\\&\Pr[\offered_t(i,S,j)|\avail_t(i)\cap\type_t(j)]\cdot\Pr[\avail_t(i)\cap\type_t(j)]. \allowdisplaybreaks
\end{align}
Given that type of customer $t$ is $j$ and assortment $S$ containing $i$ is offered to her we have $\Pr[\accept_t(i,S,j)|\Toffered_t(i,S,j)]=p_j(i,S)$. Also, it can be easily seen that $\Pr[\avail_t(i)\cap\type_t(j)]=\gamma_tq_j$. Furthermore, similar to what discussed earlier in the proof of Theorem~\ref{thm::attenFramework}, the attenuation framework ensures that $a^{\edge}_t(i,S,j)=\Pr[\Toffered_t(i,S,j)|\offered_t(i,S,j)]$. Not that we use the notation $(i,S,j)$ to emphasize that item $i$ is offered to a type $j$ customer as a part of assortment $S$ and we stress that edge attenuation is done on the combination of  each item, assortment and customer type.

It remains to analyze $\Pr[\offered_t(i,S,j)|\avail_t(i)\cap\type_t(j)]$. At time-step $t$, Algorithm~\ref{alg:full-assort-repeated} offers assortments by calling the randomized procedure from Lemma~\ref{lem:bb-assort}, with $\ell=\ell_j$,  $\cA=S_j(U^t)$, and $p(i',S')=p_j(i',S'),x(S')=\ssx_j(S')$ for all $S'\in \cA$ and $i'\in S'$.
LP constraints (\ref{constr::sellOne-assort-repeated})--(\ref{constr::01-assort-repeated}) ensure that $(x(S'))_{S'\in \cA}$ satisfies the required conditions in Lemma~\ref{lem:bb-assort}.
Therefore, by Lemma~\ref{lem:bb-assort},
given $S\in S_j(U^t)$ and $i\in S$, the probability that $S$ is offered to customer $t$ who is of type $j$ is at least $\ssx_j(S)\cdot(1-\exp({-W_t(S\ni i,j)})/W_t(S\ni i,j)$, where we use the notation $S\ni i$ to emphasize that $i$ is in $S$ and not removed due to being sold-out. $W_t(S\ni i,j)$ is a random variable which will be discussed further in the following. If for all types $j$ we have $\ell_j\ge |\cS|$, then for an assortment $S\in S_j(U^t)$ we have
\begin{align}\label{eq:case1-assort-repeated}
W_t(S,j)=\frac{1}{1-\sum_{i'\in S}p_j(i',S)\ssx_j(S)}\sum_{S'\neq S}\sum_{i'\in S'}p_j(i',S')\ssx_j(S')\bI(\avail_t(i')),\allowdisplaybreaks
\end{align}
where $W_t(S,j)$ is understood to be 1 if $\sum_{i'\in S}p_j(i',S)\ssx_j(S)=1$. Furthermore, for the case that  $\sum_{i'\in [n]}\sum_{S'\ni i'} p_j(i',S')\le 1$ for all types $j$ we have 
\begin{align}\label{eq:case2-assort-repeated}
W_t(S,j) =\frac{1}{1-\sum_{i'\in S}p_j(i',S)}\sum_{S'\neq S}\sum_{i'\in S'}p_j(i',S')\ssx_j(S')\bI(\avail_t(i')),
\end{align}
where $W_t(S,j)$ is understood to be 1 if $\sum_{i'\in S}p_j(i',S)=1$. Let us now return to finding the probability of an item $i$ being offered as part of an assortment $S$ using a procedure similar to the one used in the proof of Theorem~\ref{thm::attenFramework}
\begin{align}
\Pr[\offered_t&(i,S,j)|\avail_t(i)\cap\type_t(j)] \nonumber\\&=\bE_{W_t(S\ni i,j)|\avail_t(i)\cap\type_t(j)}\big[\Pr[\offered_t(i,S,j)|\avail_t(i)\cap\type_t(j),W_t(S\ni i,j)]\big] \nonumber \allowdisplaybreaks\\
&\ge\bE_{W_t(S\ni i,j)|\avail_t(i)\cap\type_t(j)}\left[\ssx_j(S)\cdot\frac{1-\exp(-W_t(S\ni i,j))}{W_t(S\ni i,j)}\right] \nonumber \allowdisplaybreaks\\&=\ssx_j(S)\cdot \bE_{W_t(S\ni i,j)|\avail_t(i)\cap\type_t(j)}\left[f(W_t(S\ni i,j))\right]. \nonumber \allowdisplaybreaks\\&\ge \ssx_j(S)f(\bE[W_t(S\ni i,j)|\avail_t(i)\cap\type_t(j)]). \label{eq:jensen-assort-repeated}
\end{align}
where~\eqref{eq:jensen-assort-repeated} is due to the fact that 
$f(z)=(1-e^{-z})/z$ is convex and decreasing as well as Jensen's inequality. Below we bound $f(\bE[W_t(S\ni i,j)|\avail_t(i)\cap\type_t(j)])$. Events $\type_t(j)$ and $\avail_t(i')$ are independent for any $i'\neq i$; thus,  taking conditional expectation of~\eqref{eq:case1-assort-repeated} provides
\begin{align*}
\bE[W_t(S\ni i,j)|\avail_t(i)\cap\type_t(j)] &=\frac{\sum_{S'\neq S}\sum_{i'\in S'}p_j(i',S')\ssx_j(S')\bE[\bI(\avail_t(i'))|\avail_t(i)]}{1-\sum _{i'\in S}p_j(i',S)\ssx_j(S)}.
\end{align*}
By~\citep[Lemma~3.1]{brubach2017attenuate} we have that at time-step $t$, the availability of items $i$ and $i'$ are negatively correlate; thus,
\begin{align}
\bE[W_t(S\ni i,j)|\avail_t(i)\cap\type_t(j)] &\le\frac{\sum_{S'\neq S}\sum_{i'\in S'}p_j(i',S')\ssx_j(S')\Pr[\avail_t(i')]}{1-\sum_{i'\in S}p_j(i',S)\ssx_j(S)} \nonumber \\ &=\frac{\sum_{S'\neq S}\sum_{i'\in S'}p_j(i',S')\ssx_j(S')\cdot\gamma_t}{1-\sum_{i'\in S}p_j(i',S)\ssx_j(S)} \le(1)\cdot\gamma_t, \label{eqn::expBound-assort-repeated}
\end{align}
with the final inequality coming from LP constraint~\eqref{constr::sellOne-assort-repeated}. Similarly, for the case that $\sum_{i'\in [n]}\sum_{S'\ni i'}p_j(i',S')\le 1$: 
\begin{align}
\bE[W_t(S\ni i,j)|\avail_t(i)\cap\type_t(j)]& \le \frac{\sum_{S'\neq S}\sum_{i'\in S'}p_j(i',S')\ssx_j(S')\Pr[\avail_t(i')]}{1-\sum_{i'\in S}p_j(i',S)} \nonumber\allowdisplaybreaks\\
&=\frac{\sum_{S'\neq S}\sum_{i'\in S'}p_j(i',S')\ssx_j(S')\cdot\gamma_t }{1-\sum_{i'\in S}p_j(i',S)}\le(1)\cdot\gamma_t \label{eqn::expBound2-assort-repeated}. \allowdisplaybreaks
\end{align}

Combining~\eqref{eq:jensen-assort-repeated} with~(\ref{eqn::expBound-assort-repeated}) and (\ref{eqn::expBound2-assort-repeated}), for both cases of $\ell_j\ge |\cS|$ and $\sum_{i'\in[n]}\sum_{S'\ni i'}p_j(i',S')\le 1$, we have $\Pr[\offered_t(i,S,j)|\avail_t(i)\cap\type_t(j)]\ge\ssx_j(S)\cdot(1-e^{-\gamma_t})/\gamma_t$; therefore,
\begin{align*}
\Pr[\offered_t(i,S,j)|\avail_t(i)\cap\type_t(j)]\cdot\Pr[\avail_t(i)\cap\type_t(j)] &\ge (\ssx_j(S)\cdot\frac{1-e^{-\gamma_t}}{\gamma_t})\gamma_tq_j\\&=(1-e^{-\gamma_t})q_j\ssx_j(S).
\end{align*}
Moreover, there exists an attenuation factor $a^{\edge}_t(i,S,j)\in[0,1]$ such that $a^{\edge}_t(i,S,j)\cdot\Pr[\offered_t(i,S,j)|\avail_t(i)\cap\type_t(j)]\cdot\Pr[\avail_t(i)\cap\type_t(j)]=(1-e^{-\gamma_t})q_j\ssx_j(S)$, which when plugged in (\ref{eqn::decomp-assort-repeated}) provides
\begin{align*}
\Pr[\accept_t(i,S,j)]=(1-e^{-\gamma_t})q_jp_j(i,S)\ssx_j(S).
\end{align*}
Defining such an attenuation factor for all $i\in[n]$, $j\in[m]$ and $S\in \cS$ such that $S\ni i$ in addition to LP constraint (\ref{constr::inv-assort-repeated}), it can be concluded that for any item $i$
\begin{align*}
\Pr[\avail_{t+1}(i)] &=\Pr[\avail_t(i)]-\sum_{j=1}^m\sum_{S\ni i}\Pr[\accept_t(i,S,j)] \\&=\gamma_t-(1-e^{-\gamma_t})\sum_{j=1}^m\sum_{S\ni i}q_jp_j(i,S)\ssx_j(S) \ge\gamma_t-\frac{1-e^{-\gamma_t}}{T}. \allowdisplaybreaks
\end{align*}
Thus, there exists an attenuation factor $a^{\vertex}_t(i)\in[0,1]$ such that
\begin{align*}
\Pr[\avail_{t+1}(i)]=\gamma_t-\frac{1-e^{-\gamma_t}}{T}=\gamma_{t+1}. \allowdisplaybreaks
\end{align*}
\Halmos\endproof
\proof{Proof.}[Proof of Theorem~\ref{thm:assort-0.51}] The proof of Lemma~\ref{lem::attenFramework-assort-repeated} showed that at time-step $t$, $\Pr[\accept_t(i,S,j)]=(1-e^{-\gamma_t})q_jp_j(i,S)\ssx_j(S)$. Let $\EXP[\ALG]$ denote the total expected revenue of Algorithm~\ref{alg:full-assort-repeated}; therefore,
\begin{align}
\EXP[\ALG]=\sum_{t=1}^T\sum_{j=1}^m\sum_{S\in \cS}\sum_{i\in S}  r_{ij}\Pr[\accept_t(i,S,j)]&= \sum_{t=1}^T (1-e^{-\gamma_t}) \sum_{j=1}^m q_j \sum_{S\in \cS}\sum_{i\in S}  \ssx_j(S)r_{ij}p_j(i,S).\nonumber
\end{align}
Moreover, 
\begin{align*}
\OPT= \sum_{j=1}^m Tq_j \sum_{S\in \cS}\sum_{i\in S}  \ssx_j(S)r_{ij}p_j(i,S).
\end{align*}
Therefore, the approximation ratio of Algorithm~\ref{alg:full-assort-repeated} is $\frac{\sum_{t=1}^T (1-e^{-\gamma_t})}{T}$. Theorem~\ref{thm::0.51} showed that $\frac{\sum_{t=1}^T (1-e^{-\gamma_t})}{T}$ is lower-bounded by  $(1-\ln(2-\frac{1}{e}))$; therefore, the multi-customer and multi-stage assortment optimization problem when items can be offered multiple times to customers has an approximation ration of $(1-\ln(2-\frac{1}{e}))\simeq 0.51$. 
\Halmos\endproof

\subsection{Deferred Proofs from Section~\ref{sec::generalAsst}}\label{apx-sec::generalAsst-proofs}
\proof{Proof of Lemma~\ref{lem::assort-lp-ub}.} The proof can be easily obtained by extending that of Lemma~\ref{lem::lp-ub} to the case the probability of purchasing an item is also a function of the assortment within which that item was presented, i.e., $p_j(i,S)$. Moreover, compared to the MCDLP-R, there is an additional constraint in the MCDLP-NR, that is, constraint~\eqref{constr::asst-norepeat-overlap}. As any valid algorithm does not show a previously seen item to a customer, taking the expectation, we have that any valid assortment offering strategy satisfies constraint~\eqref{constr::asst-norepeat-overlap}.
\Halmos\endproof

\proof{Proof of Theorem~\ref{thm:undeterm-patient}.} First, note that in the discussed non-deterministic version of the multi-stage multi-customer assortment optimization problem  a slight change in the patience constraint of MCDLP-NR, i.e., constraint~\eqref{constr::asst-norepeat-timeout}, is required. Namely, we replace the patience level of each customer $j$, i.e., $\ell_j$, by its expected patience level $1/p^{\text{out}}$. The modified MCDLP-NR is demonstrated below:

\begin{subequations}
	\label{lp:assort-norepeat-rand}
	\begin{align}
	\max\sum_{t=1}^T\sum_{j=1}^m q_{tj} \sum_{S\in\cS}x_j(S)\sum_{i\in S}r_{ij}p_j(i,S) \nonumber\allowdisplaybreaks\\
	\sum_{t=1}^T\sum_{j=1}^m q_{tj}\sum_{S\in\cS:S\ni i}x_j(S)p_j(i,S) &\le 1 &\forall i=1,\ldots,n \label{constr::asst-norepeat-inv-rand} \allowdisplaybreaks\\
	\sum_{S\in\cS}x_j(S)\sum_{i\in S}p_j(i,S) &\le1 &\forall j=1,\ldots,m \label{constr::asst-norepeat-sellOne-rand} \allowdisplaybreaks\\
	\sum_{S\in\cS}x_j(S) &\le \frac{1}{p^{\text{out}}_j} &\forall j=1,\ldots,m \label{constr::asst-norepeat-timeout-rand} \allowdisplaybreaks\\
	\sum_{S\ni i} x_j(S)&\le 1  &\forall i=1,\ldots,n;\ \forall j=1,\ldots,m \label{constr::asst-norepeat-overlap-rand} \allowdisplaybreaks\\
	x_j(S)&\ge 0 &\forall j=1,\ldots,m;\ \forall S\in\cS \label{constr::asst-norepeat-01-rand} \allowdisplaybreaks
	\end{align}
\end{subequations}

To prove the theorem, we need to show that Lemmas~\ref{lem:M_i}-\ref{lem:O_S-M'_S} for this variant of the problem. This can be easily checked for Lemma~\ref{lem:M_i} as it provides an upper-bound on the probability that an item $i$ is already matched when a customer of type $j$ arrives for the first time and the proof of the lemma does not depend on patience level of customers. Similarly, Lemma~\ref{lem:D_i,S} stays untouched for non-deterministic patience levels as it does not depend on patience levels either.

The only difficulty comes from Lemma~\ref{lem:O_S-M'_S}. In the proof of this lemma two cases are considered for a customer of type $j$, those are, (i) patience level of 1, and (ii) patience level of greater than 1. In non-deterministic patience levels, these two cases translate to (i) type $j$ customer leaves the platform if no purchase is made from the first offered assortment, (ii) type $j$ customer sees the second assortment if she does not make a purchase from the first assortment. Case (i) takes place with probability $p^{\text{out}}_j$ and with an argument similar to the one provided in the proof of Lemma~\ref{lem:O_S-M'_S} it can easily be seen that the probability of this case is upper-bounded by $1/(2\alpha)$. Case (ii) takes place with probability $1-p^{\text{out}}_j$. In this case, by union bound we have $\Pr[\timeout_{S}(j)\cup \cmatch_{S}(j)|\type(j)]\le \Pr[\cmatch_{S}(j)|\type(j)] + \Pr[\timeout_{S}(j)|\type(j)]$ and similar to the proof of Lemma~\ref{lem:O_S-M'_S} it can be seen that $\Pr[\cmatch_{S}(j)|\type(j)]\le 1/(2\alpha)$. In the remainder of this proof we show that $\Pr[\timeout_{S}(j)|\type(j)]\le (1-p_{\text{out}})2/(3\alpha^2)$, providing $\Pr[\timeout_{S}(j)\cup \cmatch_{S}(j)|\type(j)]\le 1/(2\alpha)+2/(3\alpha^2)$ similar to Lemma~\ref{lem:O_S-M'_S}.

To see this, suppose customer type $j$ has a patience level of $a$, that is she sees up to $a$ assortment. This happens with probability $(1-p^{\text{out}}_j)^{a-1}\cdot p^{\text{out}}_j$ as all customers see at least one assortment. Therefore we have
\begin{align}
\Pr[\timeout_{S}(j)|\type(j)] &= \sum_{a=2}^\infty \Pr[\timeout_{S}(j)|\type(j), l_j=a]\Pr[l_j=a] \nonumber\\
&\le \sum_{a=2}^\infty \frac{1}{(a+1)!}\left(\frac{a}{\alpha}\right)^a \cdot (1-p^{\text{out}}_j)^{a-1}p^{\text{out}}_j \label{eq:nondet1}\\
&\le \frac{2}{3\alpha^2} \cdot p^{\text{out}}_j \cdot \sum_{a=2}^\infty (1-p^{\text{out}}_j)^{a-1} \label{eq:nondet2}\\
&\le \frac{2}{3\alpha^2} (1-p^{\text{out}}_j) \label{eq:nondet3},
\end{align}
where~\eqref{eq:nondet1} uses~\eqref{eq:timeout-bound},~\eqref{eq:nondet2} uses bound $\frac{1}{(a+1)!}(\frac{a}{\alpha})^a\le \frac{2}{3\alpha^2}$ discussed in the proof of Lemma~\ref{lem:O_S-M'_S}, and finally~\eqref{eq:nondet3} uses sum of geometric series. In conclusion, here we showed that proofs of Lemmas~\ref{lem:M_i}-\ref{lem:O_S-M'_S} hold in case of non-deterministic patience levels, hence Algorithm~\ref{alg:asst1} has the same approximation ratio for this variant of the problem.
\Halmos\endproof

\proof{Proof of Theorem~\ref{thm:assort-sameprice}.} Let $\ssx$ be the optimal solution to the MCDLP-NRS.  As customers come their patience levels are revealed to the platform and up to their patience level or once a purchase is made by them (whichever comes first), the algorithm offers them a random permutation of assortments. For an assortment $S$ to be offered to customer $t$, items that are already seen by that customer in the previous offering stages and those that are sold-out are first removed from it and then with probability $\ssx(S)/\alpha$ it is shown to the customer. Theorem~\ref{thm:main--assort-norepeat} shows that for $\alpha=(3+\sqrt{17})/{2}$ Algorithm~\ref{alg:asst1} has a 0.093 approximation guarantee for the case that items can have different revenues across different customer types. Here, we show that the modified Algorithm~\ref{alg:asst1} has an improved approximation guarantee of 0.15 when for each item all customer types are associated with the same revenue.

Let $U^t$ denote the set of available items at the beginning of time-step $t$. The modified Algorithm~\ref{alg:asst1} offers an item $i\in U^t$ along with assortment $S\ni i$ to customer $t$ with probability at least $(1-3/(2\alpha))/\alpha \cdot \sum_{j=1}^m q_{tj}\ssx_j(S)$. Note that this is different from the guarantee presented in Lemma \ref{lem:main-with-alpha} by an additive term of $(-2/(3\alpha^2))$ as here it is assumed that item $i$ is not sold-out yet. The term $(1-3/(2\alpha))/\alpha$ is a concave function of $\alpha$ and by simple calculus it can be seen that it is maximized for $\alpha=3$, which makes $(1-3/(2\alpha))/\alpha$ equal to $1/6$. Therefore, the modified Algorithm~\ref{alg:asst1} offers an item $i\in U^t$ along assortment $S\ni i$  with probability at least $\sum_{j=1}^m q_{tj}\ssx_j(S)/6$. 

Let $p_{it}$ denote the probability that item $i$ is purchased by customer $t$ by the optimal solution of the MCDLP-NRS, that is, $p_{it}=\sum_{j=1}^m q_{tj}\sum_{S\ni i} \ssx_j(S)p_j(i,S)$. Constraint~\eqref{constr::asst-norepeat-inv} states that for each item $i$ we have $\sum_{t=1}^T\sum_{j=1}^m q_{tj}\sum_{S\in\cS:S\ni i}\ssx_j(S)p_j(i,S)\le 1$. In other words,
\begin{align}
\sum_{t=1}^T p_{it}\le 1,\label{eq:inventory-bound}
\end{align} 
which will be later used to establish how far the total expected revenue of the modified Algorithm~\ref{alg:asst1} is from the optimal value of the MCDLP-NRS.

Let $\accept_t(i)$ be the event that item $i$ is purchased by a customer $t$ and $\accept(i)$ be the event that $i$ is sold to a customer while performing the modified Algorithm~\ref{alg:asst1}. We have
\begin{align}
\Pr[\accept(i)]\ge 1-\prod_{t=1}^T (1-\Pr[\accept_t(i)]).  \label{eq:sell-prob1}
\end{align}
Since the modified Algorithm~\ref{alg:asst1} offers an item $i\in U^t$ to customer $t$ as a part of an assortment $S$ with probability at least $\sum_{j=1}^m q_{tj}\ssx_j(S)/6$, we have $\Pr[\accept_t(i)]\ge\sum_{j=1}^m q_{tj}\sum_{S\ni i} \ssx_j(S)\cdot p_j(i,S)/6$, that is, $p_{it}/6$. Thus,
\begin{align}
1-\prod_{t=1}^T (1-\Pr[\accept_t(i)]) = 1-\prod_{t=1}^T (1-\frac{p_{it}}{6}) &\ge 1-\prod_{t=1}^T \exp(-\frac{p_{it}}{6}) \label{eq:sell-prob2}\\&= 1- \exp(-\sum_{t=1}^T\frac{p_{it}}{6}),\label{eq:sell-prob3}
\end{align}
where~\eqref{eq:sell-prob2} uses the bound $1-p_{it}/6\le \exp(-p_{it}/6)$ resulted from the Taylor expansion of $\exp(-p_{it}/6)$. Combining~\eqref{eq:sell-prob1} and~\eqref{eq:sell-prob3} we have 
\begin{align}
\Pr[\accept(i)]\ge 1- \exp(-\sum_{t=1}^T\frac{p_{it}}{6}).\label{eq:sell-prob-main}
\end{align}
Now, let $\EXP[\ALG(i)]$ and $\OPT(i)$ denote the expected revenue that the modified Algorithm~\ref{alg:asst1} and the optimal solution of the MCDLP-NRS collect from item $i$, respectively. Moreover, let $\EXP[\ALG]$ and $\OPT$ denote the total expected revenue of the algorithm and the optimal value of the MCDLP-NRS, respectively. Therefore, $\EXP[\ALG]=\sum_{i=1}^n \EXP[\ALG(i)]$ and $\OPT=\sum_{i=1}^n \OPT(i)$. If we show that there exists a positive constant $c\le 1$ such that for all $i=1,\ldots,n$ we have $\EXP[\ALG(i)]\ge c\cdot \OPT(i)$, then it implies that $\EXP[\ALG]\ge c\cdot\OPT$, establishing an approximation guarantee of $c$ for the modified Algorithm~\ref{alg:asst1}. We show that such $c$ exists and it is equal to 1/6. To see that, observe that $\EXP[\ALG(i)]=r_i\Pr[\accept(i)]$ and $\OPT(i)=r_i\sum_{t=1}^T p_{it}$. This in addition to~\eqref{eq:sell-prob-main} gives
\begin{align*}
\frac{\EXP[\ALG(i)]}{\OPT(i)}\ge \frac{r_i(1- \exp(-\sum_{t=1}^T\frac{p_{it}}{6}))}{r_i\sum_{t=1}^T p_{it}}=\frac{1}{6}\cdot\frac{1- \exp(-\sum_{t=1}^T\frac{p_{it}}{6})}{\sum_{t=1}^T\frac{p_{it}}{6}}=\frac{f(\sum_{t=1}^T\frac{p_{it}}{6})}{6},
\end{align*}
where, $f(z)=(1-\exp(-z))/z$. By Lemma~\ref{lem:f(x)}, $f(z)$ is a decreasing function. Furthermore,~\eqref{eq:inventory-bound} states that $\sum_{t=1}^T p_{it}\le 1$ for all items $i=1,\ldots,n$. Therefore 
\begin{align*}
\frac{\EXP[\ALG(i)]}{\OPT(i)}\ge \frac{f(\frac{1}{6})}{6}= 1-\exp(-\frac{1}{6})\simeq 0.15,
\end{align*}
that is, $c=0.15$ and the modified Algorithm~\ref{alg:asst1} has an approximation guarantee of at least 0.15.
\Halmos\endproof

\proof{Proof of Theorem~\ref{thm::mcdlpIntegGap}.}
Fix a large even integer $M$, and consider a family of assortments $\cS'$ with $|\cS'|=M$, constructed as follows.
Every pair of assortments in $\cS'$ share exactly 1 item, so that there are $\binom{M}{2}$ items in total.
Each assortment $S\in\cS'$ contains exactly $M-1$ items, since it shares exactly one item with each of exactly $M-1$ other assortments.
The patience level of the customer is $\ell=M/2$.
The price of each item is 1.

The feasible family $\cS$ of assortments that can be offered is the downward closure of $\cS'$.
That is, an assortment $S$ can be offered if an only if $S\subseteq S'$ for some $S'\in\cS'$.
When offered any assortment $S\in\cS$, the customer chooses each item in $S$ with probability $\frac{2}{M(M-1)}$, and chooses to purchase nothing with probability $1-\frac{2|S|}{M(M-1)}$.
We remark that if $|S|=M-1$ (the maximum possible size of an assortment $S\in\cS$), then the customer makes a purchase with probability $\frac{2}{M}$.
It is easily checked that these choice probabilities satisfy the substitutability assumption.

Set $x(S)=1/2$ for each assortment $S\in\cS'$.
We claim that this is a feasible solution to the single-customer MCDLP-NRS.
Indeed, constraint~\eqref{constr::singleCust::asst-sellOne} holds because the LHS equals $M\times\frac{1}{2}\times\frac{2}{M}$ by the remark above;
constraints~\eqref{constr::singleCust::asst-timeout} and~\eqref{constr::singleCust::ass-01} hold trivially;
while
constraint~\eqref{constr::singleCust::ass-overlap} holds because each item appears in exactly two assortments.
Therefore, the optimal objective value of the MCDLP-NRS is at least the value of this solution, which is 1.

Meanwhile, consider any collection $\cO$ of assortments which an online algorithm could plan to offer the customer.
$\cO$ cannot contain more than $M/2$ assortments (because the patience level is $M/2$), and the order in which the assortments in $\cO$ are offered does not affect the revenue (because all of the items have a price of 1).
Moreover, the assortments in $\cO$ cannot overlap, by the constraint that the same item cannot be shown twice.
For any assortment $S\in\cS$, let $\oS$ denote any $S'\supseteq S$ with $S'\in\cS'$ (i.e.\ $|\oS|=|S'|=M-1$).
Then, we have
\begin{align} \label{eqn::2574}
\sum_{S\in\cO}|S|
\le\Big|\bigcup_{S\in\cO}\oS\Big|
\end{align}
where we note that the assortments $\oS$ could overlap.
Now, let $M'$ denote the number of unique values of $\oS$ over $S\in\cO$, with $M'\le|\cO|\le M/2$.
By the principle of inclusion-exclusion, the right-hand side of~\eqref{eqn::2574} equals $M'(M-1)-\binom{M'}{2}$, because $|\oS|=M-1$, the intersection of every two assortments has cardinality exactly 1, and the intersection of 3 or more assortments has cardinality 0.
This value is at most
\begin{align*}
\frac{M(M-1)}{2}-\frac{M/2(M/2-1)}{2}\le \frac{3M^2}{8},
\end{align*}
and hence $\sum_{S\in\cO}|S|\le\frac{3M^2}{8}$.

Now, since the probability of an individual assortment $S$ earning a sale is $\frac{2|S|}{M(M-1)}$, the total probability of the online algorithm earning a sale is
\begin{align*}
1-\prod_{S\in\cO}(1-\frac{2|S|}{M(M-1)})
&\le1-\prod_{S\in\cO}\left(\exp\left(-\frac{2|S|}{M(M-1)}\right)-\frac{1}{2}\left(\frac{2|S|}{M(M-1)}\right)^2\right) \\
&\le1-\prod_{S\in\cO}\exp\left(-\frac{2|S|}{M(M-1)}\right)+\sum_{S\in\cO}\frac{1}{2}\left(\frac{2|S|}{M(M-1)}\right)^2 \\
&\le1-\exp\left(-\frac{2\sum_{S\in\cO}|S|}{M(M-1)}\right)+\sum_{S\in\cO}\frac{2}{M^2} \\
&\le1-\exp\left(-\frac{3M^2}{4M(M-1)}\right)+\left(\frac{M}{2}\right)\frac{2}{M^2}
\end{align*}
where the first inequality holds because $e^{-x}\le1-x+\frac{x^2}{2}$ for $x\ge0$,
the second inequality holds because each term in the product is at most 1,
the third inequality holds because $|S|\le M-1$,
and the final inequality holds because $\sum_{S\in\cO}|S|\le\frac{3M^2}{8}$ and $|\cO|\le M/2$.
The final expression can be checked to be approach $1-\exp(-\frac{3}{4})$ as $M\to\infty$, completing the proof.
\Halmos\endproof

\section{Supplements to Section~\ref{sec:mcdlp}}\label{apx-sec:mcdlp}
In the following, we discuss the column generation technique for the case of the multi-stage and multi-customer assortment optimization problem without repeated item offerings. We later finish this section by discussing how to tackle the problem when repeated offerings of items is allowed. Suppose $\zeta$, $\gamma$, $\beta$ and $\sigma$ are the dual variables for constraints~\eqref{constr::asst-norepeat-inv}-\eqref{constr::asst-norepeat-overlap}, respectively. We refer to the dual of the MCDLP-NR by MCDLP-NR-D and for both MCDLP-NR and MCDLP-NR-D, we use the notations MCDLP-NR$(\cS')$ and MCDLP-NR-D$(\cS')$ to specify that only assortments from set $\cS'$ are considered for these linear programs. Below we present MCDLP-NR-D$(\cS')$ for a set $\cS'\subseteq \cS$.
\begin{subequations}
	\label{lp:dual-assort-norepeat}
	\begin{align}
	\min \sum_{i=1}^n \zeta_i +\sum_{j=1}^m(\gamma_j+\ell_j\beta_j)+\sum_{i=1}^n\sum_{j=1}^m \sigma_{ij} \nonumber\\
	\gamma_j\sum_{i\in S} p_j(i,S)+\beta_j+\sum_{i\in S}\sigma_{ij}+\sum_{i\in S} \zeta_i\sum_{t=1}^T q_{tj} p_j(i,S) &\ge \sum_{i\in S}r_{ij} \sum_{t=1}^T q_{tj} p_j(i,S)  &\forall j\in [m];\ \forall S\in\cS'\label{constr:dual-main}\\
	\zeta,\gamma,\beta,\sigma &\ge 0 \label{constr:dual-positive}
	\end{align}
\end{subequations}

The column generation technique starts with a set of assortments $\cS^1$ and solves MCDLP-NR-D$(\cS^1)$ to find the optimal dual variables $(\eta^1,\gamma^1,\beta^1,\sigma^1)$ for this set assortments. We then check whether the dual variables corresponding to MCDLP-NR-D$(\cS^1)$ are also feasible for MCDLP-NR-D$(\cS)$. In other words, we need to specify whether there exists any assortment in  $\cS\backslash\cS^1$ for which constraint~\eqref{constr:dual-main} is violated, i.e., has a positive reduced cost. To find any such assortment we can first solve the following problem, called the \emph{column generation subproblem} for any customer type $j$: 
\begin{align}
\max_{S\in \cS} \sum_{i\in S} \left(w_{ij}^1p_j(i,S)-\sigma^1_{ij}\right)\label{prob:subproblem},
\end{align}
where $w^1_{ij}=r_{ij}\sum_{t=1}^T q_{tj}-\zeta^1_i \sum_{t=1}^T q_{tj}-\gamma^1_j$. Let $S^1$ denote the solution of the above problem. Note that $\sum_{i\in S^1} \left(w^1_{ij}p_j(i,S^1)-\sigma^1_{ij}\right)\ge 0$ by setting $S^1=\emptyset$. The goal is to check whether  $\sum_{i\in S^1} \left(w^1_{ij}p_j(i,S^1)-\sigma^1_{ij}\right)> \beta^1_j$. In this case, we set $\cS^2=\cS^1\cup\{S^1\}$ and repeats this process again; otherwise, the process stops. Suppose the process stops at iteration $\tau$. By solving MCDLP-NR$(\cS^\tau)$ we get its optimal solution $\bar{x}$ that has the same objective value as that of MCDLP-NR$(\cS)$.

In any $\tau$-th round of this procedure, by constraint~\eqref{constr:dual-positive} of MCDLP-NR-D$(\cS^\tau)$, we have $\beta^\tau_j,\sigma^\tau_{ij}\ge0$. Therefore, subproblem~\eqref{prob:subproblem} is only interesting when $w^\tau_{ij}\ge 0$. If for the solution of this subproblem's, denoted by $S^\tau$, we have $\sum_{i\in S^\tau} \left(w^\tau_{ij}p_j(i,S^\tau)-\sigma^\tau_{ij}\right) \le \beta^\tau_j$, then $(\zeta^\tau,\gamma^\tau,\beta^\tau,\sigma^\tau)$ are dual feasible for MCDLP-NR-D$(\cS)$ and the current solution to the reduced MCDLP-NR is in fact optimal. Otherwise, $S^\tau$ has a positive reduced cost. By adding $S^\tau$ to $\cS^\tau$ we resolve the reduced MCDLP-NR of $S^\tau$, which results in a larger value of objective function of the MCDLP-NR. This process is continued until subproblem~\eqref{prob:subproblem}'s solution is the empty set, at which point the MCDLP-NR gives us the optimal solution.

In general solving column generation subproblems are NP-hard~\citep{liu-vanryzin'08}. However, there is hope to find approximate solutions for it for specific choice models. Combining an approximation algorithm with the aforementioned column generation framework provides the \emph{approximate column generation} technique. In other words, we want to find a set $S^\tau\in \cS\backslash\cS^\tau$ such that
\begin{align}
\sum_{i\in S^\tau} \left(w^\tau_{ij}p_j(i,S^\tau)-\sigma^\tau_{ij}\right) \ge \alpha\cdot  \max_{S\in \cS} \sum_{i\in S} \left(w^\tau_{ij}p_j(i,S)-\sigma^\tau_{ij}\right)\label{eq:apx-def}.
\end{align}

In approximate column generation, we still terminate when the subproblem algorithm fails to find a constraint violating dual feasibility. Since the subproblem algorithm is $\alpha$-approximate, we can guarantee that when it terminates, the dual constraints are within an $\alpha$-factor of being feasible.

Since the empty set is a feasible solution to problem~\eqref{prob:subproblem-mnl}, the optimal value of problem~\eqref{prob:subproblem-mnl} is always non-negative. Moreover, the problem at hand is a maximization problem; thus, we have $\alpha\in [0,1]$. Before discussing any such approximation algorithm, let us introduce a few notations.  For a set of assortments $\cS'\subseteq\cS$, let $\OPT$(MCDLP-NR$(\cS')$) and $\OPT$(MCDLP-NR-D$(\cS')$) denote the optimal values of MCDLP-NR$(\cS')$ and MCDLP-NR-D$(\cS')$, respectively. Moreover, we denote the optimal solution of the overall MCDLP-NR, that is MCDLP-NR$(\cS)$, by $\ssx$ and the optimal solution of the overall dual of MCDLP-NR, that is MCDLP-NR-D$(\cS)$, by $(\zeta^*,\gamma^*,\beta^*,\sigma^*)$. Furthermore, for $\bar{x}$, a feasible solution to MCDLP-NR$(\cS)$, let $\Val(\bar{x})=\sum_{j=1}^m \sum_{S\in\cS}\bar{x}_j(S)\sum_{i\in S}r_{ij}p_j(i,S)$. Lemma~\ref{lem:approx-carryover} discusses how the approximation ratio of an approximation algorithm for the column generation subproblem and that of the MCDLP-NR$(\cS)$ when the approximate column generation technique is used are related. The proof of this lemma is similar to the proof of Lemma 3.5 in~\citep{cheung-simchilevi'16} and effectively extends the result from CDLP's to our MCDLP's.

\begin{lemma}
	\label{lem:approx-carryover}
	Consider the multi-stage choice-based deterministic linear program with no repeated item offerings (MCDLP-NR) and suppose an algorithm $\mathcal{A}$ with  approximation guarantee of $\alpha$ for the column generation subproblem (i.e., subproblem~\eqref{prob:subproblem}) on the set of assortments $\cS$ and the underlying choice model $p_j(.,.)$ for all customer types $j$. Then, the approximate column generation technique returns a solution $\bar{x}$ to MCDLP-NR$(\cS)$ whose objective value is at least $\alpha$ times away from $\OPT$(MCDLP-NR$(\cS)$), that is, 
	\begin{align}
	\Val(\bar{x})\ge \alpha \Val(\ssx)
	\end{align}
\end{lemma}
\proof{Proof.} First of all, note that the approximate column generation technique halts in finite amount of time, after at most $|\cS|$ iterations. In any iteration $\tau$ of the approximate column generation technique, other than its last iteration, an assortment $S^\tau\in \cS\backslash\cS^\tau$ is added to $\cS^\tau$ to make $\cS^{\tau+1}$. To see that $\cS^\tau$ is in fact in $\cS\backslash\cS^\tau$, note that in the $\tau$-th iteration MCDLP-NR-D$(\cS^\tau)$ is solved. Therefore, for a feasible solution to MCDLP-NR-D$(\cS^\tau)$ constraint~\eqref{constr:dual-main} is satisfied for all assortments in $S^\tau$; thus, on set $S^\tau$ subproblem~\eqref{prob:subproblem} has no solution $S$.

We now show that if subproblem~\eqref{prob:subproblem} can be solved with an $\alpha$ approximation guarantee, then the approximate column generation technique returns a solution for MCDLP-NR$(\cS)$ whose value is within an $\alpha$ factor of $\OPT$(MCDLP-NR$(\cS)$). Suppose that the approximate column generation technique halts in the $\tau$-th iteration, providing $(\zeta^\tau,\gamma^\tau,\beta^\tau,\sigma^\tau)$, the optimal dual variables for MCDLP-NR-D$(\cS^\tau)$. By constraint~\eqref{constr:dual-main}, for all customer types $j$ and assortments $S^\tau$, the assortment returned by Algorithm $\mathcal{A}$ in the $\tau$-th iteration, we have 
\begin{align*}
\beta^\tau_j\ge \sum_{i\in S^\tau} (w_{ij}^\tau p_j(i,S^\tau)-\sigma^\tau_{ij}),
\end{align*}
where $w^\tau_{ij}=r_{ij}-\zeta^\tau_i-\gamma^\tau_j$. Moreover, since algorithm $\mathcal{A}$ has an $\alpha$ approximation ratio, combing with the above it gives
\begin{align*}
\beta^\tau_j\ge \sum_{i\in S^\tau} (w_{ij}^\tau p_j(i,S^\tau)-\sigma^\tau_{ij})\ge \alpha \cdot \max_{S\in \cS} \sum_{i\in S} \left(w_{ij}^\tau p_j(i,S)-\sigma^\tau _{ij}\right).
\end{align*}
Observe that not only $(\zeta^\tau,\gamma^\tau,\beta^\tau,\sigma^\tau)$ are the optimal decision variables for MCDL-NRP-D$(\cS^\tau)$, they are also feasible decision variables for MCDLP-NR-D$(\cS)$ as constraints~\eqref{constr:dual-main} and~\eqref{constr:dual-positive} are satisfied for them. Otherwise, the approximate column generation technique would have not stopped at the $\tau$-th iteration. Feasibility of $(\zeta^\tau,\gamma^\tau,\beta^\tau,\sigma^\tau)$ and the fact that $\alpha\le 1$ implies that $(\zeta^\tau,\gamma^\tau,\beta^\tau/\alpha,\sigma^\tau)$ is also a feasible to MCDLP-NR-D$(\cS)$. Thus,
\begin{align}
\sum_{i=1}^n \zeta^\tau_i +\sum_{j=1}^m(\gamma^\tau_j+\ell_j\frac{\beta^\tau_j}{\alpha})+\sum_{i=1}^n\sum_{j=1}^m \sigma^\tau_{ij}&\ge \sum_{i=1}^n \zeta^*_i +\sum_{j=1}^m(\gamma^*_j+\ell_j{\beta^*_j})+\sum_{i=1}^n\sum_{j=1}^m \sigma^*_{ij}\label{eq:carryover1}\\&=\sum_{j=1}^m \sum_{S\in\cS}\ssx_j(S)\sum_{i\in S}r_{ij}p_j(i,S)\label{eq:duality},
\end{align}
where~\eqref{eq:carryover1} is due to the fact that the dual of the MCDLP-NR is a minimization problem and $(\zeta^\tau,\gamma^\tau,\beta^\tau/\alpha,\sigma^\tau)$ is feasible solution for MCDLP-NR-D$(\cS)$, and~\eqref{eq:duality} holds by strong duality for MCDLP-NR$(\cS)$ and MCDLP-NR-D$(\cS)$. Furthermore, using strong duality for MCDLP-NR$(\cS^\tau)$ and MCDLP-NR-D$(\cS^\tau)$ we have
\begin{align}
\frac{1}{\alpha}\sum_{j=1}^m \sum_{S\in\cS^\tau}\bar{x}^\tau_j(S)\sum_{i\in S}r_{ij}p_j(i,S) &= \frac{1}{\alpha} \sum_{i=1}^n \zeta^\tau_i +\sum_{j=1}^m(\gamma^\tau_j+\ell_j{\beta^\tau_j})+\sum_{i=1}^n\sum_{j=1}^m \sigma^\tau_{ij}\nonumber\\ &\ge \sum_{i=1}^n \zeta^\tau_i +\sum_{j=1}^m(\gamma^\tau_j+\ell_j \frac{\beta^\tau_j}{\alpha})+\sum_{i=1}^n\sum_{j=1}^m \sigma^\tau_{ij},\label{eq:duality2}
\end{align}
where $\bar{x}^\tau$ is the optimal solution for MCDLP-NR$(\cS^\tau)$ that is padded with zeros for the assortments that in $\cS\backslash\cS^\tau$; therefore, it is a feasible solution for MCDLP-NR$(\cS)$. Combining~\eqref{eq:duality} and~\eqref{eq:duality2} and using the fact that as $\bar{x}^\tau$ is a feasible solution to a maximization problem we have
\begin{align*}
\frac{\Val(\ssx)}{\alpha}= \frac{1}{\alpha}\sum_{j=1}^m \sum_{S\in\cS}\ssx_j(S)\sum_{i\in S}r_{ij}p_j(i,S) &\ge \frac{1}{\alpha}\sum_{j=1}^m \sum_{S\in\cS^\tau}{x}^\tau_j(S)\sum_{i\in S}r_{ij}p_j(i,S) = \frac{\Val(\bar{x}^\tau)}{\alpha}\\ &\ge \sum_{j=1}^m \sum_{S\in\cS}\ssx_j(S)\sum_{i\in S}r_{ij}p_j(i,S)= \Val(\ssx),
\end{align*}
concluding that $\Val(\bar{x}^\tau)\ge \alpha \Val(\ssx)$, as desired.
\Halmos\endproof

Knowing how the approximation ratio of the column generation subproblem translates to that of the MCDLP-NR, it is then of interest to find an approximation algorithm for this subproblem. In the next section, we discuss an approximation algorithm for this problem in the case of the MNL choice model.  

\subsection{Specific Results for MNL Choice Model}
\label{subsec:mnl}
As discussed earlier, it is known to be NP-hard to solve the column generation subproblem for general choice models. For some restricted choice models, such as multinomial logit (MNL), the column generation subproblem has been shown to be polynomially solvable for many different CDLP's, see~\citet{liu-vanryzin'08} and~\citet{talluri-vanruzen'04}. Therefore, it is of interest to determine whether for the MCDLP-NR the column generation subproblem can be efficiently solved. Subproblem~\eqref{prob:subproblem} for multinomial logit choice model and a fixed customer type $j$ is presented below
\begin{align}
\max_{S\in \cS} \sum_{i\in S} \left(\frac{w^\tau_{ij}v_{ij}}{\sum_{k\in S}v_{kj}+1}-\sigma^\tau_{ij}\right)\label{prob:subproblem-mnl}.
\end{align}
Lemma~\ref{lem:NP-hard} addresses whether problem~\eqref{prob:subproblem-mnl}  can be solved in polynomial-time, i.e., whether the column generation subproblem is polynomially solvable when the choice model is multinomial logit. To evaluate the complexity of an optimization problem we need to study its decision problem as the complexity class NP is defined for decision problems. A decision problem is a problem with ``Yes" or ``No" answers. Thus, the decision version of problem~\eqref{prob:subproblem-mnl} can be 
\begin{align}
\max_{S\in \cS} \sum_{i\in S} \left(\frac{w^\tau_{ij}v_{ij}}{\sum_{k\in S}v_{kj}+1}-\sigma^\tau_{ij}\right)\ge \lambda \label{prob:mnl-decision}, 
\end{align}
where $\lambda$ is a real positive number.

\begin{lemma}
	\label{lem:NP-hard}
	Problem~\eqref{prob:mnl-decision} is an NP-complete problem.
\end{lemma}
\proof{Proof.} First of all, note that problem~\eqref{prob:subproblem-mnl} is in complexity class NP as a ``Yes" certificate for it can be verified in polynomial-time. A ``Yes" certificate is an assortment $S\in\cS$ for which $\sum_{i\in S} \left(w^\tau_{ij}v_{ij}/(\sum_{k\in S}v_{kj}+1)-\sigma^\tau_{ij}\right)\ge \lambda$. Hence, it remains to show that problem~\eqref{prob:mnl-decision} is an NP-hard problem. Motivated by~\citep[Theorem 1]{liu2019assortment}, we prove the NP-hardness of the problem via a reduction from the partition problem, a well-known NP-hard problem~\citep{garey-johnson'02}. 

Let us start by introducing the partition problem. In this problem we have $M$ integers, $c_1,\ldots,c_M$, such that $\sum_{k=1}^M c_k=2t$. The goal of the partition problem is to find a subset of these integers, denoted by $S$, such that the sum of integers in this set is half of the total sum of all the integers at hand, that is, $\sum_{k\in S} c_k=t$. We provide a reduction from the partition problem to problem~\eqref{prob:mnl-decision} by constructing an instance of the problem~\eqref{prob:mnl-decision} for any instance of the partition problem such the instance of problem~\eqref{prob:mnl-decision} has a solution if and only if the instance of the partition problem is partitionable.

For a given instance of the partition problem we construct the instance of problem~\eqref{prob:mnl-decision} as follows. Since problem~\eqref{prob:mnl-decision} is for a fixed customer type $j$ and iteration $\tau$, we drop the indices $j$ and $\tau$ for the rest of this proof. We have $M$ items such that for each item $1\le i\le M$ we have $w_i=1$, $v_i=c_i/t$ and $\sigma_i=c_i/(4t)$; moreover, we set $\lambda=1/4$. Therefore, problem~\eqref{prob:mnl-decision} becomes  
\begin{align*}
\max_{S\in \cS} \sum_{i\in S} \left(\frac{\frac{c_i}{t}}{\sum_{k\in S}\frac{c_i}{t}+1}-\frac{c_i}{4t}\right)\ge \frac{1}{4}.
\end{align*}
After multiplying both sides of the above by $\sum_{k\in S}\frac{c_i}{t}+1$ and simplification we have
\begin{align}
\max_{S\in \cS} 
\frac{3\sum_{i\in S}c_i}{4}-\frac{(\sum_{i\in S}c_i)^2}{4t}-\frac{\sum_{i\in S}c_i}{4}\ge\frac{t}{4}.\label{eq:mnl-simplified}
\end{align}
The left-hand side of~\eqref{eq:mnl-simplified} is quadratic in $\sum_{i\in S}c_i$; thus, it can be bounded from above. Setting $y=\sum_{i\in S}c_i$, we can write the left-hand side of~\eqref{eq:mnl-simplified} as $3y/4-y^2/(4t)-y/4$. Taking derivative with respect to $y$, it can be seen that $3y/4-y^2/(4t)-y/4$ achieves its maximum at $y=t$ and the maximum at this $y$ is $t/4$. Therefore, we have the left-hand side of~\eqref{eq:mnl-simplified} is also upper-bounded by $t/4$. Since the upper and lower-bound match, for this instance of problem~\eqref{prob:mnl-decision} it has to be that $\frac{3\sum_{i\in S}c_i}{4}-\frac{(\sum_{i\in S}c_i)^2}{4t}-\frac{\sum_{i\in S}c_i}{4}=\frac{t}{4}$, which happens only if $\sum_{i\in S}c_i=t$. Recall that the goal of the partition problem was to find a subset of integers $c_1,\ldots,c_M$ that sum to $t$. In other words, this instance of problem~\eqref{prob:mnl-decision} is solvable if and only if we can also solve the partition problem on integers $c_1,\ldots,c_M$. As partition problem is NP-hard, this proves that problem~\eqref{prob:mnl-decision} is also an NP-hard problem.
\Halmos\endproof

Despite the column generation subproblem being NP-hard even for the basic MNL choice model, we derive an FPTAS (the best possible approximation for an NP-hard problem) for it via the knapsack problem. Using the knapsack problem to design an FPTAS approximation algorithms for assortment optimization problems under the MNL choice model has been utilized previously, e.g.~\citet{Desir'14} and~\citet{liu2019assortment}. We in fact use the algorithm by~\citet{Desir'14} as a subroutine in our algorithm.

As problem~\eqref{prob:subproblem-mnl} is solved for each iteration $\tau$ and customer type $j$, in the following we omit these two indices to refer to a general instance of this problem. Suppose $S^*$ is the optimal solution to problem~\eqref{prob:subproblem-mnl}. Our algorithm is based on constructing a polynomial number of knapsack problems where for each one of them a polynomial number of dynamic programs are solved. More precisely, we first guess the value of $\sum_{i\in S^*} \sigma_i$ within a factor of $(1+\epsilon)$ for some small $\epsilon>0$. Using this we then reformulate the problem as a capacity constrained assortment optimization problem.~\citet{Desir'14} have shown that the latter problem is known to have an FPTAS approximation algorithms by using the FPTAS algorithm for the knapsack problem, which we later discuss how to utilize. 

Let $\underline{\sigma}$ (resp. $\overline{\sigma}$) be the minimum (resp. maximum) among all $\sigma_i$'s. For a given $\epsilon$ we use the following guesses for $\sum_{i\in S^*} \sigma_i$
\begin{align}
\Phi_\epsilon=\{\underline{\sigma}(1+\epsilon)^{l_1}:l_1=0,\ldots,L_1\}\label{eq:Phi},
\end{align}
where $L_1=O(\log(n\overline{\sigma}/\underline{\sigma})/\epsilon)$, ensuring that the maximum value in $\Phi_\epsilon$ is $n\overline{\sigma}$, which is an upper-bound on $\sum_{i\in S^*} \sigma_i$; hence, we cover the whole search space. Note that the number of guesses for $\sum_{i\in S^*} \sigma_i$ is polynomial in the input size $n$ and $1/\epsilon$. For a fixed $\epsilon$ and each $\phi\in \Phi_\epsilon$, problem~\eqref{prob:subproblem-mnl} turns into
\begin{align}
\max_{S\in \cS} \sum_{i\in S} \frac{w_{i}v_{i}}{\sum_{k\in S}v_{k}+1}\nonumber\\
\sum_{i\in S} \sigma_i\le \phi.\label{prob:cap-mnl}
\end{align}
For problem~\eqref{prob:cap-mnl}, we use the FPTAS for the capacity constrained assortment optimization with MNL choice model problem by~\citet{Desir'14}, presented in Algorithm~\ref{alg:desir}. We then pick the $\phi$ to which an assortment $S$ that maximizes $\frac{\sum_{i\in S}w_iv_i}{\sum_{i\in S} v_i+1}-\sum_{i\in S}\sigma_i$ is associated, see Algorithm~\ref{alg:FPTAS}.

\begin{algorithm}
	\caption{FPTAS for Problem~\eqref{prob:subproblem-mnl}}\label{alg:FPTAS}
	\textbf{INPUT:} $w_i$, $v_i$, $\sigma_i$ for $i=1,\ldots,n$ and $\epsilon$
	\begin{algorithmic}[1]
		\For{$\phi\in \Phi_\epsilon$}
		\State Solve Algorithm~\ref{alg:desir} with $\phi$ as its input and let the solution be $S_{\gamma,\delta,\phi}$.
		\EndFor
		\Return assortment $S$ that maximizes $\frac{\sum_{i\in S}w_iv_i}{\sum_{i\in S} v_i+1}-\sum_{i\in S}\sigma_i$ over all assortments $\{S_{\gamma,\delta,\phi}:\gamma \in \Gamma_\epsilon,\delta \in \Delta_\epsilon, \phi\in \Phi_\epsilon\}$.
	\end{algorithmic}
\end{algorithm}	

Here, we briefly discuss the mechanism of the FPTAS for problem~\eqref{prob:cap-mnl}, see~\citet{Desir'14} for a thorougher discussion. For a given value of $\phi$, the algorithm first makes guesses for the values of $\sum_{i\in S^*} w_iv_i$ and $\sum_{i\in S^*} v_i$. Let $\underline{w}$ (resp. $\overline{w}$) be the minimum (resp. maximum) among all $w_i$'s and $v$ (resp. $V$) be the minimum (resp. maximum) among all $v_i$'s. For a choice of $\epsilon$ we use the following guesses for $\sum_{i\in S^*} w_i v_i$ and $\sum_{i\in S^*} v_i$
\begin{align}
\Gamma_\epsilon=\{\underline{w}\underline{v}(1+\epsilon)^{l_2}:l_2=0,\ldots,L_2\},\quad\text{and}\quad \Delta_\epsilon=\{\underline{v}(1+\epsilon)^{l_3}:l_3=0,\ldots,L_3\}\label{eq:Gamma-Delta},
\end{align}
where $L_2=O(\log(n\overline{w}\overline{v}/(\underline{w}\underline{v}))/\epsilon)$ and $L_3=O(\log(n\overline{v}/\underline{v})/\epsilon)$, that is, the number of guesses for $\sum_{i\in S^*} w_iv_i$ and $\sum_{i\in S^*} v_i$ is polynomial in the input size $n$ and $1/\epsilon$. Moreover, the choice of $L_2$ and $L_3$ ensures that the maximum values in $\Gamma_\epsilon$ and $\Delta_\epsilon$ are $n\overline{w}\overline{v}$ and $n\overline{v}$, respectively, which upper-bound $\sum_{i\in S^*} w_iv_i$ and $\sum_{i\in S^*} v_i$; hence, the whole search space is covered. For each $\gamma\in \Gamma_\epsilon$ and $\delta\in \Delta_\epsilon$, the following discretizations are implemented
\begin{align}
\tilde{w}_i=\lfloor\frac{nw_iv_i}{\epsilon\gamma}\rfloor,\quad\text{and}\quad\tilde{v}_i=\lceil\frac{nv_i}{\epsilon\delta}\rceil\quad\label{eq:disc}
\end{align}

\begin{algorithm}
	\caption{FPTAS for Problem~\eqref{prob:cap-mnl}~\citep{Desir'14}}\label{alg:desir}
	\textbf{INPUT:} $w_i$, $v_i$, $\sigma_i$ for $i=1,\ldots,n$, and $\epsilon$ and $\phi$
	\begin{algorithmic}[1]
		\For{$\gamma \in \Gamma_\epsilon$}
		\For{$\delta \in \Delta_\epsilon$}
		\State Use~\eqref{eq:disc} to compute $\tilde{w_i}$ and $\tilde{v_i}$.
		\For{$(a,b,c)\in[I]\times[J]\times[n]$}
		\State Compute $V(a,b,c)$.
		\If{$V(a,b,c)\le \phi$}
		\State Let $S_{\gamma,\delta}$ be the corresponding assortment.
		\EndIf
		\EndFor
		\EndFor
		\EndFor
		\Return assortment $S$ with maximum $\frac{\sum_{i\in S}w_iv_i}{\sum_{i\in S} v_i+1}$ over all assortments $\{S_{\gamma,\delta}:\gamma \in \Gamma_\epsilon,\delta \in \Delta_\epsilon\}$.
	\end{algorithmic}
\end{algorithm}	

For each $a\in[I]$ and $b\in[J]$, a dynamic program is then used to find an assortment $S$ such that
\begin{align}
\sum_{i\in S}\tilde{w}_i\ge a,\quad\text{and}\quad\sum_{i\in S}\tilde{v}_i\le b,\quad\text{and}\quad \sum_{i\in S}\sigma_i\le \phi,\label{eq:dp-const}
\end{align}
where $I=\lfloor\frac{n}{\epsilon}\rfloor-n$ and $J=\lceil\frac{n}{\epsilon}\rceil+n$. To do this, referring to $\sigma_i$ as the ``mass" of an item $i$, for each $(a,b,c)\in[I]\times[J]\times[n]$, $V(a,b,c)$ is used to denote the minimum mass subset of $\{1,\ldots,c\}$ such that~\eqref{eq:dp-const} holds. For each $(a,b,c)\in[I]\times[J]\times[n]$, $V(a,b,c)$ can be calculated using the following recursion
\begin{align*}
V(a,b,c)&=
\begin{cases}
w_1v_1 & \text{if }0\le a\le \tilde{w}_1 \text{ and } b\ge \tilde{v}_1 \\
0 & \text{if } a\le 0 \text{ and } b\ge 0 \\
\infty & \text{otherwise}
\end{cases}\\
V(a,b,c+1)&=\min\{V(a,b,c),\sigma_{c+1}+V(a-\tilde{w}_{c+1},b-\tilde{v}_{c+1},c)\}.\nonumber
\end{align*}
The following lemma discusses the performance of Algorithm~\ref{alg:desir} and the reader is referred to~\citet{Desir'14} for its details.

\begin{lemma}[\citet{Desir'14}]
	\label{lem:desir} For a given $\phi\in \Phi_\epsilon$, Algorithm~\ref{alg:desir} returns an $(1-\epsilon)$-approximation solution to problem~\eqref{prob:cap-mnl} in $O(\log(n\overline{w}\overline{v}/(\underline{w}\underline{v}))\log(n\overline{v}/\underline{v})n^3/\epsilon^4)$ number of steps.
\end{lemma}

Let $S^*$ be the optimal solution to problem~\eqref{prob:subproblem-mnl}, we define $f^*={\sum_{i\in S^*}w_iv_i}/({\sum_{i\in S^*} v_i+1})$ and $h^*=\sum_{i\in S^*}\sigma_i$. Using the above lemma, we discuss the performance and run-time of Algorithm~\ref{alg:FPTAS} in the following. 

\begin{lemma}
	\label{lem:fptas}
	If there exists a constant $\alpha>0$ such that for all instance of problem~\eqref{prob:subproblem-mnl} we have $f^*\ge (1+2/\alpha)\cdot h^*$ then there is an FPTAS for problem~\eqref{prob:subproblem-mnl}.
\end{lemma}
\proof{Proof.} Suppose for some $\phi\in \Phi_\epsilon$ we have $\phi/(1+\epsilon)\le h^*\le \phi$. We denote the assortment returned by Algorithm~\ref{alg:desir} when $\phi$ is set as its input by $S(\phi)$ and define $f(\phi)={\sum_{i\in S(\phi)}w_iv_i}/({\sum_{i\in S(\phi)} v_i+1})$ and $h(\phi)=\sum_{i\in S(\phi)}\sigma_i$. Furthermore, let $S^*(\phi)$ be optimal solution to problem~\eqref{prob:cap-mnl} and $f^*(\phi)={\sum_{i\in S^*(\phi)}w_iv_i}/({\sum_{i\in S^*(\phi)} v_i+1})$ and $h^*(\phi)=\sum_{i\in S^*(\phi)}\sigma_i$. If $h^*\le \phi$, then problem~\eqref{prob:cap-mnl} is less constrained than the optimal solution; therefore, $f^*(\phi)\ge f^*$.

By Lemma~\ref{lem:desir} we have $f/g\ge (1-\epsilon) f^*(\phi)\ge (1-\epsilon) f^*$. Furthermore, we have $h^*\ge \phi/(1+
\epsilon)$; thus, overall we have
\begin{align*}
f-h\ge (1-\epsilon)f^*-(1+\epsilon)h^*.
\end{align*}
Suppose there exists a constant $0<\alpha<1/\epsilon-1$ such that $f^*\ge (1+2/\alpha)h^*$. Then, it can be easily checked that $f-h\ge (1-(\alpha+1)\epsilon)(f^*-h^*)$. Furthermore, Algorithm~\ref{alg:FPTAS} makes $O(\log(n\overline{\sigma}/\underline{\sigma})/\epsilon)$ calls to Algorithm~\ref{alg:desir}  and by Lemma~\ref{lem:desir} the running time of Algorithm~\ref{alg:desir} is $O(\log(n\overline{w}\overline{v}/(\underline{w}\underline{v}))\log(n\overline{v}/\underline{v})n^3/\epsilon^4)$. Therefore it takes Algorithm~\ref{alg:FPTAS} $O(\log(n\overline{\sigma}/\underline{\sigma})\log(n\overline{w}\overline{v}/(\underline{w}\underline{v}))\log(n\overline{v}/\underline{v})n^3/\epsilon^5)$ runs to find an approximate solution for problem~\eqref{prob:subproblem-mnl}, which is polynomial in the input size and $1/\epsilon$. Therefore, if there exist an $0<\alpha<1/\epsilon-1$ such that $f^*\ge (1+2/\alpha)h^*$ then Algorithm~\ref{alg:FPTAS} provides an FPTAS for problem~\eqref{prob:subproblem-mnl}.
\Halmos\endproof

\begin{remark}
	The MCDLP-R is similar to the MCDLP-NR except that constraint $\sum_{S\ni i} x_j(S)\le 1$ exists in the latter but not in the former. The dual variable associated to this constraint is $\sigma_{ij}$, which means that the dual of the MCDLP-R is similar to  MCDLP-NR-D with $\sigma_{ij}=0$ for all $i$ and $j$. As a result, Lemma~\ref{lem:approx-carryover} holds for MCDLP-R as well. Moreover, for the case of MNL choice model, when offering repeated items to customers is allowed, the column generation subproblem becomes 
	\begin{align*}
	\max_{S\in \cS} \sum_{i\in S} \frac{w^\tau_{ij}v_{ij}}{\sum_{k\in S}v_{kj}+1}.
	\end{align*}
	\citet{liu-vanryzin'08} provided a polynomial time algorithm for the above optimization problem. Consequently, when offering repeated items is permitted the column generation subproblem can solved efficiently. 
\end{remark}

\section{Supplements to Numerical Experiments in Section~\ref{sec::simul}}\label{sec:apx-sec::simul}
Figure~\ref{fig:hetero-with-pricing-sc=4} shows the performance of Greedy, Conservative, Algorithm~\ref{alg:asst1} (i.e., the 9\% algorithm) and Modified Algorithm~\ref{alg:asst1} (i.e., the 15\% algorithm) on hotel data set with heterogeneous room fares for different patience levels, sizes of permissible assortments and loading factors. However, unlike Figure~\ref{fig:hetero-with-pricing-sc=2} where the scale factor was 2, here the scale factor is set as 4. In other words, there is a wider difference between the high and low fares and different room categories.

\begin{figure}[t]
	\centering
	\begin{tabular}{c c}
		\hbox
		{\subfloat[Max. assortment size=1, patience=1]{\includegraphics[width=0.4\textwidth]{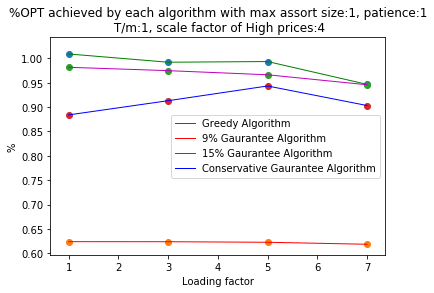}}} & 
		\subfloat[Max. assortment size=4, patience=1]{\includegraphics[width=0.4\textwidth]{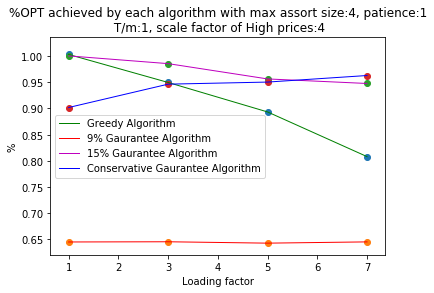}}\\
		{\subfloat[Max. assortment size=1, patience=2]{\includegraphics[width=0.4\textwidth]{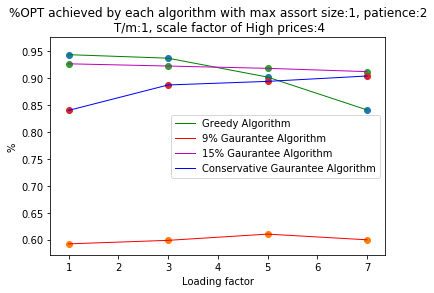}}} & 
		\subfloat[Max. assortment size=4, patience=2]{\includegraphics[width=0.4\textwidth]{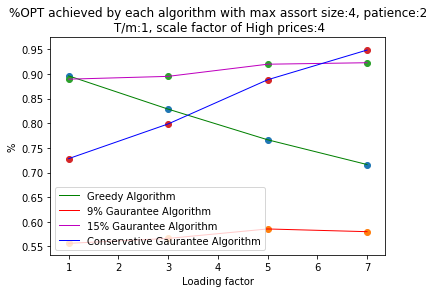}}\\
		{\subfloat[Max. assortment size=1, patience=3]{\includegraphics[width=0.4\textwidth]{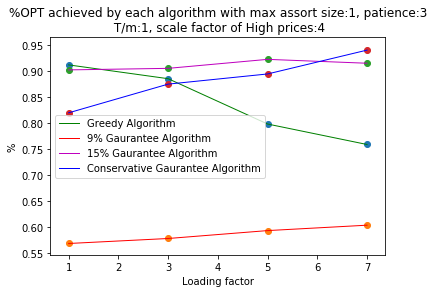}}} & 
		\subfloat[Max. assortment size=4, patience=3]{\includegraphics[width=0.4\textwidth]{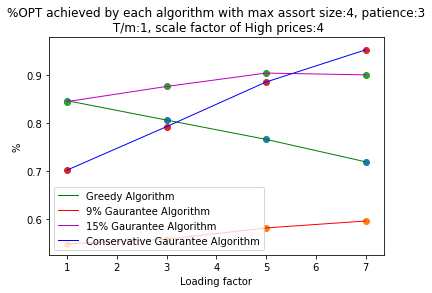}}\\
		{\subfloat[Max. assortment size=1, patience=4]{\includegraphics[width=0.4\textwidth]{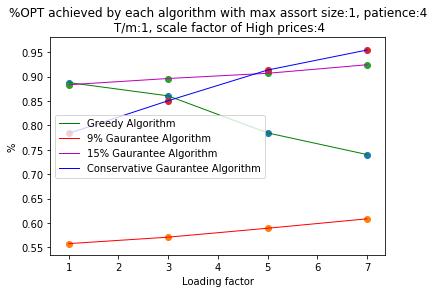}}} & 
		\subfloat[Max. assortment size=4, patience=4]{\includegraphics[width=0.4\textwidth]{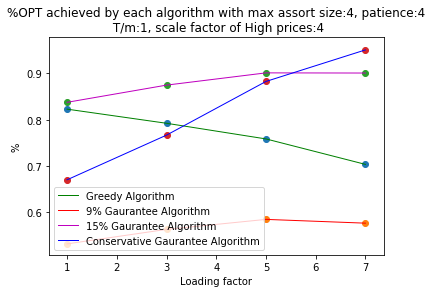}}
	\end{tabular}
	\caption{Performance of Greedy, Conservative, Algorithm~\ref{alg:asst1} (i.e., the 9\% algorithm) and Modified Algorithm~\ref{alg:asst1} (i.e., the 15\% algorithm) on hotel data set with heterogeneous room fares for different patience levels, sizes of permissible assortments for scale factor=4 over different loading factors.}
	\label{fig:hetero-with-pricing-sc=4}
\end{figure}

Similar to Figure~\ref{fig:hetero-with-pricing-sc=2}, we here see that in some cases for low loading factors the Greedy algorithm outperform the other algorithms; whereas, for large loading factors the Conservative algorithm outperform the other methods. However, in the region in-between these two extremes, the Modified Algorithm~\ref{alg:asst1} outperforms the Greedy and Conservative algorithms.

\end{APPENDICES}

\end{document}